\documentclass[12pt]{article}%
\usepackage{amsmath}
\usepackage{graphicx}
\usepackage{amsfonts}
\usepackage{color}
\usepackage{amssymb}
\usepackage{setspace}%
\providecommand{\U}[1]{\protect \rule{.1in}{.1in}}
\providecommand{\U}[1]{\protect \rule{.1in}{.1in}}
\providecommand{\U}[1]{\protect \rule{.1in}{.1in}}

\newtheorem{theorem}{Theorem}

\newtheorem{corollary}{Corollary}

\newtheorem{definition}{Definition}

\newtheorem{lemma}{Lemma}

\newtheorem{proposition}{Proposition}
\newtheorem{remark}{Remark}

\providecommand{\jelcodes}[1]{\textbf{\textit{JEL codes---}} #1}

\let \oldbibliography \thebibliography \renewcommand{\thebibliography}[1]{\oldbibliography{#1}\setlength{\itemsep}{1pt}}
\begin{document}

\title{Making Decisions under Model Misspecification\thanks{\textit{First Draft: December 2019}. We thank Attila Ambrus,
Pierpaolo Battigalli, Benjamin Brooks, Tim Christensen, Roberto Corrao,
Giacomo Lanzani, Marco Loseto, Philipp Sadowski, Todd Sarver, Jesse Shapiro as
well as the audiences at\ Advances in Decision Analysis 2019, SAET 2019, Blue
Collar Working Group 2.0, One World Mathematical Game Theory Seminar, RUD
2020, UCL-Osaka International Conference on the Mathematics for Risk and
Decisions, MUSEES 2022, Confronting Uncertainty in Climate Change, Bicocca,
Bilkent, Caltech, Duke, Glasgow, John Hopkins, LSE, Harvard, MIT, Parma, UAB,
Warwick for their very useful comments. We thank for the financial support the
Alfred P. Sloan Foundation (grant G-2018-11113), the European Research Council
(grants SDDM-TEA and INDIMACRO) and the Ministero dell'universit\`{a} e della
ricerca (grant 2017CY2NCA).}}
\author{Simone\ Cerreia--Vioglio$^{a}$, Lars Peter\ Hansen$^{b}$, Fabio
Maccheroni$^{a}$
\and and Massimo Marinacci$^{a}$\\$^{a}${\small Universit\`{a} Bocconi and Igier, }$^{b}${\small University of
Chicago}}
\date{June 2022}
\maketitle

\begin{abstract}
We use decision theory to confront uncertainty that is sufficiently broad to
incorporate \textquotedblleft models as approximations.\textquotedblright \ We
presume the existence of a featured collection of what we call
\textquotedblleft structured models\textquotedblright \ that have explicit
substantive motivations. The decision maker confronts uncertainty through the
lens of these models, but also views these models as simplifications, and
hence, as misspecified. We extend the max-min analysis under model ambiguity
to incorporate the uncertainty induced by acknowledging that the models used
in decision-making are simplified approximations. Formally, we provide an
axiomatic rationale for a decision criterion that incorporates model
misspecification concerns.

\end{abstract}

\vspace{.7cm}

\vspace{.7cm} \jelcodes{C54, D81}

\ 

\thispagestyle{empty}

\clearpage
\pagenumbering{arabic}

\textit{Come l'araba fenice:}

\textit{che vi sia, ciascun lo dice;}

\textit{dove sia, nessun lo sa.\footnote{\textquotedblleft Like the Arabian
phoenix: that it exists, everyone says; where it is, nobody
knows.\textquotedblright \ A passage from a libretto of Pietro Metastasio.}}

\section{Introduction}

The consequences of a decision may depend on exogenous contingencies and
uncertain outcomes that are outside the control of a decision maker. This
uncertainty takes on many forms. Economic applications typically feature
\emph{risk}, where the decision maker knows the correct probabilistic model
governing the contingencies but not necessarily the decision outcomes. Yet,
this is a demanding assumption. As a result, statisticians and econometricians
have long wrestled with how to confront \emph{ambiguity} over models or
unknown parameters within a model. Each model is itself a simplification or an
approximation designed to guide or enhance our understanding of some
underlying phenomenon of interest. Thus, the model, by its very nature, is
\emph{misspecified}, but in typically uncertain ways. How should a decision
maker acknowledge model misspecification in a way that guides the use of
purposefully simplified models sensibly? This concern has certainly been on
the radar screen of statisticians and control theorists, but it has been
largely absent in formal approaches to decision theory.\footnote{In Hansen
(2014)\ and Hansen and Marinacci (2016) three kinds of uncertainty are
distinguished based on the knowledge of the decision maker, the most
challenging being model misspecification viewed as uncertainty induced by the
approximate nature of the models under consideration.} Indeed, the
statisticians Box and Cox have both stated the challenge succinctly in
complementary ways:

\begin{quote}
Since all models are wrong, the scientist must be alert to what is importantly
wrong. It is inappropriate to be concerned about mice when there are tigers
abroad. Box (1976).

... it does not seem helpful just to say that all models are wrong. The very
word \textquotedblleft model\textquotedblright \ implies simplification and
idealization. The idea that complex physical, biological or sociological
systems can be exactly described by a few formulae is patently absurd. The
construction of idealized representations that capture important stable
aspects of such systems is, however, a vital part of general scientific
analysis and statistical models, especially substantive ones ... Cox (1995).
\end{quote}

\noindent While there are formulations of decision and control problems that
intend to confront model misspecification, the aim of this paper is: (i) to
develop an axiomatic approach that will provide a rigorous guide for
applications and (ii) to enrich formal decision theory when applied to
environments with uncertainty through the guise of models.

The protagonist of our analysis is a decision maker who is able to formulate
models -- for instance a policy maker having to decide a climate policy based
on existing alternative climate models -- but is concerned about their
misspecification and wants to use a decision criterion which accounts for
that. Our axiomatic analysis, which has a normative nature, aims to derive a
criterion of this kind to help the decision maker to cope with model
misspecification in a principled way. In this endeavour, we follow Hansen and
Sargent (2022) by referring to the formulated models as \textquotedblleft
structured models.\textquotedblright \ These structured models are ones that
are explicitly motivated or featured, such as ones with substantive motivation
or scientific underpinnings, consistent with the use of the term
\textquotedblleft models\textquotedblright \ by Box and Cox. They may be based
on scientific knowledge relying on empirical evidence and theoretical
arguments or on revealing parameterizations of probability models with
parameters that are interpretable to the decision maker. In posing decision
problems formally, it is often assumed, following Wald (1950), that the
correct model belongs to the set of models that decision makers posit. The
presumption that a decision maker identifies, among their hypotheses, the
correct model is often questionable -- recalling the initial quotation, the
correct model is often a decision maker phoenix. We embrace, rather than push
aside, the \textquotedblleft models are approximations\textquotedblright%
\ perspective of many applied researchers, as articulated by Box, Cox and
others. To explore misspecification formally, we introduce a potentially rich
collection of probability distributions that depict possible representations
of the data without formal substantive motivation. We refer to these as
\textquotedblleft unstructured models.\textquotedblright \ We use such
alternative models as a way to capture how models could be
misspecified.\footnote{Such a distinction is also present in earlier work by
Hansen and Sargent (2007) and Hansen and Miao (2018) but without specific
reference to the terms \textquotedblleft structured\textquotedblright \ and
\textquotedblleft unstructured.\textquotedblright}

This distinction between structured and unstructured is central to the
analysis in this paper and is used to distinguish aversion to ambiguity over
models and aversion to potential model misspecification. At a
decision-theoretic level, a proper analysis of misspecification concerns has
remained elusive so far. Indeed, many studies dealing with economic agents
confronting model misspecification still assume that they are conventional
expected utility decision makers who do not address formally potential model
misspecification concerns in their preference ordering.\footnote{See, e.g.,
Esponda and Pouzo (2016) and Fudenberg et al.\ (2017).} We extend the analysis
of Hansen and Sargent (2022) by providing an axiomatic underpinning for a
corresponding decision theory along with a representation of the implied
preferences that can guide applications. In so doing, we show an important
connection with the analysis of subjective and objective rationality of Gilboa
et al. (2010).

\paragraph{Criterion}

This paper proposes a first decision-theoretic analysis of decision making
under model misspecification. We consider a classic setup in the spirit of
Wald (1950), but relative to his seminal work we explicitly remove the
assumption that the correct model belongs to the set of posited models and we
allow for nonneutrality toward this feature. More formally, in our purely
normative approach we assume that decision makers posit a set $Q$ of
\emph{structured} (probabilistic) \emph{models} $q$ on states, motivated by
their information, but they are afraid that none of them is correct and so
face model misspecification. For this reason, decision makers contemplate what
we call \emph{unstructured models} in ranking acts $f$, according to a
conservative decision criterion\footnote{Throughout the paper $\Delta$ denotes
the set of all probabilities (Section \ref{sect:basics}).}%
\begin{equation}
V\left(  f\right)  =\min_{p\in \Delta}\left \{  \int u\left(  f\right)
dp+\min_{q\in Q}c\left(  p,q\right)  \right \}  \label{eq:repp}%
\end{equation}
To interpret this criterion, let%
\[
C\left(  p,Q\right)  =\min_{q\in Q}c\left(  p,q\right)
\]
where we presume that $C(q,Q)=0$ when $q\in Q$. In this construction,
$C\left(  p,Q\right)  $ is what we call a Hausdorff statistical set distance
between a model $p$ and the posited set $Q$ of structured models. This
distance is nonzero if and only if $p$ is unstructured, that is, $p\notin Q$.
More generally, $p$'s that are closer to the set of structured models $Q$ have
a less adverse impact on the preferences, as is evident by rewriting
(\ref{eq:repp})\ as:%
\[
V\left(  f\right)  =\min_{p\in \Delta}\left \{  \int u\left(  f\right)
dp+C\left(  p,Q\right)  \right \}
\]
This representation is a special case of the variational representation
axiomatized by Maccheroni et al.\ (2006). The unstructured models are
statistical artifacts that allow the decision maker to assess formally the
potential consequences of misspecification as captured by the construction of
$C\left(  \cdot,Q\right)  $. In this paper we provide a formal interpretation
of $C\left(  \cdot,Q\right)  $\ as an index of misspecification fear: the
lower the index, the higher the fear.\footnote{To ease terminology, we often
refer to \textquotedblleft misspecification\textquotedblright \ rather than
\textquotedblleft model misspecification.\textquotedblright}

It is because of the ability to posit a set $Q$ that the decision maker
confronts uncertainty in the guise of models, so what we may call a decision
problem under model uncertainty. In our normative approach, it is natural to
enrich the standard decision-theoretic setting by taking $Q$ as a given, a
\emph{datum} of the decision problem. For instance, in the climate policy
problem, $Q$ is the set of climate models that the policy maker considers. In
this regard, observe that we are not after detecting which choice behavior of
the decision maker may reveal model misspecification concerns, a different
revealed preference exercise that would indeed require an endogenous
$Q$.\footnote{In this exercise, the findings of Denti and Pomatto (2022) may
be useful.} In line with standard practice in applied economics, we imagine
that the substantive modeling that underlies the construction of elements of
$Q$ is simplified with an explicit structure imposed to facilitate
interpretation. Applied researchers commonly avoid reducing model building to
the construction of the complex black boxes that a purely nonparametric
exercise might well involve, especially in multivariate settings.

\paragraph{A protective belt}

When $c$ takes the entropic form $\lambda R(p||q)$, with $\lambda>0$,
criterion (\ref{eq:repp}) takes the form%
\begin{equation}
\min_{p\in \Delta}\left \{  \int u\left(  f\right)  dp+\lambda \min_{q\in
Q}R(p||q)\right \}  \label{eq:reppp}%
\end{equation}
proposed by Hansen and Sargent (2022). It is the most tractable version of
criterion (\ref{eq:repp}), which for a singleton $Q$ further reduces to a
standard multiplier criterion a la Hansen and Sargent (2001, 2008). By
exchanging orders of minimization, we preserve this tractability and provide a
revealing link to this earlier research,%
\begin{equation}
\min_{q\in Q}\left \{  \min_{p\in \Delta}\left \{  \int u\left(  f\right)
dp+\lambda R(p||q)\right \}  \right \}  \label{eq:ord-min}%
\end{equation}
The inner minimization problem gives rise to the minimization problem featured
by Hansen and Sargent (2001, 2008) to confront the potential misspecification
of a given probability model $q$.\footnote{The Hansen and Sargent (2001, 2008)
formulation of preferences builds on extensive literature in control theory
starting with Jacobson (1973)'s deterministic robustness criterion and a
stochastic extension given by Petersen et al. (2000), among several others.}
Unstructured models lack the substantive motivation of structured models, yet
in (\ref{eq:repp}) they act as a protective belt against model
misspecification. The importance of their role is proportional -- as
quantified by $\lambda$ in (\ref{eq:reppp}) -- to their proximity to the set
$Q$, a measure of their plausibility in view of the decision maker
information. The outer minimization over structured models is the counterpart
to the Wald (1950) and the more general Gilboa and Schmeidler (1989) max-min criterion.

Our analysis provides a decision-theoretic underpinning for incorporating
misspecification concerns in a distinct way from ambiguity aversion. Observe
that misspecification fear is absent when the index $\min_{q\in Q}c\left(
p,q\right)  $ equals the indicator function $\delta_{Q}$ of the set of
structured models $Q$, that is,%
\[
\min_{q\in Q}c\left(  p,q\right)  =\left \{
\begin{tabular}
[c]{ll}%
$0$ & if $p\in Q$\\
$+\infty$ & else
\end{tabular}
\  \  \  \  \  \  \  \  \  \  \  \  \  \  \  \  \  \right.
\]
In this case, which corresponds to $\lambda=+\infty$ in (\ref{eq:reppp}),
criterion (\ref{eq:repp}) takes a max-min form:
\[
V\left(  f\right)  =\min_{q\in Q}\int u\left(  f\right)  dq
\]
This max-min criterion thus characterizes decision makers who confront model
misspecification but are not concerned by it, so are misspecification neutral
(see Section \ref{sect:inter}). The criterion in (\ref{eq:repp}) may thus be
viewed as representing decision makers who use a more prudential variational
criterion (\ref{eq:repp}) than if they were to max-minimize over the set of
structured models which they posited. In particular, the farther away an
unstructured model is from the set $Q$ (so the less plausible it is), the less
it is weighted in the minimization.

\paragraph{Axiomatics}

We use the entropic case (\ref{eq:reppp}) to outline our axiomatic approach.
Start with a singleton $Q=\left \{  q\right \}  $. Decision makers, being afraid
that the reference model $q$ might not be correct, contemplate also
unstructured models $p\in \Delta$ and rank acts $f$ according to the multiplier
criterion%
\begin{equation}
V_{\lambda,q}\left(  f\right)  =\min_{p\in \Delta}\left \{  \int u\left(
f\right)  dp+\lambda R(p||q)\right \}  \label{eq:int-mul}%
\end{equation}
Here the positive scalar $\lambda$ is interpreted as an index of
misspecification fear. When decision makers posit a nonsingleton set $Q$ of
structured models, but are concerned that none of them is correct, the
multiplier criterion (\ref{eq:int-mul}) then gives only an incomplete
\emph{dominance relation}:%
\begin{equation}
f\succsim^{\ast}g\iff V_{\lambda,q}\left(  f\right)  \geq V_{\lambda,q}\left(
g\right)  \qquad \forall q\in Q \label{eq:dom-rel}%
\end{equation}
With (\ref{eq:dom-rel}), decision makers can safely regard $f$ better than
$g$. Through this ranking, the dominance relation provides a preferential
account of the probabilistic information that $Q$ represents. The dominance
relation thus naturally arises when the set $Q$ is posited.

Yet, the ranking (\ref{eq:dom-rel}) has little traction because of the
incomplete nature of $\succsim^{\ast}$. Nonetheless, the burden of choice will
have decision makers select between alternatives, be they rankable by
$\succsim^{\ast}$ or not. A cautious way to complete the binary relation
$\succsim^{\ast}$ is given by the preference $\succsim$ represented by
(\ref{eq:reppp}), or equivalently by (\ref{eq:ord-min}). This criterion thus
emerges in our analysis as a cautious completion of a multiplier dominance
relation $\succsim^{\ast}$. In this way, the probabilistic information gets
embedded in the behavioral preference. Suitably extended to a general
preference pair $\left(  \succsim^{\ast},\succsim \right)  $, we support this
approach by axiomatizing criterion (\ref{eq:repp}) as the representation of
the behavioral preference $\succsim$ and the unanimity criterion%
\[
f\succsim^{\ast}g\Longleftrightarrow \min_{p\in \Delta}\left \{  \int u\left(
f\right)  dp+c\left(  p,q\right)  \right \}  \geq \min_{p\in \Delta}\left \{  \int
u\left(  g\right)  dp+c\left(  p,q\right)  \right \}  \qquad \forall q\in Q
\]
as the representation of the incomplete dominance relation $\succsim^{\ast}$.

To sum up, our two-preference approach is motivated by the natural way with
which the dominance relation arises when the set $Q$ is posited. In this
approach, we then connect the dominance and behavioral preferences to derive
their desired representations.

\section{Preliminaries\label{sec:pre}}

\subsection{Mathematics}

\paragraph{Basic notions\label{sect:basics}}

We consider a non-trivial \emph{event} $\sigma$\emph{-algebra} $\Sigma$ in a
\emph{state space }$S$. We denote by $\Delta$ the set of finitely additive
probabilities and endow $\Delta$ and any of its subsets with the weak*
topology (see Appendix \ref{app:EM}\ for further details). In particular,
$\Delta^{\sigma}$ denotes the subset of $\Delta$ formed by the countably
additive probability measures. Given a subset $Q$ in $\Delta$, we denote by
$\Delta^{\ll}\left(  Q\right)  $ the collection of all probabilities $p$ which
are absolutely continuous with respect to $Q$, that is, if $A\in \Sigma$ and
$q\left(  A\right)  =0$ for all $q\in Q$, then $p\left(  A\right)  =0$.
Moreover, $\Delta^{\sigma}\left(  q\right)  $ denotes the set of elements of
$\Delta^{\sigma}$ which are absolutely continuous with respect to a single
$q\in \Delta^{\sigma}$, i.e., $\Delta^{\sigma}\left(  q\right)  =\left \{
p\in \Delta^{\sigma}:p\ll q\right \}  $. Unless otherwise specified, all the
subsets of $\Delta$ are to be intended non-empty.

The (convex analysis) indicator function $\delta_{C}:\Delta \rightarrow \left[
0,\infty \right]  $ of a convex subset $C$ of $\Delta$ is defined by%
\[
\delta_{C}\left(  p\right)  =\left \{
\begin{tabular}
[c]{ll}%
$0$ & if $p\in C$\\
$+\infty$ & else
\end{tabular}
\  \  \  \  \right.
\]
Throughout we adopt the convention $0\cdot \pm \infty=0$.

The \emph{effective domain} of $f:C\rightarrow \left(  -\infty,\infty \right]
$, denoted by $\operatorname*{dom}f$, is the set $\left \{  p\in C:f\left(
p\right)  <\infty \right \}  $ where $f$ takes on a finite value. The function
$f\ $is \emph{grounded} if the infimum of its image is $0$, i.e., $\inf_{p\in
C}f\left(  p\right)  =0$.

\paragraph{Statistical distances}

Consider a given collection $\mathcal{Q}$ of compact subsets $Q\, \ $of
$\Delta^{\sigma}$ that contains, as singletons $\left \{  q\right \}  $, all the
elements $q$ of the sets $Q$. Denote by $\mathcal{S}$ the set of all these
singletons.\footnote{That is, $\mathcal{S}=%
{\displaystyle \bigcup \limits_{Q\in \mathcal{Q}}}
Q$. Different collections $\mathcal{Q}$ may share the same set $\mathcal{S}$.
The smallest such collection is $\left \{  \left \{  q\right \}  :q\in
\mathcal{S}\right \}  $.} For instance, we have $\mathcal{S}=\Delta^{\sigma}$
when $\mathcal{Q}$ covers the space $\Delta^{\sigma}$.

We say that a function $C:\Delta \times \mathcal{Q}\rightarrow \left[
0,\infty \right]  $ is a \emph{statistical set distance} if:

\begin{enumerate}
\item[(C.i)] for each $Q\in \mathcal{Q}$,
\[
C\left(  p,Q\right)  =0\Longleftrightarrow p\in Q
\]

\item[(C.ii)] for each $Q,Q^{\prime}\in \mathcal{Q}$,
\[
Q\supseteq Q^{\prime}\Longrightarrow C\left(  \cdot,Q\right)  \leq C\left(
\cdot,Q^{\prime}\right)
\]

\item[(C.iii)] $C\left(  \cdot,\left \{  q\right \}  \right)  $ is lower
semicontinuous for all $q\in \mathcal{S}$.
\end{enumerate}

The first two properties make possible to interpret the quantity $C\left(
p,Q\right)  $ as a distance between the probability $p$ and the set $Q$ of
probabilities. The last property is a basic regularity condition whose
usefulness will become clear momentarily.

A statistical set distance $C$ induces a function $c:\Delta \times
\mathcal{S}\rightarrow \left[  0,\infty \right]  $ defined by
\[
c\left(  p,q\right)  =C\left(  p,\left \{  q\right \}  \right)
\]
The value $c\left(  p,q\right)  $ of this function is a distance between two
probabilities $p$ and $q$.\footnote{This distance is not a metric as, for
instance, symmetry is not required.} The induced function $c$ has the
following properties:

\begin{enumerate}
\item[(c.i)] for each $q\in \mathcal{S}$,
\[
c\left(  p,q\right)  =0\Longleftrightarrow p=q
\]

\item[(c.ii)] $c\left(  \cdot,q\right)  :\Delta \rightarrow \left[
0,\infty \right]  $ is lower semicontinuous for all $q\in \mathcal{S}$.
\end{enumerate}

Through the function $c$ we can characterize an important class of statistical
set distances. Specifically, we say that a statistical set distance
$C:\Delta \times \mathcal{Q}\rightarrow \left[  0,\infty \right]  $ is
\emph{Hausdorff} if:

\begin{enumerate}
\item[(C.iv)] $C\left(  \cdot,Q\right)  =\min_{q\in Q}c\left(  \cdot,q\right)
$ for all $Q\in \mathcal{Q}$.
\end{enumerate}

This property defines a Hausdorff-type distance between $p$ and $Q$, in which
the distance between points and sets is subsumed by that between points. This
important class of statistical set distances will be the protagonist of our
analysis. In this class, there is a duality between $c$ and $C$. Indeed, we
say that a function $c:\Delta \times \mathcal{S}\rightarrow \left[
0,\infty \right]  $ is a \emph{statistical distance}\ if it satisfies (c.i) and
(c.ii), now taken as defining properties, and if it induces a well-defined
Hausdorff statistical set distance $C:\Delta \times \mathcal{Q}\rightarrow
\left[  0,\infty \right]  $ given by
\[
C\left(  \cdot,Q\right)  =\min_{q\in Q}c\left(  \cdot,q\right)
\]
This final property is automatically satisfied when $c$ is jointly lower
semicontinuous, which is an important case in our analysis.

Statistical distances and Hausdorff statistical set distances are thus dual
notions that can be defined one in terms of the other. It is sometimes
convenient to denote by $c_{Q}$ the section $C\left(  \cdot,Q\right)
:\Delta \rightarrow \left[  0,\infty \right]  $ at $Q$ of the Hausdorff
statistical set distance $C$ induced by $c$, that is,
\[
c_{Q}\left(  \cdot \right)  =\min_{q\in Q}c\left(  \cdot,q\right)
\]
An important special case is when $\mathcal{Q}$ consists of a fixed set $Q$
along with its elements (as singletons), so that $\mathcal{S}=Q$. In this
case, to ease notation we just write
\[
c:\Delta \times Q\rightarrow \left[  0,\infty \right]
\]
This function is, for instance, the protagonist of Theorem \ref{thm:mai-rep}.
In this result we also consider a \emph{pseudo-statistical distance}, which is
a function $c:\Delta \times Q\rightarrow \left[  0,\infty \right]  $ that
satisfies all the properties of a statistical distance except (c.i), which is
weakened to: for each $q\in \mathcal{S}$ there is $p\in \Delta$ such that
$c\left(  p,q\right)  =0$.\footnote{In other words, (C.i) is satisifed for
sets which are not singletons while, when $Q$ is a singleton, (C.i) is
weakened to groundedness.}

\paragraph{Divergences}

We say that a statistical distance $c:\Delta \times \mathcal{S}\rightarrow
\left[  0,\infty \right]  $ is a \emph{divergence} if:

\begin{itemize}
\item[(c.iii)] for each $q\in \mathcal{S}$,%
\[
c\left(  p,q\right)  <\infty \Longrightarrow p\ll q
\]

\end{itemize}

Divergences thus assign an infinite penalty when $p$ is not absolutely
continuous with respect to $q$. In the important \textquotedblleft
universal\textquotedblright \ case $\mathcal{S}=\Delta^{\sigma}$, there is a
well-known class of divergences. To introduce it, given a continuous strictly
convex function $\phi:\left[  0,\infty \right)  \rightarrow \left[
0,\infty \right)  $, with $\phi \left(  1\right)  =0$ and $\lim_{t\rightarrow
\infty}\phi \left(  t\right)  /t=+\infty$, define $D_{\phi}:\Delta \times
\Delta^{\sigma}\rightarrow \left[  0,\infty \right]  $ by%
\begin{equation}
D_{\phi}\left(  p||q\right)  =\left \{
\begin{array}
[c]{ll}%
\int \phi \left(  \dfrac{dp}{dq}\right)  dq & \text{if }p\in \Delta^{\sigma
}\left(  q\right)  \\
+\infty & \text{otherwise}%
\end{array}
\right.\label{eq:div}
\end{equation}
under the conventions $0/0=0$ and $\ln0=-\infty$.\footnote{The function
$dp/dq$ is any version of the Radon-Nikodym derivative of $p$ with respect to
$q$.} It can be proved that $D_{\phi}:\Delta \times \Delta^{\sigma} \rightarrow \left[
0,\infty \right]  $ is a convex divergence, called $\phi$\emph{-divergence}%
.\footnote{See Chapter 1 of Liese and Vajda (1987). We refer to this book for
properties of $\phi$-divergences.} The most important example of $\phi
$-divergence is the \emph{relative entropy} given by$\  \phi \left(  t\right)
=t\ln t-t+1$ and denoted by $R\left(  p||q\right)  $.\footnote{Given the
conventions $0/0=0\cdot \pm \infty=0$, it holds $\phi \left(  0\right)  =1$%
.}\ Another important example is the \emph{Gini relative index} given by the
quadratic function $\phi \left(  t\right)  =\left(  t-1\right)  ^{2}/2$ and
denoted by $\chi^{2}\left(  p||q\right)  $.

Given a coefficient $\lambda \in \left(  0,\infty \right]  $, the function
$\lambda D_{\phi}:\Delta \times \Delta^{\sigma}\rightarrow \left[  0,\infty
\right]  $ is also a convex divergence.\ In particular, when $\lambda=\infty$
we have%
\[
\left(  \infty \right)  D_{\phi}\left(  p||q\right)  =\delta_{\left \{
q\right \}  }\left(  p\right)  =\left \{
\begin{tabular}
[c]{ll}%
$0$ & if $p=q$\\
$\infty$ & else
\end{tabular}
\  \  \  \  \  \  \  \  \  \  \  \  \  \  \  \  \  \  \  \  \  \right.
\]
because of the convention $0\cdot \infty=0$.

\paragraph{Variational statistical distances}

We say that a statistical set distance $C:\Delta \times \mathcal{Q}%
\rightarrow \left[  0,\infty \right]  $ is \emph{variational} if:

\begin{enumerate}
\item[(C.v)] $C\left(  \cdot,Q\right)  $ is lower semicontinuous and convex
for all $Q\in \mathcal{Q}$.
\end{enumerate}

This is a regularity condition that, when assumed, strengthens property
(C.iii). We say that a (pseudo-)statistical distance $c$ is \emph{variational}
when it induces a variational Hausdorff statistical set distance. For
instance, when $\mathcal{Q}$ consists of compact and convex subsets of
$\Delta^{\sigma}$, a statistical distance is variational if it is convex and
lower semicontinuous (see Lemma \ref{lem:app-suff-div}). Thus, $\phi
$-divergences are variational with such a $\mathcal{Q}$.

\subsection{Decision theory}

\paragraph{Setup}

We consider a generalized\ Anscombe and Aumann (1963)\ setup where a decision
maker chooses among uncertain alternatives described by (simple) acts
$f:S\rightarrow X$, which are $\Sigma$-measurable simple
(i.e.,\ finite-valued) functions from a \emph{state space} $S$ to a
\emph{consequence space} $X$. This latter set is assumed to be a
non-empty\ convex subset of a vector space (for instance, $X$ is the set of
all simple lotteries defined on a prize space). The triple
\begin{equation}
\left(  S,\Sigma,X\right)  \label{eq:standard-dp}%
\end{equation}
forms an (Anscombe-Aumann) \emph{decision framework}.

Let us denote by $\mathcal{F}$ the set of all acts. Given any consequence
$x\in X$, we denote by $x\in \mathcal{F}$ also the constant act that takes
value $x$. Thus, with a standard abuse of notation, we identify $X$ with the
subset of constant acts in $\mathcal{F}$. Given a function $u:X\rightarrow
\mathbb{R}$, we denote by $\operatorname*{Im}u$ its image. Observe that
$u\circ f$ is a simple\ real-valued $\Sigma$-measurable function.

A preference $\succsim$ is a binary relation on $\mathcal{F}$ that satisfies
the so-called \emph{basic conditions} (cf. Gilboa et al., 2010), i.e., it is:

\begin{enumerate}
\item[(i)] \emph{reflexive }and\emph{ transitive};

\item[(ii)] \emph{monotone}: for all $f,g\in \mathcal{F}$, if $f\left(
s\right)  \succsim g\left(  s\right)  $ for all $s\in S$, then $f\succsim g$;

\item[(iii)] \emph{continuous}: for all\ $f,g,h\in \mathcal{F}$, the sets
\[
\left \{  \alpha \in \left[  0,1\right]  :\alpha f+\left(  1-\alpha \right)
g\succsim h\right \}  \text{\quad and\quad}\left \{  \alpha \in \left[
0,1\right]  :h\succsim \alpha f+\left(  1-\alpha \right)  g\right \}
\]
are closed;

\item[(iv)] \emph{non-trivial}: there exist $f,g\in \mathcal{F}$ such that
$f\succ g$.
\end{enumerate}

Moreover, a preference $\succsim$ is \emph{unbounded} if, for each $x,y\in X$
with $x\succ y$, there exist $z,z^{\prime}\in X$ such that%
\[
\frac{1}{2}z+\frac{1}{2}y\succsim x\succ y\succsim \frac{1}{2}x+\frac{1}%
{2}z^{\prime}%
\]
Bets are binary acts that play a key role in decision theory. Formally, given
any two prizes $x\succ y$, a bet on an event $A$ is the act $xAy$ defined by%
\[
xAy\left(  s\right)  =\left \{
\begin{tabular}
[c]{ll}%
$x$ & if $s\in A$\\
$y$ & else
\end{tabular}
\  \  \  \  \  \  \  \right.
\]
In words, a bet on event $A$ is a binary\ act that yields a more preferred
consequence if $A$ obtains.

\paragraph{Comparative uncertainty aversion}

As in Ghirardato and Marinacci (2002), given two preferences $\succsim_{1}$
and $\succsim_{2}$ on $\mathcal{F}$, we say that $\succsim_{1}$ is \emph{more
uncertainty averse than} $\succsim_{2}$ if, for each consequence $x\in X$ and
act $f\in \mathcal{F}$,%
\[
f\succsim_{1}x\implies f\succsim_{2}x
\]
In words, a preference is more uncertainty averse than another one if,
whenever this preference is \textquotedblleft bold enough\textquotedblright%
\ to prefer an uncertain alternative over a sure one, so does the other one.

\paragraph{Decision criteria}

A complete preference $\succsim$ on $\mathcal{F}$ is \emph{variational} if it
is represented by a decision criterion $V:\mathcal{F}\rightarrow \mathbb{R}$
given by%
\begin{equation}
V\left(  f\right)  =\min_{p\in \Delta}\left \{  \int u\left(  f\right)
dp+c\left(  p\right)  \right \}  \label{eq:criterion-vp}%
\end{equation}
where the affine utility function $u$ is non-constant and the index of
uncertainty aversion $c:\Delta \rightarrow \left[  0,\infty \right]  $ is
grounded, lower semicontinuous and convex. In particular, given two unbounded
variational preferences $\succsim_{1}$ and $\succsim_{2}$ on $\mathcal{F}$
that share the same $u$, but different indexes $c_{1}$ and $c_{2}$, we have
that $\succsim_{1}$ is more uncertainty averse than $\succsim_{2}$ if and only
if $c_{1}\leq c_{2}$ (see Maccheroni et al., 2006, Propositions 6 and\ 8).

When\ the function $c$ has the entropic form $c\left(  p,q\right)  =\lambda
R\left(  p||q\right)  $ with respect to a reference probability $q\in
\Delta^{\sigma}$, criterion (\ref{eq:criterion-vp}) takes the
\emph{multiplier} form%
\[
V_{\lambda,q}\left(  f\right)  =\min_{p\in \Delta}\left \{  \int u\left(
f\right)  dp+\lambda R(p||q)\right \}
\]
analyzed by Hansen and Sargent (2001, 2008).\footnote{Strzalecki (2011)
provides the behavioral assumptions\ that characterize multiplier preferences
among variational preferences.} If, instead, the function $c$ has the
indicator form $\delta_{C}$, with $C$ compact and convex, criterion
(\ref{eq:criterion-vp}) takes the \emph{max-min} form%
\[
V\left(  f\right)  =\min_{p\in C}\int u\left(  f\right)  dp
\]
axiomatized by Gilboa and Schmeidler (1989).

All these criteria are here considered in their classical interpretation, so
Waldean for the max-min criterion, in which the elements of $\Delta$ are
interpreted as models.

\section{Models and preferences\label{sect:main}}

\subsection{Models\label{sect:models}}

The consequences of the acts among which decision makers have to choose depend
on exogenous states that are outside their control. They know that states
obtain according to a probabilistic model described by a probability measure
in $\Delta$, the so-called \emph{true }or \emph{correct model}. If decision
makers knew the true model, they would confront only risk, which is the
randomness inherent to the probabilistic nature of the model. Our decision
makers, unfortunately, may not know the true model. Yet, they are able to
posit a set of \emph{structured} probabilistic models $Q$, based on their
information (which might well include existing scientific theories, say
economic or physical), that form a set of alternative hypotheses regarding the
true model. It is a classical assumption, in the spirit of Wald (1950), in
which $Q$ is a set of posited hypotheses about the probabilistic behavior of
a, natural or social, phenomenon of interest.

A \emph{classical decision framework} is described by a quartet:%
\begin{equation}
\left(  S,\Sigma,X,Q\right)  \label{eq:dform}%
\end{equation}
in which a set $Q$ of models is added to a standard decision framework
(\ref{eq:standard-dp}), as discussed in the Introduction. The true model might
not be in $Q$, that is, the decision makers information may be unable to pin
it down. Throughout the paper we assume that decision makers know this
limitation of their information and so confront model
misspecification.\footnote{Aydogan et al. (2018) propose an experimental
setting that reveals the relevance of model misspecification for decision
making.} This is in contrast with Wald (1950) and most of the subsequent
decision-theoretic literature, which assumes that decision makers either know
the true model and so face risk or, at least, know that the true model belongs
to $Q$ and so face model ambiguity.\footnote{The model ambiguity (or
uncertainty) literature is reviewed in Marinacci (2015).}

In Theorem \ref{thm:mai-rep}, but not in Theorem \ref{thm:mai-rep-inc-fin}, we
assume that $Q$ is a convex subset of $\Delta^{\sigma}$. As usual, convexity
significantly simplifies the analysis. Yet, conceptually it is not an
innocuous property: a hybrid model that mixes two structured models can only
be less well motivated than either of them. Decision criterion (\ref{eq:repp}%
), however, accounts for the lower appeal of hybrid models when $c\left(
p,q\right)  $ is also convex in $q$ (as, for instance, when $c$ is a $\phi
$-divergence). To see why, observe that $\min_{p\in \Delta}\left \{  \int
u\left(  f\right)  dp+c\left(  p,q\right)  \right \}  $ is, for each act $f\,
$, convex in $q$. In turn, this implies that hybrid models negatively
affect\ criterion (\ref{eq:repp}). This negative impact of mixing thus
features an \textquotedblleft aversion to model
hybridization\textquotedblright \ attitude, behaviorally captured by axiom A.9.
Remarkably, the relative entropy criterion (\ref{eq:reppp}) turns out to be
neutral to model hybridization. In this important special case, the assumption
of convexity of $Q$ is actually without any loss of generality (as Appendix
\ref{app:non-cvx} clarifies).

Convexity of $Q$ can be also justified in a robust Bayesian interpretation of
our analysis that regards $Q$ as the set of the so-called predictive
distributions, which are combinations of \textquotedblleft
primitive\textquotedblright \ models (typically extreme points of $Q$) weighted
according to alternative priors $\mu$ over them. For instance, if the
primitive models describe states through i.i.d. processes, the elements of $Q$
describe them via exchangeable processes that combine primitive models and
priors (as in the Hewitt and Savage, 1955, version of the de Finetti
Representation Theorem). Under this interpretation, the $p$'s are introduced
to provide a protective shield for each of the predictive distributions
constructed from the alternative priors that are considered.

In this robust Bayesian vein, a second approach to constructing a convex $Q$
is based on the potential misspecification of the likelihood along with
uncertainty over the choice of a prior distribution. Let $Q^{0}=\left \{
q_{\theta}\right \}  _{\theta \in \Theta}$ be a parameterized family where each
$\theta$ denotes an alternative structured model. For simplicity, we consider
the case of a finite collection of such models, that is, $\Theta$ has $n$
elements. When needed, we identify $Q^{o}$ with the vector $\left(  q_{\theta
}\right)  _{\theta \in \Theta}$ of the Cartesian product $\Delta^{n}$ of
simplexes. Each prior $\mu$ on $\Theta$ induces a predictive probability:
\begin{equation}
q=\sum_{\theta \in \Theta}q_{\theta}\mu_{\theta} \label{predict_q}%
\end{equation}
We consider a convex family of priors $\mu \in \Pi$. The set of predictive
probabilities, $Q$, formed in this manner inherits the convexity of $\Pi$. One
possible choice of $\Pi$ is the set of all possible prior distributions over
$\Theta$ giving rise to $Q=\operatorname{co}Q^{o}$ (we include our more
general construction of $Q\subseteq \operatorname{co}Q^{o}$ to capture the
perspective of a robust Bayesian with prior uncertainty). Assume that the
reductive map $\mu \mapsto \sum_{\theta \in \Theta}q_{\theta}\mu_{\theta}$ is
injective (so bijective).\footnote{For instance, this is the case when the
structured models $q_{\theta}$ are suitably orthogonal. Lemma 1 of
Cerreia-Vioglio et al. (2013) characterizes the injectivity of the reductive
map.} Each predictive $q$ is thus represented with a unique prior $\mu^{q}%
\in \Pi$ that quantifies a possible belief of the decision maker over the
structured models of substantive interest.

The decision maker entertains misspecified likelihoods denoted $p_{\theta}$
over $\theta$ using, to fix ideas, a $\phi$-divergence $D_{\phi}(p_{\theta
}\Vert q_{\theta})$. This calculation depends on $\theta$. Alternative priors
in $\Pi$ provide alternative ways to average across the $\theta$-specific
divergences. For a given prior $\mu$, we construct the composite divergence
as:
\begin{equation}
\sum_{\theta \in \Theta}D_{\phi}(p_{\theta}\Vert q_{\theta})\mu_{\theta}
\label{weighted_divergence}%
\end{equation}
This formula, considered in Hansen and Sargent (2022, 2022b), gives a measure
of potential likelihood misspecification for a pre-specified predictive $q$
given by (\ref{predict_q}). For implementing the decision formulation in
practice, it would suffice to stop here while letting the decision maker
investigating prior sensitivity by searching over misspecified likelihoods and
priors in the set $\Pi$. Moreover, notice that for a given family of
$p_{\theta}$, minimizing (\ref{weighted_divergence}) over all possible priors
will result in a degenerate prior putting all the weight on the structured
model $\theta$ that is least misspecified according to $D_{\phi}(p_{\theta
}\Vert q_{\theta})$. Only when multiple models have the same low measure of
misspecification will the minimization include non-degenerate priors over models.

The weighted divergence (\ref{weighted_divergence}) implies a divergence
between $q$ and a predictive $p$ formed with the same prior $\mu^{q}$ as $q$:
\[
p=\sum_{\theta \in \Theta}p_{\theta}\mu_{\theta}^{q}%
\]
There will be multiple ways to represent this predictive distribution. For
instance, $p_{\theta}=p$ for all $\theta \in \Theta$ is one obvious choice.
Thus, there is an induced distance between predictive distributions given, for
each $\left(  p,q\right)  \in \Delta \times Q$, by
\begin{equation}
c(p,q)=\min_{\left(  p_{\theta}\right)  _{\theta \in \Theta}\in \Delta^{n}%
:p=\sum_{\theta \in \Theta}p_{\theta}\mu_{\theta}^{q}}\sum_{\theta \in \Theta
}D_{\phi}(p_{\theta}\Vert q_{\theta})\mu_{\theta}^{q} \label{ex:diverge}%
\end{equation}
This distance is a lower semicontinuous and convex variational divergence, as
detailed in Lemma \ref{lem:Lar-dis} of Appendix \ref{app:sta-dis} with a
general statistical distance playing the role of $D_{\phi}$ here.

\subsection{Preferences}

We consider a two-preference setup, as in Gilboa et al. (2010), with a mental
preference $\succsim^{\ast}$ and a behavioral preference $\succsim$.

\begin{definition}
A preference $\succsim$ is (\emph{subjectively}) \emph{rational} if it is:

\begin{enumerate}
\item[a.] complete;

\item[b.] risk independent: for all\ $x,y,z\in X$ and $\alpha \in \left(
0,1\right)  $, if\ $x\sim y$ then\ $\alpha x+\left(  1-\alpha \right)
z\sim \alpha y+\left(  1-\alpha \right)  z$.
\end{enumerate}
\end{definition}

The behavioral preference $\succsim$ governs the decision maker choice
behavior and so it is natural to require it to be complete because,
eventually, the decision maker has to choose between alternatives (burden of
choice). It is subjectively rational because, in an \textquotedblleft
argumentative\textquotedblright \ perspective, the decision maker cannot be
convinced that it leads to incorrect\ choices. Risk independence ensures that
$\succsim$ is represented on the space of consequences $X$ by an affine
utility function $u:X\rightarrow \mathbb{R}$, for instance an expected utility
functional when $X$ is the set of simple lotteries. So, risk is addressed in a
standard way and we abstract from non-expected utility issues.

The mental preference $\succsim^{\ast}$ on $\mathcal{F}$ represents the
decision maker's \textquotedblleft genuine\textquotedblright \ preference over
acts, so it has the nature of a dominance relation for the decision maker. As
such, it might well not be complete because of the decision maker inability to
compare some pairs of acts.

\begin{definition}
A preference $\succsim^{\ast}$ is a \emph{dominance relation} (or is
\emph{objectively rational}) if it is:

\begin{enumerate}
\item[a.] c-complete: for all\ $x,y\in X$, $x\succsim^{\ast}y$ or
$y\succsim^{\ast}x$;

\item[b.] completeness: when $Q$ is a singleton, for all $f,g\in \mathcal{F}$,
$f\succsim^{\ast}g$ or $g\succsim^{\ast}f$;

\item[c.] weak c-independent: for all\ $f,g\in \mathcal{F}$, $x,y\in X$ and
$\alpha \in \left(  0,1\right)  $,%
\[
\alpha f+(1-\alpha)x\succsim^{\ast}\alpha g+(1-\alpha)x\implies \alpha
f+(1-\alpha)y\succsim^{\ast}\alpha g+(1-\alpha)y
\]

\item[d.] convex: for all\ $f,g,h\in \mathcal{F}$ and $\alpha \in \left(
0,1\right)  $,%
\[
f\succsim^{\ast}h\text{\emph{ }and\emph{ }}g\succsim^{\ast}h\implies \alpha
f+\left(  1-\alpha \right)  g\succsim^{\ast}h
\]
\noindent
\end{enumerate}
\end{definition}

If $f\succsim^{\ast}g$ we say that $f$ \emph{dominates} $g$ (\emph{strictly}
if $f\succ^{\ast}g$). It is objectively rational because the decision maker
can convince others of its reasonableness, for instance through arguments
based on scientific theories (a case especially relevant for our purposes).
Momentarily, axiom A.3\ will further clarify its nature. The dominance
relation is, axiomatically, a variational preference which is not required to
be complete, unless $Q$ is a singleton.\footnote{Convexity is stronger than
uncertainty aversion a la Schmeidler (1989), which merely requires that
$f\sim^{\ast}g$ implies $\alpha f+\left(  1-\alpha \right)  g\succsim^{\ast}g$
for all $\alpha \in \left(  0,1\right)  $.\ Yet, convexity and uncertainty
aversion coincide under completeness (see, e.g., Lemma 56 of Cerreia-Vioglio
et al., 2011b). Nascimento and Riella (2011) study incomplete variational
preferences, but their result is not applicable to our setting because their
axioms are over lotteries of acts (and their state space is finite).} When $Q$
is a singleton, the dominance relation is complete and yet, because of model
misspecification, satisfies only a weak form of independence. In other words,
in our approach model misspecification may cause violations of the
independence axiom for the dominance relation. Later in the paper, Proposition
\ref{pro:c-ind} will show that relaxing independence to weak c-independence is
conceptually necessary as, otherwise, the behavioral preference would
be\ misspecification neutral. This is a key observation for our analysis.

\bigskip

Along with the classical decision framework\ (\ref{eq:dform}), the preferences
$\succsim^{\ast}$ and $\succsim$ form a \emph{two-preference classical
decision environment}%
\begin{equation}
\left(  S,\Sigma,X,Q,\succsim^{\ast},\succsim \right)  \label{eq:dform-pref}%
\end{equation}
The next two assumptions, which we take from Gilboa et al. (2010), connect the
two preferences $\succsim^{\ast}$ and $\succsim$.

\begin{enumerate}
\item[A.1] \emph{Consistency}: for all $f,g\in \mathcal{F}$,%
\[
f\succsim^{\ast}g\Longrightarrow f\succsim g
\]

\end{enumerate}

\noindent Consistency asserts that, whenever possible, the mental ranking
informs the behavioral one. The next condition says that the decision maker
opts, by default, for a sure alternative $x$ over an uncertain one $f$, unless
the dominance relation says otherwise.

\begin{enumerate}
\item[A.2] \emph{Caution}: for all $x\in X$ and $f\in \mathcal{F}$,%
\[
f\not \succsim ^{\ast}x\Longrightarrow x\succsim f
\]

\end{enumerate}

Unlike the previous assumptions, the next two are peculiar to our analysis.
They both link the posited set $Q$ to the two preferences $\succsim^{\ast}$
and $\succsim$ of the decision maker. We begin with the dominance relation
$\succsim^{\ast}$. Here we write $f\overset{Q}{=}g$ when $q\left(  f=g\right)
=1$ for all $q\in Q$, i.e., $f$ and $g$ are equal almost everywhere according
to each structured model.

\begin{enumerate}
\item[A.3] \emph{Objective }$Q$\emph{-coherence}:\ for all $f,g\in \mathcal{F}%
$,%
\[
f\overset{Q}{=}g\implies f\sim^{\ast}g
\]

\end{enumerate}

\noindent This axiom\ provides a preferential translation of the special
status of structured models over unstructured ones: if they all regard two
acts to be almost surely identical, the decision maker's \textquotedblleft
genuine\textquotedblright \ preference $\succsim^{\ast}$ follows suit and ranks
them indifferent.

\bigskip

{Previously, we noted that for some applications it may be important to allow
the set of structured models, $Q$, not to be convex. Nevertheless, the closed
convex hull, $\overline{\operatorname*{co}}\,Q$, of $Q$ will play an important
role in our next axiom.\footnote{The need to consider the weak$^{\text{*}}%
$-closure of the convex hull is a technical detail (with a finite set $Q$ we
can just consider convex hulls).} Even when $Q$ is not convex, we assign a
special role to the probabilities in its convex hull relative to other
unstructured models. Our rationale is that hybrid models retain an epistemic
status and are more than just statistical artifacts used to assess model
misspecification.\footnote{In the robust Bayesian perspective previously
discussed, the elements of $\overline{\operatorname*{co}}\,Q$ are the
predictive distributions determined by alternative priors over $Q$.}}

{To introduce our next axiom, recall that a rational preference $\succsim$
satisfies risk independence and thus admits an affine utility function
$u:X\rightarrow \mathbb{R}$ that can be used to represent it over consequences
{as an expected utility}.\footnote{{Under the usual identification\ of
constant acts with consequences.}} }Given a model $p\in \Delta$ and an act $f$,
we define an indifference class $X_{f}^{p}\subseteq X$ of consequences
$x_{f}^{p}$ via the equality%
\begin{equation}
u(x_{f}^{p})=\int u\left(  f\right)  dp \label{eq:uce}%
\end{equation}
We can interpret each $x_{f}^{p}$ as a consequence that would be indifferent,
so equivalent, to act $f$ if $p$ were the correct model. By constructing these
equivalent consequences for alternative acts and models, our next axiom
relates the\ posited set of models $Q$ with the behavioral preference
$\succsim$.

\begin{enumerate}
\item[A.4] {\emph{Subjective }$Q$\emph{-coherence}:\emph{\ }for all
$f\in \mathcal{F}$ and $x\in X$, we have%
\[
x\succ^{\ast}x_{f}^{p}\Longrightarrow x\succ f
\]
if and only if $p\in \overline{\operatorname*{co}}\,Q$.}
\end{enumerate}

{In words, $p\in \Delta$ is a structured or hybrid model, so belongs to
$\overline{\operatorname*{co}}\,Q$, if and only if decision makers take it
seriously, that is, they never choose an act $f$ that would be strictly
dominated if $p$ were the correct model. Such a salience of $p$ for the
decision makers'\ preference is the preferential footprint of a structured or
hybrid model that decision makers take seriously under consideration because
of its epistemic status -- as opposed to a purely unstructured model, which
they regard as a mere statistical artifact with no epistemic content.}

More can be said in the original\ Anscombe-Aumann setting with lottery-valued
acts. For a given model $p\in \Delta$ and act $f$, we construct the integral
$\int fdp$, which is a lottery that describes the prize distribution induced
by act $f$ when states are generated by model $p\in \Delta$.\footnote{For the
simple act $f=\sum_{i}1_{A_{i}}x_{i}$, by definition $\left(  \int fdp\right)
\left(  z\right)  $ is the probability $\sum_{i}p\left(  A_{i}\right)
x_{i}\left(  z\right)  $ of obtaining prize $z$ by choosing $f$ under $p$.} If
$u:X\rightarrow \mathbb{R}$ is any affine utility function that represents
$\succsim$ on $X$, then this integral obviously satisfies (\ref{eq:uce}). This
particular construction adds further clarity to axiom A.4 because it
identifies one lottery in the indifference class $X_{f}^{p}$ that depends
directly on the model $p$. This axiom can now be written as%
\[
x\succ^{\ast}\int fdp\implies x\succ f
\]
As an additional benefit, this formulation makes it clear that the definition
of $x_{f}^{p}$ is independent of the choice in (\ref{eq:uce}) of the specific
utility $u$ that represents $\succsim$ on $X$.

To conclude, observe that in the traditional purely subjective
axiomatizations, there is no way (actually, no language) to embed the
probabilistic information that $Q$ represents in the decision maker
preference.\footnote{For instance, in the Gilboa and Schmeidler (1989) seminal
axiomatization the derived set of probabilities $C$ is purely subjective.
There is no formal connection with any underlying probabilistic information,
something left to the decision maker personal, unmodelled, elaborations. A
notable exception is Gajdos et al. (2008), which considers probabilistic
information. Its analysis proceeds along lines very different from ours.} The
last two axioms provide the needed embedding, as the representation theorems
will show momentarily.

\section{Representation with given structured information\label{sec:giv-inf}}

We now show how the assumptions on the mental and behavioral preferences
permit to characterize criterion (\ref{eq:repp}) for a given set $Q$ in
$\Delta^{\sigma}$, that is, for a DM's given structured information.

To this end, throughout this section we assume that $Q$ is a compact and
convex set and we say that a function $c:\Delta \times Q\rightarrow \left[
0,\infty \right]  $ is \emph{uniquely null} if, for all\ $\left(  p,q\right)
\in \Delta \times Q$, the sets $c_{p}^{-1}\left(  0\right)  $ and $c_{q}%
^{-1}\left(  0\right)  $ are at most singletons. For instance, statistical
distances are uniquely null because of the distance property (c.i).

We are now ready to state our first representation result.

\begin{theorem}
\label{thm:mai-rep}Let $\left(  S,\Sigma,X,Q,\succsim^{\ast},\succsim \right)
$ be a two-preference classical decision environment,\ where $\left(
S,\Sigma \right)  $ is a standard Borel space. The following statements are equivalent:

\begin{enumerate}
\item[(i)] $\succsim^{\ast}$ is an unbounded dominance relation and $\succsim$
is a rational preference that are both $Q$-coherent and jointly satisfy
consistency and caution;

\item[(ii)] there exist an onto affine function $u:X\rightarrow \mathbb{R}$ and
a variational pseudo-statistical distance $c:\Delta \times Q\rightarrow \left[
0,\infty \right]  $, with $\operatorname*{dom}c_{Q}\subseteq \Delta^{\ll}\left(
Q\right)  $, such that, for all acts $f,g\in \mathcal{F}$,%
\begin{equation}
f\succsim^{\ast}g\Longleftrightarrow \min_{p\in \Delta}\left \{  \int u\left(
f\right)  dp+c\left(  p,q\right)  \right \}  \geq \min_{p\in \Delta}\left \{  \int
u\left(  g\right)  dp+c\left(  p,q\right)  \right \}  \qquad \forall q\in Q
\label{eq:Bew-var-mai}%
\end{equation}
and%
\begin{equation}
f\succsim g\Longleftrightarrow \min_{p\in \Delta}\left \{  \int u\left(
f\right)  dp+\min_{q\in Q}c\left(  p,q\right)  \right \}  \geq \min_{p\in \Delta
}\left \{  \int u\left(  g\right)  dp+\min_{q\in Q}c\left(  p,q\right)
\right \}  \label{eq:var-var-mai}%
\end{equation}

\end{enumerate}

\noindent If, in addition, $c$ is uniquely null, then it can be chosen to be a
variational statistical distance.
\end{theorem}

This result identifies, in particular, the main preferential assumptions
underlying\ a representation of the type%
\begin{equation}
V\left(  f\right)  =\min_{p\in \Delta}\left \{  \int u\left(  f\right)
dp+\min_{q\in Q}c\left(  p,q\right)  \right \}  \label{eq:rep-bis}%
\end{equation}
for the preference $\succsim$ when a set $Q$ of structured models is given.
{While this representation is of interest for a general variational
pseudo-statistical distance $c$, it is of particular interest when $c$ is a
variational statistical distance. In this case, the partial ordering
$\succsim^{\ast}$ is more easily interpreted.} Though a technical condition of
\textquotedblleft unique nullity\textquotedblright \ is imposed to pin down
statistical distances, our representation arguably{ has more general
applicability} and captures the preferential underpinning of criterion
(\ref{eq:rep-bis}).

The Hausdorff statistical set distance $\min_{q\in Q}c\left(  p,q\right)  $
between $p$ and $Q$ is strictly positive if and only if $p$ is an unstructured
model, i.e., $p\notin Q$. In particular, the more distant from $Q$ is an
unstructured model, the more it is penalized {as reflected in} the
minimization problem that criterion (\ref{eq:rep-bis}) features. In terms of
uniqueness of the representation, the variational representation $\left(
u,c_{Q}\right)  $ is unique, up to scaling, as in Maccheroni et al. (2006). As
to the uniqueness of $c$, it will be established in the richer framework of
Theorem \ref{thm:mai-rep-inc-fin}.

\paragraph{A misspecification index}

A behavioral preference $\succsim$ represented by (\ref{eq:rep-bis}) is
variational with index $\min_{q\in Q}c\left(  p,q\right)  $. So, if two
unbounded preferences $\succsim_{1}$ and $\succsim_{2}$ represented by
(\ref{eq:rep-bis}) share the same $u$ but feature different statistical
distances $\min_{q\in Q}c_{1}\left(  p,q\right)  $ and $\min_{q\in Q}%
c_{2}\left(  p,q\right)  $, then $\succsim_{1}$ is more uncertainty averse
than $\succsim_{2}$ if and only if
\[
\min_{q\in Q}c_{1}\left(  p,q\right)  \leq \min_{q\in Q}c_{2}\left(
p,q\right)
\]
In the present \textquotedblleft classical\textquotedblright \ setting we
interpret this comparative result as saying that the lower is $\min_{q\in
Q}c\left(  p,q\right)  $, the higher is the fear of misspecification.
Indeed,\ $Q$ is fixed and the differences in behavior cannot be due to model
ambiguity. We thus regard the section $c_{Q}$, i.e., the map%
\begin{equation}
p\mapsto \min_{q\in Q}c\left(  p,q\right)  \label{eq:index-mis}%
\end{equation}
as an index of aversion to model misspecification and we call it, for short, a
\emph{misspecification index}. The lower is this index, the higher is the fear
of misspecification. The index is maximal when%
\[
c_{Q}\left(  p\right)  =\delta_{Q}\left(  p\right)  =\left \{
\begin{tabular}
[c]{ll}%
$0$ & if $p\in Q$\\
$+\infty$ & else
\end{tabular}
\  \  \  \  \  \  \right.
\]
Later we will interpret this maximal case as representing a neutral attitude
toward model misspecification (cf. Definition \ref{def:missp-n}). In this
case, the decision maker does not care about unstructured models and maximally
penalizes them, so they play no role in the decision criterion. In contrast,
unstructured models are penalized less, so play a bigger role in the
criterion, when the decision maker wants to keep them on the table to express
a concern about model misspecification. Comparing two indexes, when
\[
c_{1,Q}\leq c_{2,Q}%
\]
we interpret the lower penalization of unstructured models in $c_{1,Q}$ as
modelling a higher concern for model misspecification.

\paragraph{Specifications and computability}

Two specifications of our representation are noteworthy. First, when $c$ is
the entropic statistical distance $\lambda R(p||q)$, with $\lambda \in \left(
0,\infty \right]  $, we have the following important special case of our
representation%
\begin{equation}
V\left(  f\right)  =\min_{p\in \Delta}\left \{  \int u\left(  f\right)
dp+\lambda \min_{q\in Q}R(p||q)\right \}  \label{eq:entropic-criterion}%
\end{equation}
which gives tractability to our decision criterion under model
misspecification. Specifically, for $\lambda \in \left(  0,\infty \right)
$,\footnote{When $\lambda=\infty$, we have $\min_{p\in \Delta}\left \{  \int
u\left(  f\right)  dp+\lambda \min_{q\in Q}R(p||q)\right \}  =\min_{q\in Q}\int
u\left(  f\right)  dq$. See Appendix \ref{app:non-cvx} for the simple proof of
(\ref{eq:dual-entropy}).}%
\begin{equation}
\min_{p\in \Delta}\left \{  \int u\left(  f\right)  dp+\lambda \min_{q\in
Q}R(p||q)\right \}  =\min_{q\in Q}-\lambda \log \int e^{-\frac{u\left(  f\right)
}{\lambda}}dq \label{eq:dual-entropy}%
\end{equation}
This result is well known when $Q$ is a singleton, that is, when
(\ref{eq:entropic-criterion}) is a standard multiplier criterion.

A second noteworthy special case of our representation is the Gini criterion
\begin{equation}
V\left(  f\right)  =\min_{p\in \Delta}\left \{  \int u\left(  f\right)
dp+\lambda \min_{q\in Q}\chi^{2}(p||q)\right \}  \label{eq:rob-mea-var}%
\end{equation}
Remarkably, we have
\begin{equation}
\min_{p\in \Delta}\left \{  \int u\left(  f\right)  dp+\lambda \min_{q\in Q}%
\chi^{2}(p||q)\right \}  =\min_{q\in Q}\left \{  \int u\left(  f\right)
dq-\frac{1}{2\lambda}\mathrm{Var}_{q}\left(  u\left(  f\right)  \right)
\right \}  \label{eq:rob-mea-var-bis}%
\end{equation}
for all acts $f$ for which the \emph{mean-variance} (in utils) criteria\ on
the r.h.s. are monotone. So, the Gini criterion is a monotone version of the
max-min mean-variance criterion.\footnote{At the end of Appendix
\ref{app:non-cvx} we further discuss this point.}

As to computability, in the important case when criterion (\ref{eq:repp})
features a $\phi$-divergence, like the specifications just discussed, we need
only to know the set $Q$ to compute it, no integral with respect to
unstructured models is needed. This is proved in the next result which is a
consequence of a duality formula of Ben-Tal and Teboulle (2007).\footnote{Here
$\phi^{\ast}$ denotes the convex Fenchel conjugate of $\phi$, once extended to
$\mathbb{R}$\ by setting $\phi \left(  t\right)  =+\infty$ if $t<0$. In
particular, $\phi^{\ast}$ is increasing.}

\begin{proposition}
\label{prop:comput}Given $Q\subseteq \Delta^{\sigma}$ and $\lambda \in
(0,\infty)$, for each act $f\in \mathcal{F}$ it holds%
\[
V\left(  f\right)  =\min_{p\in \Delta}\left \{  \int u\left(  f\right)
dp+\lambda \min_{q\in Q}D_{\phi}(p||q)\right \}  =\lambda \min_{q\in Q}\sup
_{\eta \in \mathbb{R}}\left \{  \eta-\int \phi^{\ast}\left(  \eta-\frac{u\left(
f\right)  }{\lambda}\right)  dq\right \}
\]

\end{proposition}

The r.h.s. formula computes criterion (\ref{eq:repp}) for $\phi$-divergences
by using only integrals with respect to structured models. This formula
substantially simplifies computations and thus confirms the analytical
tractability of the previous specifications.

\subsection{Interpretation of the decision criterion\label{sect:inter}}

In the Introduction we outlined a \textquotedblleft protective
belt\textquotedblright \ interpretation of decision criterion (\ref{eq:rep-bis}%
), i.e.,%
\[
V\left(  f\right)  =\min_{p\in \Delta}\left \{  \int u\left(  f\right)
dp+\min_{q\in Q}c\left(  p,q\right)  \right \}
\]
To elaborate, we begin by observing that the misspecification index
(\ref{eq:index-mis}) has the following bounds%
\begin{equation}
0\leq \min_{q\in Q}c\left(  p,q\right)  \leq \delta_{Q}\left(  p\right)
\qquad \forall p\in \Delta \label{eq:com-amb}%
\end{equation}
So, fear of misspecification is absent when the misspecification index is
$\delta_{Q}$ -- e.g., when $\lambda=+\infty$ in (\ref{eq:entropic-criterion})
-- in which case criterion (\ref{eq:rep-bis}) takes a Wald (1950) max-min form%
\begin{equation}
V\left(  f\right)  =\min_{q\in Q}\int u\left(  f\right)  dq \label{eq:rep-GS}%
\end{equation}
This max-min criterion characterizes a decision maker who confronts model
misspecification, but is not concerned by it, and exhibits\ only aversion to
model ambiguity. In other words, this Waldean decision maker is a natural
candidate to be (model) misspecification neutral. The next limit result
further corroborates this insight by showing that, when the fear of
misspecification vanishes, the decision maker becomes Waldean.\footnote{To
ease matters, we state the result in terms of criterion
(\ref{eq:entropic-criterion}). A general version can be easily established via
an increasing sequence of misspecification indexes.}

\begin{proposition}
\label{prop:lim}For each act $f\in \mathcal{F}$, we have%
\[
\lim_{\lambda \uparrow \infty}\min_{p\in \Delta}\left \{  \int u\left(  f\right)
dp+\lambda \min_{q\in Q}R\left(  p||q\right)  \right \}  =\min_{q\in Q}\int
u\left(  f\right)  dq
\]

\end{proposition}

These observations, via bounds and limits, call for a proper
decision-theoretic analysis of misspecification neutrality. To this end, note
that structured models may be incorrect, yet useful as Box (1976) famously
remarked. This motivates the next notion. Recall that act $xAy$, with $x\succ
y$, represents a bet on event $A$.

\begin{definition}
A preference $\succsim$ is \emph{bet-consistent} if, given any $x\succ y$,%
\[
q\left(  A\right)  \geq q\left(  B\right)  \quad \forall q\in Q\Longrightarrow
xAy\succsim xBy
\]
for all events $A,B\in \Sigma$.
\end{definition}

Under bet-consistency, a decision maker may fear model misspecification yet
regards structured models as good enough to choose to bet on events that they
unanimously rank as more likely. Preferences that are bet-consistent can be
classified as exhibiting a mild form of fear of model misspecification. The
following result shows that an important class of preferences, which includes
the ones represented by criterion (\ref{eq:entropic-criterion}), are bet-consistent.

\begin{proposition}
\label{prop:divergence-models}If $\lambda \in \left(  0,\infty \right]  $ and
$c=\lambda D_{\phi}$, then a\ preference $\succsim$ represented by
(\ref{eq:rep-bis}) is bet-consistent.
\end{proposition}

Next we substantially strengthen bet-consistency by considering all acts, not
just bets.

\begin{definition}
\label{def:missp-n}A preference $\succsim$ is (\emph{model})
\emph{misspecification neutral} if%
\[
\int u\left(  f\right)  dq\geq \int u\left(  g\right)  dq\quad \forall q\in
Q\Longrightarrow f\succsim g
\]
for all acts $f,g\in \mathcal{F}$.
\end{definition}

In this case, a decision maker trusts models enough so to follow them when
they unanimously rank pairs of acts. Fear of misspecification thus plays no
role in the decision maker preference, so it is decision-theoretically
irrelevant. For this reason, the decision maker attitude toward model
misspecification can be classified as neutral. The next result shows that this
may happen if and only if the decision maker adopts the max-min criterion
(\ref{eq:rep-GS}).

\begin{proposition}
\label{prop:model-amb}A preference $\succsim$ represented by criterion
(\ref{eq:rep-bis}) is misspecification neutral if and only if it is
represented by the max-min criterion (\ref{eq:rep-GS}).
\end{proposition}

This result provides the sought-after decision-theoretic argument for the
interpretation of the max-min criterion as the special case of decision
criterion (\ref{eq:rep-bis}) that corresponds to aversion to model ambiguity,
with no fear of misspecification.\footnote{This result actually holds without
any convexity assumption on $Q$. The same applies to Propositions
\ref{prop:comput}, \ref{prop:divergence-models} and \ref{prop:strong-dom-ch}
of this section.} As remarked in the Introduction, it suggests that a decision
maker using criterion (\ref{eq:rep-bis}) may be viewed as a decision maker
who, under model ambiguity, would max-minimize over the set of structured
models which she\ posited but that, for fear of misspecification, ends up
using the more prudential variational criterion (\ref{eq:rep-bis}).
Unstructured models lack the informational status of structured models, yet in
the criterion (\ref{eq:rep-bis}) they act as a \textquotedblleft protective
belt\textquotedblright \ against model misspecification.

Under this interpretation of the criterion (\ref{eq:rep-bis}), the special
multiplier case of a singleton $Q=\left \{  q\right \}  $ corresponds to a
decision maker who, with no fear of misspecification, would adopt the expected
utility criterion $\int u\left(  f\right)  dq$ to confront the risk inherent
to $q$. In other words, a singleton $Q$ in (\ref{eq:rep-bis}) corresponds to
an expected utility decision maker who fears misspecification.

Summing up, in our analysis decision makers adopt the max-min criterion
(\ref{eq:rep-GS}) if they either confront only model ambiguity (an information
trait) or are averse to model ambiguity with no fear of model misspecification
(a taste trait).

\subsection{Interpretation of the dominance relation\label{sect:dom}}

As just argued, the singleton $Q=\left \{  q\right \}  $ special case%
\begin{equation}
\min_{p\in \Delta}\left \{  \int u\left(  f\right)  dp+c\left(  p,q\right)
\right \}  \label{eq:eum}%
\end{equation}
of decision criterion (\ref{eq:rep-bis}) is an expected utility criterion
under fear of misspecification (of the unique posited $q$). Via the relation%
\begin{equation}
f\succsim^{\ast}g\Longleftrightarrow \min_{p\in \Delta}\left \{  \int u\left(
f\right)  dp+c\left(  p,q\right)  \right \}  \geq \min_{p\in \Delta}\left \{  \int
u\left(  g\right)  dp+c\left(  p,q\right)  \right \}  \qquad \forall q\in Q
\label{eq:strict-dom-str}%
\end{equation}
the representation theorem thus clarifies the interpretation of $\succsim
^{\ast}$ as a dominance relation under model misspecification by showing that
it amounts to uniform dominance across all structured models with respect to
criterion (\ref{eq:eum}). The preference $\succsim^{\ast}$ thus arises
naturally when a set $Q$ is posited by providing a preferential account of the
decision maker's probabilistic information that this set represents. In the
two-preference setting that we adopted, the axiomatic connections between
$\succsim^{\ast}$ and $\succsim$, via consistency and caution, then allow us
to embed this information in the behavioral preference.

It is easy to see that strict dominance amounts to (\ref{eq:strict-dom-str}),
with strict inequality for some $q\in Q$. This observation raises a question:
is there a notion of dominance that corresponds to strict inequality for all
$q\in Q$? To address this question, we introduce a \emph{strong dominance}
relation by writing $f\succ \hspace{-5pt}\succ^{\ast}g$ if, for all acts
$h,l\in \mathcal{F}$,%
\[
\left(  1-\delta \right)  f+\delta h\succ^{\ast}\left(  1-\delta \right)
g+\delta l
\]
for all small enough $\delta \in \left[  0,1\right]  $.\footnote{Strong
dominance has been introduced by Cerreia-Vioglio et al. (2020).} By taking
$h=f$ and $l=g$, we have the basic implication%
\[
f\succ \hspace{-5pt}\succ^{\ast}g\Longrightarrow f\succ^{\ast}g
\]
Strong dominance is a strengthening of strict dominance in which the decision
maker can convince others \textquotedblleft beyond reasonable
doubt.\textquotedblright \ The next characterization corroborates this
interpretation and, at the same time, answers the previous question in the
positive.\footnote{Up to an $\varepsilon$ that ensures a needed uniformity of
the strict inequality across structured models.}

\begin{proposition}
\label{prop:strong-dom-ch}Let $c:\Delta \times Q\rightarrow \left[
0,\infty \right]  $ be\ a variational statistical distance, $u:X\rightarrow
\mathbb{R}$ an onto and affine function and $\succsim^{\ast}$ an
unbounded\ dominance relation represented by (\ref{eq:strict-dom-str}). For
all acts $f,g\in \mathcal{F}$, we have $f\succ \hspace{-5pt}\succ^{\ast}g$ if
and only if there exists $\varepsilon>0$ such that$\ $%
\begin{equation}
\min_{p\in \Delta}\left \{  \int u\left(  f\right)  dp+c\left(  p,q\right)
\right \}  \geq \min_{p\in \Delta}\left \{  \int u\left(  g\right)  dp+c\left(
p,q\right)  \right \}  +\varepsilon \qquad \forall q\in Q \label{eq:epsilon-dom}%
\end{equation}

\end{proposition}

This characterization shows that $\succ^{\ast}$ and $\succ \hspace{-5pt}%
\succ^{\ast}$ agree on consequences and, more importantly, that%
\[
f\succ \hspace{-5pt}\succ^{\ast}g\Longrightarrow \min_{p\in \Delta}\left \{  \int
u\left(  f\right)  dp+c\left(  p,q\right)  \right \}  >\min_{p\in \Delta
}\left \{  \int u\left(  g\right)  dp+c\left(  p,q\right)  \right \}
\qquad \forall q\in Q
\]
At the same time, (\ref{eq:epsilon-dom})\ implies
\begin{equation}
f\succ \hspace{-5pt}\succ^{\ast}g\Longrightarrow f\succ g \label{eq:dom-rat}%
\end{equation}

We can diagram the relationships among the different dominance notions as
follows:%
\[%
\begin{tabular}
[c]{lllll}%
$\succ \hspace{-5pt}\succ^{\ast}$ & $\Longrightarrow$ & $\succ^{\ast}$ &
$\not \Longrightarrow $ & $\succ$\\
$\Downarrow$ &  & $\Downarrow$ &  & \\
$\succ$ & $\Longrightarrow$ & $\succsim$ &  &
\end{tabular}
\  \  \  \  \  \  \  \  \
\]
An instance when%
\begin{equation}
f\succ^{\ast}g\implies f\succ g \label{eq:strict-pref}%
\end{equation}
may fail is the max-min criterion (\ref{eq:rep-GS}).

We close by discussing misspecification neutrality, which in view of
Proposition \ref{prop:model-amb} is characterized by the misspecification
index $\min_{q\in Q}c\left(  p,q\right)  =\delta_{Q}\left(  p\right)  $.

\begin{lemma}
\label{lm:stat-dist-max}Let $c$\ be a variational statistical distance
$c:\Delta \times Q\rightarrow \left[  0,\infty \right]  $.\ We have $\min_{q\in
Q}c\left(  p,q\right)  =\delta_{Q}\left(  p\right)  $ if and only if, for each
$q\in Q$, $c\left(  p,q\right)  =\infty$ for all $p\notin Q$.
\end{lemma}

Misspecification neutrality is thus characterized by a statistical distance
that maximally penalizes unstructured models, which end up playing no role.
From a statistical distance angle, this confirms that misspecification
neutrality is the attitude of a decision maker who confronts model
misspecification, but does not care about it (and so has no use for
unstructured models).

This angle becomes relevant here because it shows that, under misspecification
neutrality, the representation (\ref{eq:strict-dom-str}) of the dominance
relation becomes%
\begin{equation}
f\succsim^{\ast}g\Longleftrightarrow \min_{q^{\prime}\in Q}\left \{  \int
u\left(  f\right)  dq^{\prime}+c\left(  q^{\prime},q\right)  \right \}
\geq \min_{q^{\prime}\in Q}\left \{  \int u\left(  g\right)  dq^{\prime
}+c\left(  q^{\prime},q\right)  \right \}  \qquad \forall q\in Q
\label{eq:Bew-var-cer-ind-bis}%
\end{equation}
Unstructured models play no role here. This is shown by the next result which
also demonstrates how relaxing independence to weak c-independence is
conceptually necessary. For, if the dominance relation $\succsim^{\ast}$
satisfies the stronger assumption of c-independence, then the behavioral
preference $\succsim$ is necessarily misspecification neutral.

\begin{enumerate}
\item[A.5] \emph{C-independence.} For all\ $f\in \mathcal{F}$, $x,y\in X$ and
$\alpha \in \left(  0,1\right]  $,%
\[
f\succsim^{\ast}x\iff \alpha f+\left(  1-\alpha \right)  y\succsim^{\ast}\alpha
x+\left(  1-\alpha \right)  y
\]

\end{enumerate}

When the dominance relation $\succsim^{\ast}$ is complete, our version is
equivalent to the original version of Gilboa and Schmeidler (1989). Otherwise,
ours is weaker.

\begin{proposition}
\label{pro:c-ind}Let $\left(  S,\Sigma,X,Q,\succsim^{\ast},\succsim \right)  $
be a two-preference classical decision environment,\ where $\left(
S,\Sigma \right)  $ is a standard Borel space. The following statements are equivalent:

\begin{enumerate}
\item[(i)] $\succsim^{\ast}$ is an unbounded dominance relation that satisfies
c-independence and $\succsim$ is a rational preference that are both
$Q$-coherent and jointly satisfy consistency and caution;

\item[(ii)] there exist an onto affine function $u:X\rightarrow \mathbb{R}$ and
a variational pseudo-statistical distance $c:\Delta \times Q\rightarrow \left[
0,\infty \right]  $, with $c_{Q}=\delta_{Q}$, such that, for all acts
$f,g\in \mathcal{F}$, it holds (\ref{eq:Bew-var-cer-ind-bis}) and%
\begin{equation}
f\succsim g\iff \min_{q\in Q}\int u\left(  f\right)  dq\geq \min_{q\in Q}\int
u\left(  g\right)  dq \label{eq:var-var-cer-ind}%
\end{equation}

\end{enumerate}

\noindent Moreover, $\succsim^{\ast}$ satisfies independence if and only if
$c:\Delta \times Q\rightarrow \left[  0,\infty \right]  $ can be chosen to be the
variational statistical distance $c\left(  p,q\right)  =\delta_{\left \{
q\right \}  }\left(  p\right)  $ for all $\left(  p,q\right)  \in \Delta \times
Q$.
\end{proposition}

To sum up, only a genuine variational dominance relation can accommodate fear
of model misspecification and an approach where structured models always have
a different and more relevant status than unstructured models.

The last part of the statement,\footnote{In proving this last part, we can
dispense with the assumption of $\left(  S,\Sigma \right)  $ being a standard
Borel space. Similarly, (i) would still imply (\ref{eq:var-var-cer-ind}),
again\ without any assumption on $\left(  S,\Sigma \right)  $.} which is the
version for our setting of the main result of Gilboa et al. (2010), shows that
also statistical distances play no role, so (\ref{eq:Bew-var-cer-ind-bis})
reduces to%
\[
f\succsim^{\ast}g\Longleftrightarrow \int u\left(  f\right)  dq\geq \int
u\left(  g\right)  dq\qquad \forall q\in Q
\]
when the dominance relation satisfies the independence axiom.

\section{Representation with varying structured information\label{sec:var-str}%
}

So far,\ we carried out our analysis for a given set $Q$ of structured models.
Indeed, a two-preference classical decision environment (\ref{eq:dform-pref})
should be more properly written as%
\[
\left(  S,\Sigma,X,Q,\succsim_{Q}^{\ast},\succsim_{Q}\right)
\]
with the dependence of preferences on $Q$ highlighted. Decision environments,
however, may share common state and consequence spaces, but differ on the
posited sets of structured models because of different information that
decision makers may have. It then becomes important to ensure that decision
makers use decision criteria that, across such environments, are consistent.

To address this issue, in this section we consider a family%
\[
\left \{  \left(  S,\Sigma,X,Q,\succsim_{Q}^{\ast},\succsim_{Q}\right)
\right \}  _{Q\in \mathcal{Q}}%
\]
of classical decision environments that differ in the set $Q$ of posited
models and we introduce axioms on the family $\left \{  \succsim_{Q}^{\ast
}\right \}  _{Q\in \mathcal{Q}}$ that connect these environments. We assume that
$\mathcal{Q}$ is a collection of compact subsets of $\Delta^{\sigma}$\ that
contains all singletons and that covers all doubletons, that is, for each
$q,q^{\prime}\in \Delta^{\sigma}$ there exists some $Q\in \mathcal{Q}$ such that
$\left \{  q,q^{\prime}\right \}  \subseteq Q$.\ These assumptions are
satisfied, for example, by the collection of finite sets of $\Delta^{\sigma}%
$\ as well as by the collection $\mathcal{K}$ of its compact and convex sets.

\begin{enumerate}
\item[A.6] \emph{Monotonicity} (\emph{in model ambiguity}): for all
$f,g\in \mathcal{F}$, if $Q^{\prime}\subseteq Q$ then%
\[
f\succsim_{Q}^{\ast}g\Longrightarrow f\succsim_{Q^{\prime}}^{\ast}g
\]

\end{enumerate}

\noindent According to this axiom, if the \textquotedblleft
structured\textquotedblright \ information underlying a set $Q$ is good enough
for the decision maker to establish that an act dominates another one, a
better information which decreases model ambiguity can only confirm such
judgement. Its reversal would be, indeed, at odds with the objective
rationality spirit of the dominance relation.

Next we consider a separability assumption.

\begin{enumerate}
\item[A.7] $Q$\emph{-separability}: for all $f,g\in \mathcal{F}$,%
\[
f\succsim_{q}^{\ast}g\quad \forall q\in Q\implies f\succsim_{Q}^{\ast}g
\]

\end{enumerate}

In words, an act dominates another one when it does, separately, through the
lenses of each structured model. In this axiom the incompleteness of
$\succsim_{Q}^{\ast}$ arises as that of a Paretian order over the, complete
but possibly misspecification averse, preferences $\succsim_{q}^{\ast}$
determined by the elements of $Q$.

We close with a continuity axiom. To state it, we need a last piece of
notation: we denote by $x_{f,q}$\ the consequence indifferent to act\ $f$\ for
preference $\succsim_{q}^{\ast}$.\footnote{In symbols, $f\sim_{q}^{\ast
}x_{f,q}$. In particular, $x_{f,q}$ should not be confused with $x_{f}^{q}$ as
in (\ref{eq:uce}).}

\begin{enumerate}
\item[A.8] \emph{Lower semicontinuity}: for all $x\in X$ and $f\in \mathcal{F}%
$, the set $\left \{  q\in \Delta^{\sigma}:x\succsim_{q}^{\ast}x_{f,q}\right \}
$ is closed.
\end{enumerate}

The next class of two-preference families $P_{\mathcal{Q}}=\left \{  \left(
\succsim_{Q}^{\ast},\succsim_{Q}\right)  \right \}  _{Q\in \mathcal{Q}}$ builds
on the properties that we have introduced.

\begin{definition}
\label{def:rob}A two-preference family $P_{\mathcal{Q}}$ is
(\emph{misspecification}) \emph{robust} if:

\begin{enumerate}
\item[(i)] $\left \{  \succsim_{Q}^{\ast}\right \}  _{Q\in \mathcal{Q}}$ is
monotone, separable, and lower semicontinuous;

\item[(ii)] for each $Q\in \mathcal{Q}$, $\succsim_{Q}^{\ast}$ is an
unbounded\ dominance relation, $\succsim_{Q}$ is a rational preference, both
are $Q$-coherent and jointly satisfy caution and consistency.
\end{enumerate}
\end{definition}

We can now state our first representation result.

\begin{theorem}
\label{thm:mai-rep-inc-fin}Let $P_{\mathcal{Q}}$ be a two-preference family.
The following statements are equivalent:

\begin{enumerate}
\item[(i)] $P_{\mathcal{Q}}$ is robust;

\item[(ii)] there exist an onto affine $u:X\rightarrow \mathbb{R}$ and a lower
semicontinuous divergence $c:\Delta \times \Delta^{\sigma}\rightarrow \left[
0,\infty \right]  $, convex in $p$, such that, for each $Q\in \mathcal{Q}$,%
\[
f\succsim_{Q}^{\ast}g\Longleftrightarrow \min_{p\in \Delta}\left \{  \int
u\left(  f\right)  dp+c\left(  p,q\right)  \right \}  \geq \min_{p\in \Delta
}\left \{  \int u\left(  g\right)  dp+c\left(  p,q\right)  \right \}
\qquad \forall q\in Q
\]
and%
\[
f\succsim_{Q}g\Longleftrightarrow \min_{p\in \Delta}\left \{  \int u\left(
f\right)  dp+\min_{q\in Q}c\left(  p,q\right)  \right \}  \geq \min_{p\in \Delta
}\left \{  \int u\left(  g\right)  dp+\min_{q\in Q}c\left(  p,q\right)
\right \}
\]
for all acts $f,g\in \mathcal{F}$.
\end{enumerate}

\noindent Moreover, $u$ is cardinal and, given $u$, $c$ is unique.
\end{theorem}

A robust $P_{\mathcal{Q}}$ is thus characterized by a utility and divergence
pair $\left(  u,c\right)  $ that, consistently across decision environments,
represents each $\succsim_{Q}^{\ast}$ via the unanimity rule
(\ref{eq:Bew-var-mai}) and each $\succsim_{Q}$ via the decision criterion
\begin{equation}
V_{Q}\left(  f\right)  =\min_{p\in \Delta}\left \{  \int u\left(  f\right)
dp+\min_{q\in Q}c\left(  p,q\right)  \right \}  \label{eq:var-rep-two-pre}%
\end{equation}
An unstructured model $p$ may play a role in this criterion when $c\left(
p,q\right)  <\infty$ for some structured model $q$, that is, when it has a
finite distance from a structured model.

In this representation theorem we do not make any convexity assumption on the
sets of structured models. Next we sharpen this result by assuming that they
are compact and convex subsets of $\Delta^{\sigma}$. We introduce a new axiom
based on this added structure on sets of models. Under the hypotheses of
Theorem \ref{thm:mai-rep-inc-fin}, all dominance relations $\succsim_{Q}%
^{\ast}$ agree on $X$ and so we can just write $\succsim^{\ast}$,\ dropping
the subscript\ $Q$.

\begin{enumerate}
\item[A.9] \emph{Model hybridization aversion}: for all $q,q^{\prime}\in
\Delta^{\sigma}$, $\lambda \in \left(  0,1\right)  $ and $f\in \mathcal{F}$,%
\[
\lambda x_{f,q}+\left(  1-\lambda \right)  x_{f,q^{\prime}}\succsim^{\ast
}x_{f,\lambda q+\left(  1-\lambda \right)  q^{\prime}}%
\]

\end{enumerate}

According to this axiom, the decision maker dislikes, \emph{ceteris paribus},
facing a hybrid structured model $\lambda q+\left(  1-\lambda \right)
q^{\prime}$ that, by mixing two structured models $q$ and $q^{\prime}$, could
only have a less substantive motivation (cf. Section \ref{sect:models}).

The next result extends Theorem \ref{thm:mai-rep} to families of decision
environments. It also sharpens Theorem \ref{thm:mai-rep-inc-fin} by dealing
with sets of structured models that are also convex; in particular, here we
get\ a variational divergence.

\begin{proposition}
\label{pro:mai-rep-inc}Let $P_{\mathcal{K}}$ be a two-preference family. The
following statements are equivalent:

\begin{enumerate}
\item[(i)] $P_{\mathcal{K}}$ is robust and model hybridization averse;

\item[(ii)] there exist an onto affine $u:X\rightarrow \mathbb{R}$ and a lower
semicontinuous and convex variational divergence $c:\Delta \times \Delta
^{\sigma}\rightarrow \left[  0,\infty \right]  $ such that, for each
$Q\in \mathcal{K}$,%
\[
f\succsim_{Q}^{\ast}g\Longleftrightarrow \min_{p\in \Delta}\left \{  \int
u\left(  f\right)  dp+c\left(  p,q\right)  \right \}  \geq \min_{p\in \Delta
}\left \{  \int u\left(  g\right)  dp+c\left(  p,q\right)  \right \}
\qquad \forall q\in Q
\]
and%
\[
f\succsim_{Q}g\Longleftrightarrow \min_{p\in \Delta}\left \{  \int u\left(
f\right)  dp+\min_{q\in Q}c\left(  p,q\right)  \right \}  \geq \min_{p\in \Delta
}\left \{  \int u\left(  g\right)  dp+\min_{q\in Q}c\left(  p,q\right)
\right \}
\]
for all acts $f,g\in \mathcal{F}$.
\end{enumerate}

\noindent Moreover, $u$ is cardinal and, given $u$, $c$ is unique.
\end{proposition}

This result ensures that the decision maker uses consistently criterion
(\ref{eq:repp}) across decision environments. In particular, the same
statistical distance function is used (e.g., the relative entropy). Moreover,
axioms A.6-A.9\ further clarify the nature of structured models and their
connection with the dominance relation.

Besides its broader scope, Proposition \ref{pro:mai-rep-inc} improves Theorem
\ref{thm:mai-rep} on two counts. First, it features a statistical distance
without the need of a unique nullity condition. Second, it contains a sharp
uniqueness part. The cost of these improvements is a less parsimonious setting
in which the set $Q$ is permitted to vary across the collection $\mathcal{K}$
of compact and convex subsets of $\Delta^{\sigma}$.

\section{Admissibility\label{sect:dp}}

A \emph{two-preference classical decision problem} is a septet%
\begin{equation}
\left(  F,S,\Sigma,X,Q,\succsim_{Q}^{\ast},\succsim_{Q}\right)
\label{eq:dp-q}%
\end{equation}
where $F\subseteq \mathcal{F}$ is a non-empty choice set formed by the acts
among which a decision maker has actually to choose, and the preferences
$\succsim_{Q}^{\ast}$ and $\succsim_{Q}$ are represented as in Theorem
\ref{thm:mai-rep-inc-fin}-(ii).

Given a set $Q$ in $\mathcal{Q}$, the decision maker chooses the best act in
$F$ according to $\succsim_{Q}$. In particular, the \emph{value function}
$v:\mathcal{Q}\rightarrow \left(  -\infty,\infty \right]  $ is given by%
\begin{equation}
v\left(  Q\right)  =\sup_{f\in F}\min_{p\in \Delta}\left \{  \int u\left(
f\right)  dp+\min_{q\in Q}c\left(  p,q\right)  \right \}  \label{eq:dec-pro}%
\end{equation}
Yet, it is the dominance relation $\succsim_{Q}^{\ast}$ that permits to
introduce admissibility.

\begin{definition}
An act $f\in F$ is (\emph{weakly}) \emph{admissible} if there is no act $g\in
F$ that (strongly) strictly dominates $f$.
\end{definition}

To relate this notion to the usual notion of admissibility,\footnote{See,
e.g., Ferguson (1967) p. 54. Weak admissibility is, \emph{mutatis mutandis},
related via formula (\ref{eq:epsilon-dom}) to the notion of extended
admissibility studied in Blackwell and Girschick (1954), Heath and Sudderth
(1978) and, more recently, in Duanmu and Roy (2021). This connection was
pointed out to us by Jesse Shapiro. A statistical risk version of Proposition
\ref{prop:strong-dom-ch} provides a preferential foundation for extended
admissibility.} observe that $g\succ_{Q}^{\ast}f$ amounts to
\[
\min_{p\in \Delta}\left \{  \int u\left(  g\right)  dp+c\left(  p,q\right)
\right \}  \geq \min_{p\in \Delta}\left \{  \int u\left(  f\right)  dp+c\left(
p,q\right)  \right \}  \qquad \forall q\in Q
\]
with strict inequality for some $q\in Q$. We are thus purposefully defining
admissibility in terms of the structured models $Q$, not the larger class of
models $\Delta$, with a model-by-model adjustment for misspecification that
makes our notion different from the usual one.

The next result relates optimality and admissibility.

\begin{proposition}
\label{pro:adm}Consider a decision problem (\ref{eq:dp-q}).

\begin{enumerate}
\item[(i)] Optimal acts are weakly admissible. They are admissible provided
(\ref{eq:strict-pref}) holds.

\item[(ii)] Unique optimal acts are admissible.
\end{enumerate}
\end{proposition}

Optimal acts (if exist) might not be admissible because the max-min nature of
decision criterion (\ref{eq:repp}) may lead to violations of
(\ref{eq:strict-pref}). Yet, the last result ensures that they belong to the
collection of weakly admissible acts%
\[
F_{Q}^{\ast}=\left \{  f\in F:\nexists g\in F,g\succ \hspace{-5pt}\succ
_{Q}^{\ast}f\right \}
\]

Next we build on this property to establish a comparative statics exercise
across decision problems (\ref{eq:dp-q}) that differ on the posited set $Q$ of
structured models.

\begin{proposition}
\label{pro:com-sta-val}We have%
\[
Q\subseteq Q^{\prime}\Longrightarrow v\left(  Q\right)  \geq v\left(
Q^{\prime}\right)
\]
and%
\[
v\left(  Q\right)  =\max_{f\in F_{Q}^{\ast}}\min_{p\in \Delta}\left \{  \int
u\left(  f\right)  dp+\min_{q\in Q}c\left(  p,q\right)  \right \}
\]
provided the $\sup$ in (\ref{eq:dec-pro}) is achieved.
\end{proposition}

Smaller sets of structured models are, thus, more valuable. Indeed, in
decision problems that feature a larger set of structured models -- so, a more
discordant information -- the decision maker exhibits, \emph{ceteris paribus},
a higher uncertainty aversion due to a larger model ambiguity:%
\begin{equation}
Q\subseteq Q^{\prime}\Longrightarrow \min_{q\in Q}c\left(  p,q\right)  \geq
\min_{q\in Q^{\prime}}c\left(  p,q\right)  \label{eq:admin-comp}%
\end{equation}
In turn, this easily implies $v\left(  Q\right)  \geq v\left(  Q^{\prime
}\right)  $, as the proof shows.

In the comparison (\ref{eq:admin-comp}), the divergence $c$ is invariant as we
change the set of structured models. For this reason, in Proposition
\ref{pro:com-sta-val} a larger set of structured models implies a higher
uncertainty aversion due to model ambiguity and aversion to it (as is the case
for max-min utility).\footnote{See Ghirardato and Marinacci (2002).} This
invariance, however, is not an innocuous assumption as it rules out the
possibility that the divergence becomes larger when an enlarged set of
structured models reduces misspecification concerns.\footnote{We thank Tim
Christensen for having alerted us on this issue.} For instance, the entropic
divergence may feature a higher $\lambda$\ when $Q$ gets larger, something
that may reverse the inequality (\ref{eq:admin-comp}) by making more valuable
larger sets of structured models. Nevertheless, with an invariant $c$ any
probability measure outside the set of structured models will necessarily be
closer to a larger set of such models, as captured by the divergence. In this
sense, increasing the set of structured models may diminish misspecification
concerns even under the maintained invariance.

\section{Beyond caution}

Caution is the axiom behind the prudential nature of our representations
results: Theorems \ref{thm:mai-rep}\ and \ref{thm:mai-rep-inc-fin}. It is
natural to wonder about what happens when we remove this assumption. We
formally establish a representation result that extends Theorem
\ref{thm:mai-rep-inc-fin}. A similar version can be discussed within the setup
of Section \ref{sec:giv-inf}.\ To discuss our more general result, we
introduce a new class of two-preference families.

\begin{definition}
A two-preference family $P_{\mathcal{Q}}$ is (\emph{misspecification})
\emph{sensitive} if:

\begin{enumerate}
\item[(i)] $\left \{  \succsim_{Q}^{\ast}\right \}  _{Q\in \mathcal{Q}}$ is
monotone, separable, and lower semicontinuous;

\item[(ii)] for each $Q\in \mathcal{Q}$, $\succsim_{Q}^{\ast}$ is an
unbounded\ dominance relation, $\succsim_{Q}$ is a monotone\ binary relation,
both are $Q$-coherent when restricted to singletons and jointly satisfy consistency.
\end{enumerate}
\end{definition}

Compared to the notion of robust family (cf. Definition \ref{def:rob}), we
made three changes. The most important is that we removed caution. Moreover,
we require $\succsim_{Q}$ to be only a monotone\ binary relation and
$Q$-coherence to hold only if $Q$ is a singleton.\footnote{A binary relation
$\succsim$ over acts $\mathcal{F}$ is a monotone binary relation if it is a
non-trivial complete preorder which satisfies monotonicity, continuity and
independence over $X$, and is solvable, that is, for each $f\in \mathcal{F}$
there exists $x\in X$ such that $f\sim x$. In particular, a monotone binary
relation is a rational preference if and only if it satisfies continuity over
$\mathcal{F}$.} These two latter changes are immaterial as we will later
discuss, when caution is present.\ Thus, given a sensitive $P_{\mathcal{Q}}$
and $Q$, this implies that the dominance relation $\succsim_{Q}^{\ast}$ keeps
on being represented as before, that is,%
\[
f\succsim_{Q}^{\ast}g\iff \min_{p\in \Delta}\left \{  \int u\left(  f\right)
dp+c\left(  p,q\right)  \right \}  \geq \min_{p\in \Delta}\left \{  \int u\left(
g\right)  dp+c\left(  p,q\right)  \right \}  \quad \forall q\in Q
\]
In particular, given an act $f$, we have an evaluation\ map $q\mapsto
\min_{p\in \Delta}\left \{  \int u\left(  f\right)  dp+c\left(  p,q\right)
\right \}  $ which belongs to $B\left(  Q\right)  $: the collection of all
real-valued bounded functions on $Q$. Our criterion (\ref{eq:var-rep-two-pre}%
)\ emerges when these evaluations are aggregated via the minimum on $Q$. But,
a priori, less extreme stances are conceivable. This would require dropping
caution as the next two results\ shows.

\begin{proposition}
\label{pro:com-no-cau}Let $P_{\mathcal{Q}}$ be a two-preference family. The
following statements are equivalent:

\begin{enumerate}
\item[(i)] $P_{\mathcal{Q}}$ is sensitive;

\item[(ii)] there exist an onto affine $u:X\rightarrow \mathbb{R}$, a lower
semicontinuous divergence $c:\Delta \times \Delta^{\sigma}\rightarrow \left[
0,\infty \right]  $, convex in $p$, and a normalized and\ monotone functional
$J_{Q}:B\left(  Q\right)  \rightarrow \mathbb{R}$ such that for each
$Q\in \mathcal{Q}$,%
\begin{equation}
f\succsim_{Q}^{\ast}g\Longleftrightarrow \min_{p\in \Delta}\left \{  \int
u\left(  f\right)  dp+c\left(  p,q\right)  \right \}  \geq \min_{p\in \Delta
}\left \{  \int u\left(  g\right)  dp+c\left(  p,q\right)  \right \}
\qquad \forall q\in Q \label{eq:Bew-uti-rep}%
\end{equation}
and%
\begin{equation}
f\succsim_{Q}g\iff J_{Q}\left(  \min_{p\in \Delta}\left \{  \int u\left(
f\right)  dp+c\left(  p,\cdot \right)  \right \}  \right)  \geq J_{Q}\left(
\min_{p\in \Delta}\left \{  \int u\left(  g\right)  dp+c\left(  p,\cdot \right)
\right \}  \right)  \label{eq:uti-rep}%
\end{equation}
for all acts $f,g\in \mathcal{F}$.
\end{enumerate}

\noindent Moreover, $u$ is cardinal and, given $u$, $c$ is unique.
\end{proposition}

Decision-theoretically, Theorem \ref{thm:mai-rep-inc-fin} is the special case
of this result when $\succsim_{Q}^{\ast}$ and $\succsim_{Q}$\ jointly satisfy
caution for all $Q\in \mathcal{Q}$, as Corollary \ref{cor:com-no-cau+cau} below
shows. Analytically, it corresponds to the special case where $J_{Q}$\ is the
minimum over $Q$ of the maps%
\begin{equation}
q\longmapsto \min_{p\in \Delta}\left \{  \int u\left(  f\right)  dp+c\left(
p,q\right)  \right \}  \label{eq:maps-no-caution}%
\end{equation}
In this case, by exchanging the order of minima, (\ref{eq:uti-rep}) reduces to
the decision criterion (\ref{eq:var-rep-two-pre}).

An altogether different case is when $J_{Q}$ is a quasi-arithmetic mean over
the maps (\ref{eq:maps-no-caution}), so that (\ref{eq:uti-rep}) now becomes%
\begin{equation}
V_{Q}\left(  f\right)  =\phi_{Q}^{-1}\left(  \int_{Q}\phi_{Q}\left(
\min_{p\in \Delta}\left \{  \int_{S}u\left(  f\left(  s\right)  \right)
dp\left(  s\right)  +c\left(  p,q\right)  \right \}  \right)  d\mu_{Q}\left(
q\right)  \right)  \label{eq:quasi}%
\end{equation}
where $\phi_{Q}:\mathbb{R}\rightarrow \mathbb{R}$ is strictly increasing and
continuous and $\mu_{Q}\in \Delta \left(  Q\right)  $.\footnote{Here, with a
small abuse of notation, the set $\Delta \left(  Q\right)  $ denotes the set of
all Borel probability \textit{measures} over $Q$. In particular, in order to
discuss this functional form, we need the maps defined as
in\ (\ref{eq:maps-no-caution}), to be Borel measurable: a property which is
guaranteed by the joint lower semicontinuity of $c$.} Momentarily, this
criterion will be the protagonist of the next section. We conclude by
observing that the leading assumption driving our representation results is
indeed caution while continuity of $\succsim_{Q}$ and $Q$-coherence could have
been dispensed with.

\begin{corollary}
\label{cor:com-no-cau+cau}Let $P_{\mathcal{Q}}$ be a two-preference family.
The following statements are equivalent:

\begin{enumerate}
\item[(i)] $P_{\mathcal{Q}}$ is robust;

\item[(ii)] $P_{\mathcal{Q}}$ is sensitive and $\succsim_{Q}^{\ast}$ and
$\succsim_{Q}$ jointly satisfy caution for all $Q\in \mathcal{Q}$;

\item[(iii)] there exist an onto affine $u:X\rightarrow \mathbb{R}$ and a lower
semicontinuous divergence $c:\Delta \times \Delta^{\sigma}\rightarrow \left[
0,\infty \right]  $, convex in $p$, such that, for each $Q\in \mathcal{Q}$,%
\[
f\succsim_{Q}^{\ast}g\Longleftrightarrow \min_{p\in \Delta}\left \{  \int
u\left(  f\right)  dp+c\left(  p,q\right)  \right \}  \geq \min_{p\in \Delta
}\left \{  \int u\left(  g\right)  dp+c\left(  p,q\right)  \right \}
\qquad \forall q\in Q
\]
and%
\[
f\succsim_{Q}g\Longleftrightarrow \min_{p\in \Delta}\left \{  \int u\left(
f\right)  dp+\min_{q\in Q}c\left(  p,q\right)  \right \}  \geq \min_{p\in \Delta
}\left \{  \int u\left(  g\right)  dp+\min_{q\in Q}c\left(  p,q\right)
\right \}
\]
for all acts $f,g\in \mathcal{F}$.
\end{enumerate}

\noindent Moreover, $u$ is cardinal and, given $u$, $c$ is unique.
\end{corollary}

\section{A Bayesian approach}

With the exception of the robust Bayesian interpretation of models outlined to
interpret convex sets of models (Section \ref{sect:models}), so far we
conducted our analysis in a classic Waldean setting where the DM's beliefs
over the likelihood of models, in particular their quantification via prior
probabilities, play no role. In contrast, in this section we outline a
Bayesian approach\ based on them.

Under model ambiguity, the DM has a prior probability $\mu_{Q}$\ over the set
of structured models $Q$. In particular, the prior probability $\mu_{Q}\left(
q\right)  $ of a structured model $q\in Q$\ quantifies the DM belief that $q$
is the correct model. Under model misspecification, this interpretation is no
longer possible because the DM no longer regards structured models as
alternative correct models, one of them being correct. So, they no longer form
an exhaustive collection of mutually exclusive uncertain alternatives -- a
logical partition -- over which a meaningful belief can be expressed.

We thus face two possibilities. The first one is to content ourselves with the
interpretation of the prior $\mu_{Q}$\ as an averaging device which specifies
the quasi-arithmetic aggregator in (\ref{eq:quasi}). The second, better, one
is to find a meaningful logical partition that gives $\mu_{Q}$ a proper
Bayesian interpretation. In both cases, one could argue that the prior
$\mu_{Q}$, compared to the standard case, might quantify a fragile belief
which might need to be robustified (cf. Hansen and Sargent, 2007). We begin by
introducing a Bayesian criterion under the average view of the prior $\mu_{Q}%
$. We then discuss a possible interpretation of this prior that gives the
criterion a genuine Bayesian flavor.

\subsection{A Bayesian criterion}

Consider the quasi-arithmetic specification (\ref{eq:quasi}), that is,
\begin{equation}
V_{Q}\left(  f\right)  =\phi_{Q}^{-1}\left(  \int_{Q}\phi_{Q}\left(
\min_{p\in \Delta}\left \{  \int_{S}u\left(  f\left(  s\right)  \right)
dp\left(  s\right)  +c\left(  p,q\right)  \right \}  \right)  d\mu_{Q}\left(
q\right)  \right)  \label{eq:quasi-bis}%
\end{equation}
where $\mu_{Q}\in \Delta \left(  Q\right)  $ and $\phi_{Q}:\mathbb{R}%
\rightarrow \mathbb{R}$ is strictly increasing and continuous. This is,
formally, a Bayesian criterion with the prior probability $\mu_{Q}$
interpreted as an averaging device over the structured models. The variational
criteria, indexed by $Q$,%
\[
\min_{p\in \Delta}\left \{  \int_{S}u\left(  f\left(  s\right)  \right)
dq\left(  s\right)  +c\left(  p,q\right)  \right \}
\]
and the corresponding dominance relation account for fear of model
misspecification about the posited models $q$, while the function $\phi_{Q}%
$\ addresses the fear of prior misspecification.

The Bayesian criterion (\ref{eq:quasi-bis}) generalizes to model
misspecification the smooth ambiguity criterion%
\[
V_{Q}\left(  f\right)  =\phi_{Q}^{-1}\left(  \int_{Q}\phi_{Q}\left(  \int
_{S}u\left(  f\left(  s\right)  \right)  dq\left(  s\right)  \right)  d\mu
_{Q}\left(  q\right)  \right)
\]
under model ambiguity, which is the special case $c\left(  p,q\right)
=\delta_{\left \{  q\right \}  }\left(  p\right)  $ for all $\left(  p,q\right)
\in \Delta \times \Delta^{\sigma}$ that corresponds to model misspecification
neutrality. When each $\phi_{Q}$\ is the identity, the criterion
(\ref{eq:quasi-bis}) further specializes to a standard subjective expected
utility criterion%
\[
V_{Q}\left(  f\right)  =\int_{Q}\left(  \int_{S}u\left(  f\left(  s\right)
\right)  dq\left(  s\right)  \right)  d\mu_{Q}\left(  q\right)
\]
An important entropic specification of criterion (\ref{eq:quasi-bis}) is%
\begin{equation}
V_{Q}^{\lambda,\xi}\left(  f\right)  =\phi_{\xi}^{-1}\left(  \int_{Q}\phi
_{\xi}\left(  \min_{p\in \Delta}\left \{  \int_{S}u\left(  f\left(  s\right)
\right)  dp\left(  s\right)  +\lambda R\left(  p||q\right)  \right \}  \right)
d\mu_{Q}\left(  q\right)  \right)  \label{eq:entrop-bayes}%
\end{equation}
where $\mu_{Q}\in \Delta \left(  Q\right)  $ and $\phi_{\xi}\left(  t\right)
=-e^{-\frac{1}{\xi}t}$\ has an exponential form (common across the sets of
models $Q$). The parameter $\xi>0$ captures fear of (reference) prior
misspecification, while the parameter $\lambda>0$ is a fear of model
misspecification index. The lower the parameter, the higher the fear. Next we
show that, as fear of either model or prior misspecification vanishes or
explodes, we get the criteria that one would expect. This provides an
analytical consistency check for criterion (\ref{eq:quasi-bis}). In deriving,
this result we focus on the entropic formulation, but the result can be
generalized in different directions, for example, by replacing the relative
entropy with a general divergence as in (\ref{eq:div}) and by replacing the
conditions on $\xi$ with similar conditions on the Arrow-Pratt index of
$\phi_{Q}$.

\begin{proposition}
\label{pro:lim-con}If $\operatorname*{supp}\mu_{Q}=Q$ and\ $f\in \mathcal{F}$,
then%
\begin{equation}
\lim_{\xi \rightarrow0^{+}}V_{Q}^{\lambda,\xi}\left(  f\right)  =\min
_{p\in \Delta}\left \{  \int_{S}u\left(  f\left(  s\right)  \right)  dp\left(
s\right)  +\lambda \min_{q\in Q}R\left(  p||q\right)  \right \}  \quad
\forall \lambda \in \left(  0,\infty \right]  \label{eq:bayes-lim1}%
\end{equation}
and%
\begin{equation}
\lim_{\xi \rightarrow \infty}V_{Q}^{\lambda,\xi}\left(  f\right)  =\int
_{Q}\left(  \min_{p\in \Delta}\left \{  \int_{S}u\left(  f\left(  s\right)
\right)  dp\left(  s\right)  +\lambda R\left(  p||q\right)  \right \}  \right)
d\mu_{Q}\left(  q\right)  \quad \forall \lambda \in \left(  0,\infty \right]
\label{eq:bayes-lim2}%
\end{equation}
Moreover,%
\begin{equation}
\lim_{\xi \rightarrow \infty}\lim_{\lambda \rightarrow \infty}V_{Q}^{\lambda,\xi
}\left(  f\right)  =\lim_{\lambda \rightarrow \infty}\lim_{\xi \rightarrow \infty
}V_{Q}^{\lambda,\xi}\left(  f\right)  =\int_{Q}\left(  \int_{S}u\left(
f\left(  s\right)  \right)  dq\left(  s\right)  \right)  d\mu_{Q}\left(
q\right)  \label{eq:bayes-lim3}%
\end{equation}

\end{proposition}

In words, the limit (\ref{eq:bayes-lim1}) shows that, as fear of prior
misspecification explodes, criterion (\ref{eq:quasi-bis}) gets closer and
closer to our criterion (\ref{eq:entropic-criterion}). In contrast, the limit
(\ref{eq:bayes-lim2}) shows that, when such fear vanishes, we end up with a
criterion that averages, via the prior $\mu_{Q}$, multiplier criteria (one per
structured model $q$). Finally, the limit (\ref{eq:bayes-lim3}) shows that, when
both fear vanish, at the limit we have a standard subjective expected utility criterion.

\subsection{On the interpretation of priors}

As we previously remarked, under model misspecification a set $Q$\ of
structured models is no longer a set of exhaustive and mutually exclusive
alternatives, so a logical partition upon which to define a prior probability.
What might be a new partition of this kind?

To address this question, denote by $p^{\ast}\in \Delta$ the correct model. The
agents\ do not know whether or not it belongs to $Q$. Let $q^{\ast}$ be the
structured model, assumed to uniquely exist, such that%
\[
c\left(  p^{\ast},q^{\ast}\right)  =\min_{q\in Q}c\left(  p^{\ast},q\right)
\]
Model $q^{\ast}$ best approximates, or best fits, the correct model $p^{\ast}$
according to the variational statistical distance $c$ that decision makers
adopt. If they know that $p^{\ast}$ is in $Q$\ (model ambiguity), we have
$p^{\ast}=q^{\ast}$ and so $q^{\ast}$ itself is the correct model.

Decision makers\ are uncertain about $q^{\ast}$, that is, about which
structured model $q\in Q$\ best fits the correct model. But, they know that
one of them is, indeed, the best fit. Under this interpretation of its
elements, $Q$\ thus forms a collection of exhaustive and mutually exclusive
alternatives. Decision makers\ now regard each element $q$ of $Q$\ as a
\textquotedblleft candidate best fitting model\textquotedblright: this is how
they interpret $q$\ and what they are uncertain about. The meaning of prior
$\mu_{Q}\left(  q\right)  $ is then clear: it quantifies the DM belief that
$q$ is the best fit of the correct model (see Walker, 2013, for an insightful discussion).

This interpretation of $\mu_{Q}$ reduces to the standard one under model
ambiguity because, as previously remarked, in this case the best fit coincides
with the correct model itself. In the rest of the section, we make more
rigorous this discussion.

To this end, let $c:\Delta \times \Delta^{\sigma}\rightarrow \left[
0,\infty \right]  $ be a lower semicontinuous statistical distance. Consider a
compact set of structured models $Q$. For each $q\in Q$,$\ $define the set%

\[
B_{c}\left(  q,Q\right)  =\left \{  p\in \Delta:c\left(  p,q\right)
=\min_{\tilde{q}\in Q}c(p,\tilde{q})\right \}
\]
If decision makers believe that a structured model $q\in Q$\ best fits the
correct model, then they consistently believe that the correct model belongs
to a subset $B_{c}\left(  q,Q\right)  $ of $\Delta$. So, for a structured
model $q\in Q$, the set $B_{c}\left(  q,Q\right)  $ consists of all
unstructured models that, if correct, make $q$ their best fit.

We can thus regard $B_{c}\left(  q,Q\right)  $ as the partial identification
set that corresponds to the agents belief that the structured model $q$\ best
fits the correct one. They can construct, at least in principle, this set by
solving the minimization problem that it features. Next we report some basic
properties of these partial identification sets. Let%
\[
\Delta_{c,Q}=\left \{  p\in \Delta:c\left(  p,\cdot \right)  \text{ is proper and
strictly convex on }Q\right \}
\]
For instance, by\ Lemma \ref{lem:str-cvx} we have $\Delta_{D_{\phi}%
,Q}\supseteq \left \{  p\in \Delta^{\sigma}:p\sim Q\right \}  $ for a $\phi
$-divergence $D_{\phi}$ when $Q\in \mathcal{K}$.

\begin{lemma}
\label{lem:bay-pri}If $Q\in \mathcal{K}$, then

\begin{enumerate}
\item[(i)] $\Delta=%
{\displaystyle \bigcup \limits_{q\in Q}}
B_{c}\left(  q,Q\right)  $;

\item[(ii)] $B_{c}\left(  q,Q\right)  \cap Q=\left \{  q\right \}  $ for all
$q\in Q$;

\item[(iii)] $B_{c}\left(  q,Q\right)  \cap B_{c}\left(  q^{\prime},Q\right)
\cap \left(  Q\cup \Delta_{c,Q}\right)  =\emptyset$ for all distinct
$q,q^{\prime}\in Q$.
\end{enumerate}
\end{lemma}

Properties (i) and (iii) ensure that the family of the partial identification
sets%
\[
\left \{  B_{c}\left(  q,Q\right)  \right \}  _{q\in Q}%
\]
forms a partition of $Q\cup \Delta_{c,Q}$. As long as the correct model belongs
to $Q\cup \Delta_{c,Q}$, this permits to interpret $\mu_{Q}\left(  q\right)  $
as the probability that the structured model $q\in Q$ is the best fit of the
correct model. In particular, we can interpret in this way the prior $\mu_{Q}$
that the Bayesian criterion (\ref{eq:quasi-bis}) features, thus giving this
criterion a genuine Bayesian status. Property (ii) ensures that under model
ambiguity we go back to the traditional interpretation of priors.

\section{Conclusion\label{sect:concl}}

Quantitative researchers use models to enhance their understanding of economic
phenomena and to make policy assessments. In essence, each model tells its own
quantitative story. We refer to such models as\  \textquotedblleft structured
models.\textquotedblright \ Typically, there are more than just one such type
of model, with each giving rise to a different quantitative story. Statistical
and economic decision theories have addressed how best to confront the
ambiguity among structured models. Such structured models are, by their very
nature, misspecified. Nevertheless, the decision maker seeks to use such
models in sensible ways. This problem is well recognized by applied
researchers, but it is typically not part of formal decision theory. In this
paper, we extend decision theory to confront model misspecification concerns.
In so doing, we recover a variational representation of preferences that
includes penalization based on discrepancy measures between \textquotedblleft
unstructured alternatives\textquotedblright \ and the set of structured
probability models.

In terms of future research, a natural generalization of our criterion is%
\[
V\left(  f\right)  =\min_{p\in \Delta}\left \{  \int u\left(  f\right)
dp+C\left(  p,Q\right)  \right \}
\]
where $C$ is a general statistical set distance, not necessarily Hausdorff (so
not necessarily characterized by an underlying statistical distance). This
variational criterion still represents a preference that is more uncertainty
averse than the corresponding max-min one. It may also easily accommodate
reversals of the inequality (\ref{eq:admin-comp}), along the lines previously
discussed. Though the analysis of this general criterion is beyond the scope
of this paper and left for future research, we close our exposition with it as
its form should help to put our exercise in a final perspective.

\appendix

\section{Proofs and related analysis}

In this appendix, we provide the proofs of our main results. We relegate to
the Appendix \ref{app:EM} the proofs of most of our ancillary results\ (e.g.,
Propositions \ref{prop:comput}, \ref{prop:lim}, \ref{pro:adm} and
\ref{pro:com-sta-val}). In the same appendix, we also formally discuss few
results about statistical distances and divergences (Lemmas \ref{lem:min-c}%
--\ref{lem:str-cvx}). Appendix \ref{app:pro-mai} contains the proofs of our
representation results (Theorems \ref{thm:mai-rep} and
\ref{thm:mai-rep-inc-fin}, and Proposition \ref{pro:mai-rep-inc}). Appendix
\ref{app:pro-oth} contains the proofs of the remaining analysis. In both
appendices, we denote by $B_{0}\left(  \Sigma \right)  $ the space of $\Sigma
$-measurable simple functions $\varphi:S\rightarrow \mathbb{R}$, endowed with
the supnorm $\left \Vert \text{ }\right \Vert _{\infty}$. The dual of
$B_{0}\left(  \Sigma \right)  $ can be identified with the space $ba\left(
\Sigma \right)  $ of all bounded finitely additive measures on $\left(
S,\Sigma \right)  $.

\subsection{Representation results\label{app:pro-mai}}

The proof of Theorem \ref{thm:mai-rep} is based on three\ key steps. We first
provide two\ results regarding\ variational preferences which will help
isolate the set of structured models $Q$ in the main representation (their
routine proof is confined to Appendix \ref{app:EM}). Second, we provide a
representation\ for an unbounded and objectively $Q$-coherent dominance
relation $\succsim^{\ast}$ (Appendix \ref{app:Bew-rep}). Third, we prove
Theorem \ref{thm:mai-rep}\ (Appendix \ref{app:fin-pro-mai}). The proof of
Theorem \ref{thm:mai-rep-inc-fin}\ and\ Proposition \ref{pro:mai-rep-inc}
instead is presented as one result (Appendix \ref{app:thm-var-str}). In what
follows, given a function $c:\Delta \times Q\rightarrow \left[  0,\infty \right]
$, where $Q$ is a compact and convex subset of $\Delta^{\sigma}$, we say that
$c$ is \emph{variational\ }if $c_{q}$ is grounded, lower semicontinuous and
convex and $c_{Q}$($=\min_{q\in Q}c\left(  \cdot,q\right)  $) is well defined,
grounded, lower semicontinuous and convex. The next two lemmas, proved in
Appendix \ref{app:anc-rep}, are key in characterizing subjective and objective
$Q$-coherence.

\begin{lemma}
\label{lm:zero-set}Let $\succsim$ be\ a variational preference represented by
$V:\mathcal{F}\rightarrow \mathbb{R}$ defined by%
\[
V\left(  f\right)  =\min_{p\in \Delta}\left \{  \int u\left(  f\right)
dp+c\left(  p\right)  \right \}  \qquad \forall f\in \mathcal{F}%
\]
and let $\bar{p}\in \Delta$. If $\succsim$ is unbounded, then the following
conditions are equivalent:

\begin{enumerate}
\item[(i)] $c\left(  \bar{p}\right)  =0$;

\item[(ii)] $x_{f}^{\bar{p}}\succsim f$\ for all $f\in \mathcal{F}$;

\item[(iii)] for each $f\in \mathcal{F}$ and for each $x\in X$%
\[
x\succ x_{f}^{\bar{p}}\implies x\succ f
\]

\end{enumerate}
\end{lemma}

\begin{lemma}
\label{lem:del-Q}Let $\succsim$ be\ a variational preference represented by
$V:\mathcal{F}\rightarrow \mathbb{R}$ defined by%
\[
V\left(  f\right)  =\min_{p\in \Delta}\left \{  \int u\left(  f\right)
dp+c\left(  p\right)  \right \}  \qquad \forall f\in \mathcal{F}%
\]
If $\succsim$ is unbounded, then the following conditions are equivalent:

\begin{enumerate}
\item[(i)] For each $f,g\in \mathcal{F}$%
\[
f\overset{Q}{=}g\implies f\sim g
\]

\item[(ii)] $\operatorname*{dom}c\subseteq \Delta^{\ll}\left(  Q\right)  $.
\end{enumerate}
\end{lemma}

\subsubsection{A Bewley-type representation\label{app:Bew-rep}}

The next result is a multi-utility (variational) representation for unbounded
dominance relations.

\begin{lemma}
\label{lem:inc-var}Let $\succsim^{\ast}$ be a binary relation on $\mathcal{F}%
$, where $\left(  S,\Sigma \right)  $ is a standard Borel space. The following
statements are equivalent:

\begin{enumerate}
\item[(i)] $\succsim^{\ast}$ is an unbounded\ dominance relation which
satisfies objective $Q$-coherence;

\item[(ii)] there exist an onto affine function $u:X\rightarrow \mathbb{R}$ and
a variational\ $c:\Delta \times Q\rightarrow \left[  0,\infty \right]  $ such
that $\operatorname*{dom}c\left(  \cdot,q\right)  \subseteq \Delta^{\ll}\left(
Q\right)  $ for all $q\in Q$ and%
\begin{equation}
f\succsim^{\ast}g\iff \min_{p\in \Delta}\left \{  \int u\left(  f\right)
dp+c\left(  p,q\right)  \right \}  \geq \min_{p\in \Delta}\left \{  \int u\left(
g\right)  dp+c\left(  p,q\right)  \right \}  \quad \forall q\in Q
\label{eq:Bew-var}%
\end{equation}

\end{enumerate}
\end{lemma}

To prove this result, we need to introduce one mathematical object. Let
$\succeq^{\ast}$ be a binary relation on $B_{0}\left(  \Sigma \right)  $. We
say that $\succeq^{\ast}$ is \textit{convex niveloidal} if and only if
$\succeq^{\ast}$ is a preorder that satisfies the following five properties:

\begin{enumerate}
\item For each $\varphi,\psi \in B_{0}\left(  \Sigma \right)  $ and for each
$k\in \mathbb{R}$%
\[
\varphi \succeq^{\ast}\psi \implies \varphi+k\succeq^{\ast}\psi+k
\]

\item If $\varphi,\psi \in B_{0}\left(  \Sigma \right)  $ and $\left \{
k_{n}\right \}  _{n\in%
\mathbb{N}
}\subseteq \mathbb{R}$ are such that\ $k_{n}\uparrow k$ and $\varphi
-k_{n}\succeq^{\ast}\psi$ for all $n\in%
\mathbb{N}
$, then $\varphi-k\succeq^{\ast}\psi$;

\item For each $\varphi,\psi \in B_{0}\left(  \Sigma \right)  $%
\[
\varphi \geq \psi \implies \varphi \succeq^{\ast}\psi
\]

\item For each $k,h\in \mathbb{R}$ and for each $\varphi \in B_{0}\left(
\Sigma \right)  $%
\[
k>h\implies \varphi+k\succ^{\ast}\varphi+h
\]

\item For each $\varphi,\psi,\xi \in B_{0}\left(  \Sigma \right)  $ and for each
$\lambda \in \left(  0,1\right)  $%
\[
\varphi \succeq^{\ast}\xi \text{ and }\psi \succeq^{\ast}\xi \implies
\lambda \varphi+\left(  1-\lambda \right)  \psi \succeq^{\ast}\xi
\]

\end{enumerate}

\begin{lemma}
\label{lem:HM}If $\succsim^{\ast}$ is an unbounded\ dominance relation, then
there exists an onto affine function $u:X\rightarrow \mathbb{R}$ such that%
\begin{equation}
x\succsim^{\ast}y\iff u\left(  x\right)  \geq u\left(  y\right)
\label{eq:u-rep}%
\end{equation}

\end{lemma}

\noindent \textbf{Proof }Since $\succsim^{\ast}$ is a non-trivial preorder on
$\mathcal{F}$ that satisfies c-completeness, continuity and weak
c-independence, it is immediate to conclude that $\succsim^{\ast}$ restricted
to $X$ satisfies weak order, continuity and risk independence.\footnote{To
prove that $\succsim^{\ast}$ satisfies risk independence, it suffices to
deploy the same technique of Lemma 28 of Maccheroni et al. (2006) and observe
that $\succsim^{\ast}$ is a complete preorder on $X$. This yields that%
\[
x\sim^{\ast}y\implies \frac{1}{2}x+\frac{1}{2}z\sim^{\ast}\frac{1}{2}y+\frac
{1}{2}z\quad \forall z\in X
\]
By Theorem 2 of Herstein and Milnor (1953) and since $\succsim^{\ast}$
satisfies continuity,\ we can conclude that $\succsim^{\ast}$ satisfies risk
independence.} By Herstein and Milnor (1953), it follows that there exists an
affine function $u:X\rightarrow \mathbb{R}$ that satisfies (\ref{eq:u-rep}).
Since $\succsim^{\ast}$ is a non-trivial c-complete preorder on $\mathcal{F}$
that satisfies monotonicity, we have that $\succsim^{\ast}$ is non-trivial on
$X$. By Lemma 59 of Cerreia-Vioglio et al. (2011b) and since $\succsim^{\ast}$
is non-trivial on $X$ and satisfies unboundedness, we can conclude that $u$ is
onto.\hfill$\blacksquare$

\smallskip

Since $u$ is affine and onto, note that $\left \{  u\left(  f\right)
:f\in \mathcal{F}\right \}  =B_{0}\left(  \Sigma \right)  $. In light of this
observation, we can define a binary relation $\succeq^{\ast}$ on $B_{0}\left(
\Sigma \right)  $ by%
\begin{equation}
\varphi \succeq^{\ast}\psi \iff f\succsim^{\ast}g\text{ where }u\left(
f\right)  =\varphi \text{ and }u\left(  g\right)  =\psi \label{eq:def-sta}%
\end{equation}

\begin{lemma}
\label{lem:def-sta}If $\succsim^{\ast}$ is an unbounded\ dominance relation,
then $\succeq^{\ast}$, defined as in (\ref{eq:def-sta}), is a well defined
convex niveloidal binary relation. Moreover, if $\succsim^{\ast}$ is
objectively $Q$-coherent, then $\varphi \overset{Q}{=}\psi$ implies
$\varphi \sim^{\ast}\psi$.
\end{lemma}

We confine the routine proof to Appendix \ref{app:EM}. The next three results
(Lemmas \ref{lem:upp-con} and \ref{lem:upp-niv} as well as Proposition
\ref{pro:rep-sta}) will help us representing $\succeq^{\ast}$. This paired
with Lemma \ref{lem:HM}\ and Proposition\  \ref{pro:car}\ will yield the proof
of Lemma \ref{lem:inc-var}.

\begin{lemma}
\label{lem:upp-con}Let $\succeq^{\ast}$ be a convex niveloidal binary
relation. If\ $\psi \in B_{0}\left(  \Sigma \right)  $, then $U\left(
\psi \right)  =\left \{  \varphi \in B_{0}\left(  \Sigma \right)  :\varphi
\succeq^{\ast}\psi \right \}  $ is a non-empty convex set such that:

\begin{enumerate}
\item $\psi \in U\left(  \psi \right)  $;

\item if $\varphi \in B_{0}\left(  \Sigma \right)  $ and $\left \{
k_{n}\right \}  _{n\in%
\mathbb{N}
}\subseteq \mathbb{R}$ are\ such that $k_{n}\uparrow k$ and $\varphi-k_{n}\in
U\left(  \psi \right)  $ for all $n\in%
\mathbb{N}
$, then $\varphi-k\in U\left(  \psi \right)  $;

\item if $k>0$, then $\psi-k\not \in U\left(  \psi \right)  $;

\item if $\varphi_{1}\geq \varphi_{2}$ and $\varphi_{2}\in U\left(
\psi \right)  $, then $\varphi_{1}\in U\left(  \psi \right)  $;

\item if $k\geq0$ and $\varphi_{2}\in U\left(  \psi \right)  $, then
$\varphi_{2}+k\in U\left(  \psi \right)  $.
\end{enumerate}
\end{lemma}

\noindent \textbf{Proof }Since $\succeq^{\ast}$ is reflexive, we have that
$\psi \in U\left(  \psi \right)  $, proving that $U\left(  \psi \right)  $ is
non-empty and point 1. Consider $\varphi_{1},\varphi_{2}\in U\left(
\psi \right)  $ and $\lambda \in \left(  0,1\right)  $. By definition, we have
that $\varphi_{1}\succeq^{\ast}\psi$ and $\varphi_{2}\succeq^{\ast}\psi$.
Since $\succeq^{\ast}$ satisfies convexity, we have that $\lambda \varphi
_{1}+\left(  1-\lambda \right)  \varphi_{2}\succeq^{\ast}\psi$, proving
convexity of $U\left(  \psi \right)  $. Consider $\varphi \in B_{0}\left(
\Sigma \right)  $\ and $\left \{  k_{n}\right \}  _{n\in%
\mathbb{N}
}\subseteq \mathbb{R}$ such that $k_{n}\uparrow k$ and $\varphi-k_{n}\in
U\left(  \psi \right)  $ for all $n\in%
\mathbb{N}
$. It follows that $\varphi-k_{n}\succeq^{\ast}\psi$ for all $n\in%
\mathbb{N}
$, then $\varphi-k\succeq^{\ast}\psi$, that is, $\varphi-k\in U\left(
\psi \right)  $, proving point 2. If $k>0$, then $0>-k$ and $\psi=\psi
+0\succ^{\ast}\psi-k$, that is, $\psi-k\not \in U\left(  \psi \right)  $,
proving point 3. Consider\ $\varphi_{1}\geq \varphi_{2}$ such that $\varphi
_{2}\in U\left(  \psi \right)  $, then $\varphi_{1}\succeq^{\ast}\varphi_{2}%
\ $and $\varphi_{2}\succeq^{\ast}\psi$, yielding that $\varphi_{1}%
\succeq^{\ast}\psi$ and, in particular, $\varphi_{1}\in U\left(  \psi \right)
$, proving point 4. Finally, to prove point 5, it is enough to set
$\varphi_{1}=\varphi_{2}+k$ in point 4.\hfill$\blacksquare$

\smallskip

Before stating the next result, we define few properties that will turn out to
be useful later on. A functional $I:B_{0}\left(  \Sigma \right)  \rightarrow
\mathbb{R}$ is:

\begin{enumerate}
\item a niveloid if $I\left(  \varphi \right)  -I\left(  \psi \right)  \leq
\sup_{s\in S}\left(  \varphi \left(  s\right)  -\psi \left(  s\right)  \right)
$ for all $\varphi,\psi \in B_{0}\left(  \Sigma \right)  $;

\item normalized if $I\left(  k\right)  =k$ for all $k\in \mathbb{R}%
$;\footnote{With the usual abuse of notation, we denote by $k$ both the real
number and the constant function taking value $k$.}

\item monotone if for each $\varphi,\psi \in B_{0}\left(  \Sigma \right)  $%
\[
\varphi \geq \psi \implies I\left(  \varphi \right)  \geq I\left(  \psi \right)
\]

\item $\succeq^{\ast}$-consistent if for each $\varphi,\psi \in B_{0}\left(
\Sigma \right)  $%
\[
\varphi \succeq^{\ast}\psi \implies I\left(  \varphi \right)  \geq I\left(
\psi \right)
\]

\item concave if for each $\varphi,\psi \in B_{0}\left(  \Sigma \right)  $ and
$\lambda \in \left(  0,1\right)  $%
\[
I\left(  \lambda \varphi+\left(  1-\lambda \right)  \psi \right)  \geq \lambda
I\left(  \varphi \right)  +\left(  1-\lambda \right)  I\left(  \psi \right)
\]

\item translation invariant if for each $\varphi \in B_{0}\left(
\Sigma \right)  $ and $k\in \mathbb{R}$%
\[
I\left(  \varphi+k\right)  =I\left(  \varphi \right)  +k
\]

\end{enumerate}

\begin{lemma}
\label{lem:upp-niv}Let $\succeq^{\ast}\ $be a convex niveloidal binary
relation. If $\psi \in B_{0}\left(  \Sigma \right)  $, then the functional
$I_{\psi}:B_{0}\left(  \Sigma \right)  \rightarrow \mathbb{R}$, defined by%
\[
I_{\psi}\left(  \varphi \right)  =\max \left \{  k\in \mathbb{R}:\varphi-k\in
U\left(  \psi \right)  \right \}  \qquad \forall \varphi \in B_{0}\left(
\Sigma \right)
\]
is a concave niveloid which is $\succeq^{\ast}$-consistent and such that
$I_{\psi}\left(  \psi \right)  =0$. Moreover, we have that:

\begin{enumerate}
\item The functional $\bar{I}_{\psi}=I_{\psi}-I_{\psi}\left(  0\right)  $ is a
normalized concave niveloid which is $\succeq^{\ast}$-consistent.

\item If $\succeq^{\ast}$ satisfies%
\[
\psi \overset{Q}{=}\psi^{\prime}\implies \psi \sim^{\ast}\psi^{\prime}%
\]
then%
\[
\psi \overset{Q}{=}\psi^{\prime}\implies I_{\psi}=I_{\psi^{\prime}}\text{ and
}\bar{I}_{\psi}=\bar{I}_{\psi^{\prime}}%
\]

\end{enumerate}
\end{lemma}

We confine the routine proof of the previous lemma to Appendix \ref{app:EM}.

\begin{proposition}
\label{pro:rep-sta}Let $\succeq^{\ast}$ be a binary relation on $B_{0}\left(
\Sigma \right)  $. The following statements are equivalent:

\begin{enumerate}
\item[(i)] $\succeq^{\ast}$\ is convex niveloidal;

\item[(ii)] there exists a family of concave niveloids $\left \{  I_{\alpha
}\right \}  _{\alpha \in A}$ on $B_{0}\left(  \Sigma \right)  $\ such that%
\begin{equation}
\varphi \succeq^{\ast}\psi \iff I_{\alpha}\left(  \varphi \right)  \geq
I_{\alpha}\left(  \psi \right)  \qquad \forall \alpha \in A \label{eq:rep-can}%
\end{equation}

\item[(iii)] there exists a family of normalized concave niveloids $\left \{
\bar{I}_{\alpha}\right \}  _{\alpha \in A}$ on $B_{0}\left(  \Sigma \right)
$\ such that%
\begin{equation}
\varphi \succeq^{\ast}\psi \iff \bar{I}_{\alpha}\left(  \varphi \right)  \geq
\bar{I}_{\alpha}\left(  \psi \right)  \qquad \forall \alpha \in A
\label{eq:rep-can-nor}%
\end{equation}

\end{enumerate}
\end{proposition}

\noindent \textbf{Proof }(iii) implies (i). It is trivial.

\smallskip

(i) implies (ii). Let $A=B_{0}\left(  \Sigma \right)  $. We next show that%
\[
\varphi_{1}\succeq^{\ast}\varphi_{2}\iff I_{\psi}\left(  \varphi_{1}\right)
\geq I_{\psi}\left(  \varphi_{2}\right)  \qquad \forall \psi \in B_{0}\left(
\Sigma \right)
\]
where $I_{\psi}$ is defined as in Lemma \ref{lem:upp-niv} for all $\psi \in
B_{0}\left(  \Sigma \right)  $. By Lemma \ref{lem:upp-niv}, we have that
$I_{\psi}$ is $\succeq^{\ast}$-consistent for all $\psi \in B_{0}\left(
\Sigma \right)  $. This implies that%
\[
\varphi_{1}\succeq^{\ast}\varphi_{2}\implies I_{\psi}\left(  \varphi
_{1}\right)  \geq I_{\psi}\left(  \varphi_{2}\right)  \qquad \forall \psi \in
B_{0}\left(  \Sigma \right)
\]
Vice versa, consider $\varphi_{1},\varphi_{2}\in B_{0}\left(  \Sigma \right)
$. Assume that $I_{\psi}\left(  \varphi_{1}\right)  \geq I_{\psi}\left(
\varphi_{2}\right)  $\ for all$\  \psi \in B_{0}\left(  \Sigma \right)  $. Let
$\psi=\varphi_{2}$. By Lemma \ref{lem:upp-niv}, we have that $I_{\varphi_{2}%
}\left(  \varphi_{1}\right)  \geq I_{\varphi_{2}}\left(  \varphi_{2}\right)
=0$, yielding that $\varphi_{1}\geq \varphi_{1}-I_{\varphi_{2}}\left(
\varphi_{1}\right)  \in U\left(  \varphi_{2}\right)  $. By point 4 of Lemma
\ref{lem:upp-con}, this implies\ that $\varphi_{1}\in U\left(  \varphi
_{2}\right)  $, that is, $\varphi_{1}\succeq^{\ast}\varphi_{2}$.

\smallskip

(ii) implies (iii). Given a family of concave niveloids $\left \{  I_{\alpha
}\right \}  _{\alpha \in A}$, define $\bar{I}_{\alpha}=I_{\alpha}-I_{\alpha
}\left(  0\right)  $ for all $\alpha \in A$. It is immediate to verify that
$\bar{I}_{\alpha}$ is a normalized concave niveloid for all $\alpha \in A$. It
is also immediate to observe that%
\[
I_{\alpha}\left(  \varphi_{1}\right)  \geq I_{\alpha}\left(  \varphi
_{2}\right)  \quad \forall \alpha \in A\iff \bar{I}_{\alpha}\left(  \varphi
_{1}\right)  \geq \bar{I}_{\alpha}\left(  \varphi_{2}\right)  \quad
\forall \alpha \in A
\]
proving the implication.\hfill$\blacksquare$

\begin{remark}
\label{rmk:can-rep}\emph{Given a convex niveloidal binary relation }%
$\succeq^{\ast}$\emph{ on }$B_{0}\left(  \Sigma \right)  $\emph{, we call
}canonical\emph{\ (resp., }canonical normalized\emph{) the representation
}$\left \{  I_{\psi}\right \}  _{\psi \in B_{0}\left(  \Sigma \right)  }$
\emph{(resp., }$\left \{  \bar{I}_{\psi}\right \}  _{\psi \in B_{0}\left(
\Sigma \right)  }$\emph{) obtained from Lemma \ref{lem:upp-niv}\ and the proof
of Proposition \ref{pro:rep-sta}. By the previous proof, clearly,\ }$\left \{
I_{\psi}\right \}  _{\psi \in B_{0}\left(  \Sigma \right)  }$ \emph{and}%
\ $\left \{  \bar{I}_{\psi}\right \}  _{\psi \in B_{0}\left(  \Sigma \right)  }%
$\emph{ satisfy (\ref{eq:rep-can}) and (\ref{eq:rep-can-nor}) respectively.}
\end{remark}

The next result clarifies what is the relation between any representation of
$\succeq^{\ast}$ and the canonical ones. This will be useful in establishing
an extra property of $\left \{  \bar{I}_{\psi}\right \}  _{\psi \in B_{0}\left(
\Sigma \right)  }$ in Corollary \ref{cor:pro-can-rep-min}.

\begin{lemma}
\label{lem:can-rep-rel}Let $\succeq^{\ast}$ be a convex niveloidal binary
relation. If $B$ is an index\ set and $\left \{  J_{\beta}\right \}  _{\beta \in
B}$ is a family of normalized concave niveloids such that%
\[
\varphi \succeq^{\ast}\psi \iff J_{\beta}\left(  \varphi \right)  \geq J_{\beta
}\left(  \psi \right)  \qquad \forall \beta \in B
\]
then for each $\psi \in B_{0}\left(  \Sigma \right)  $%
\begin{equation}
I_{\psi}\left(  \varphi \right)  =\inf_{\beta \in B}\left(  J_{\beta}\left(
\varphi \right)  -J_{\beta}\left(  \psi \right)  \right)  \qquad \forall
\varphi \in B_{0}\left(  \Sigma \right)  \label{eq:I-psi-rep-rel}%
\end{equation}
and%
\begin{equation}
\bar{I}_{\psi}\left(  \varphi \right)  =\inf_{\beta \in B}\left(  J_{\beta
}\left(  \varphi \right)  -J_{\beta}\left(  \psi \right)  \right)  +\sup
_{\beta \in B}J_{\beta}\left(  \psi \right)  \qquad \forall \varphi \in
B_{0}\left(  \Sigma \right)  \label{eq:I-bar-psi-rep-rel}%
\end{equation}

\end{lemma}

\noindent \textbf{Proof }Fix $\varphi \in B_{0}\left(  \Sigma \right)  $ and
$\psi \in B_{0}\left(  \Sigma \right)  $. By definition, we have that%
\[
I_{\psi}\left(  \varphi \right)  =\max \left \{  k\in \mathbb{R}:\varphi-k\in
U\left(  \psi \right)  \right \}
\]
Since $\left \{  J_{\beta}\right \}  _{\beta \in B}$ represents $\succeq^{\ast}$
and each $J_{\beta}$ is translation invariant, note that for each
$k\in \mathbb{R}$%
\begin{align*}
\varphi-k  &  \in U\left(  \psi \right)  \iff \varphi-k\succeq^{\ast}\psi \iff
J_{\beta}\left(  \varphi-k\right)  \geq J_{\beta}\left(  \psi \right)
\quad \forall \beta \in B\\
&  \iff J_{\beta}\left(  \varphi \right)  -k\geq J_{\beta}\left(  \psi \right)
\quad \forall \beta \in B\iff J_{\beta}\left(  \varphi \right)  -J_{\beta}\left(
\psi \right)  \geq k\quad \forall \beta \in B\\
&  \iff \inf_{\beta \in B}\left(  J_{\beta}\left(  \varphi \right)  -J_{\beta
}\left(  \psi \right)  \right)  \geq k
\end{align*}
By definition of $I_{\psi}$ and since $\varphi-I_{\psi}\left(  \varphi \right)
\in U\left(  \psi \right)  $, this implies that $I_{\psi}\left(  \varphi
\right)  =\inf_{\beta \in B}\left(  J_{\beta}\left(  \varphi \right)  -J_{\beta
}\left(  \psi \right)  \right)  $. Since $\varphi$ and $\psi$ were arbitrarily
chosen, (\ref{eq:I-psi-rep-rel}) follows. Since $\bar{I}_{\psi}=I_{\psi
}-I_{\psi}\left(  0\right)  $, we only need to compute $-I_{\psi}\left(
0\right)  $. Since each $J_{\beta}$ is normalized, we have that $-I_{\psi
}\left(  0\right)  =-\inf_{\beta \in B}\left(  J_{\beta}\left(  0\right)
-J_{\beta}\left(  \psi \right)  \right)  =-\inf_{\beta \in B}\left(  -J_{\beta
}\left(  \psi \right)  \right)  =\sup_{\beta \in B}J_{\beta}\left(  \psi \right)
$, proving (\ref{eq:I-bar-psi-rep-rel}).\hfill$\blacksquare$

\begin{corollary}
\label{cor:pro-can-rep-min}If $\succeq^{\ast}$ is a convex niveloidal binary
relation, then $\bar{I}_{0}\leq \bar{I}_{\psi}$\ for all $\psi \in B_{0}\left(
\Sigma \right)  $.
\end{corollary}

\noindent \textbf{Proof }By Lemma \ref{lem:can-rep-rel} and Remark
\ref{rmk:can-rep} and since each $\bar{I}_{\psi^{\prime}}$ is a normalized
concave niveloid, we have that%
\[
\bar{I}_{0}\left(  \varphi \right)  =\inf_{\psi^{\prime}\in B_{0}\left(
\Sigma \right)  }\left(  \bar{I}_{\psi^{\prime}}\left(  \varphi \right)
-\bar{I}_{\psi^{\prime}}\left(  0\right)  \right)  +\sup_{\psi^{\prime}\in
B_{0}\left(  \Sigma \right)  }\bar{I}_{\psi^{\prime}}\left(  0\right)
=\inf_{\psi^{\prime}\in B_{0}\left(  \Sigma \right)  }\bar{I}_{\psi^{\prime}%
}\left(  \varphi \right)  \leq \bar{I}_{\psi}\left(  \varphi \right)
\quad \forall \varphi \in B_{0}\left(  \Sigma \right)
\]
for all $\psi \in B_{0}\left(  \Sigma \right)  $, proving the statement.\hfill
$\blacksquare$

\smallskip

The next result will be instrumental in providing a niveloidal
multi-representation of $\succsim^{\ast}$ when $\left \vert Q\right \vert \geq
2$.\ In order to discuss it, we need a piece of terminology. We denote by $V$
the quotient space $B_{0}\left(  \Sigma \right)  /M$ where $M$ is the vector
subspace $\left \{  \varphi \in B_{0}\left(  \Sigma \right)  :\varphi \overset
{Q}{=}0\right \}  $. Recall that the elements of $V$ are equivalence classes
$\left[  \psi \right]  $ with $\psi \in B_{0}\left(  \Sigma \right)  $ where
$\psi^{\prime},\psi^{\prime \prime}\in \left[  \psi \right]  $ if and only if
$\psi \overset{Q}{=}\psi^{\prime}\overset{Q}{=}\psi^{\prime \prime}$. Recall
that $Q$ is convex.

\begin{proposition}
\label{pro:car}If $\left(  S,\Sigma \right)  $ is a standard Borel space and
$\left \vert Q\right \vert \geq2$, then there exists a bijection $f:V\rightarrow
Q$.
\end{proposition}

The routine proof of the previous result is relegated to Appendix
\ref{app:EM}. We next prove our representation result for incomplete
variational preferences.

\smallskip

\noindent \textbf{Proof of Lemma\  \ref{lem:inc-var} }(ii) implies (i). It is trivial.

\smallskip

(i) implies (ii). Since $\succsim^{\ast}$ is a dominance relation, if
$\left \vert Q\right \vert =1$, that is $Q=\left \{  \bar{q}\right \}  $, then
$\succsim^{\ast}$ is complete. By Maccheroni et al. (2006) and since
$\succsim^{\ast}$ is unbounded, it follows that there exists an onto and
affine $u:X\rightarrow \mathbb{R}$ and a grounded, lower semicontinuous and
convex $c_{\bar{q}}:\Delta \rightarrow \left[  0,\infty \right]  $ such that
$V:\mathcal{F}\rightarrow \mathbb{R}$ defined by%
\[
V\left(  f\right)  =\min_{p\in \Delta}\left \{  \int u\left(  f\right)
dp+c_{\bar{q}}\left(  p\right)  \right \}  \qquad \forall f\in \mathcal{F}%
\]
represents $\succsim^{\ast}$. If we define $c:\Delta \times Q\rightarrow \left[
0,\infty \right]  $\ by $c\left(  p,q\right)  =c_{\bar{q}}\left(  p\right)  $
for all $\left(  p,q\right)  \in \Delta \times Q$, then we have that $c$ is
variational. By Lemma \ref{lem:del-Q} and since $\succsim^{\ast}$ is
objectively $Q$-coherent, it follows that $\operatorname*{dom}c\left(
\cdot,q\right)  \subseteq \Delta^{\ll}\left(  Q\right)  $ for all $q\in Q$,
proving the implication. Assume $\left \vert Q\right \vert >1$. By Lemma
\ref{lem:HM}, there exists an onto affine function $u:X\rightarrow \mathbb{R}$
which represents $\succsim^{\ast}$ on $X$. By Lemma \ref{lem:def-sta}, this
implies that we can consider the convex niveloidal binary relation
$\succeq^{\ast}$ defined as in (\ref{eq:def-sta}). By definition of
$\succeq^{\ast}$\ and Proposition \ref{pro:rep-sta} (and Remark
\ref{rmk:can-rep}), we have that%
\[
f\succsim^{\ast}g\iff u\left(  f\right)  \succeq^{\ast}u\left(  g\right)
\iff \bar{I}_{\psi}\left(  u\left(  f\right)  \right)  \geq \bar{I}_{\psi
}\left(  u\left(  g\right)  \right)  \quad \forall \psi \in B_{0}\left(
\Sigma \right)
\]
where each $\bar{I}_{\psi}$ is a normalized concave niveloid. As before,
consider $V=B_{0}\left(  \Sigma \right)  /M$ where $M$ is the vector subspace
$\left \{  \varphi \in B_{0}\left(  \Sigma \right)  :\varphi \overset{Q}%
{=}0\right \}  $. For each equivalence class $\left[  \psi \right]  $, select
exactly one $\psi^{\prime}\in B_{0}\left(  \Sigma \right)  $ such that
$\psi^{\prime}\in \left[  \psi \right]  $. In particular, let $\psi^{\prime}=0$
when $\left[  \psi \right]  =\left[  0\right]  $.\ We denote this subset of
$B_{0}\left(  \Sigma \right)  $ by $\tilde{V}$. Clearly, we have that%
\[
\bar{I}_{\psi}\left(  u\left(  f\right)  \right)  \geq \bar{I}_{\psi}\left(
u\left(  g\right)  \right)  \quad \forall \psi \in B_{0}\left(  \Sigma \right)
\implies \bar{I}_{\psi}\left(  u\left(  f\right)  \right)  \geq \bar{I}_{\psi
}\left(  u\left(  g\right)  \right)  \quad \forall \psi \in \tilde{V}%
\]
Vice versa, assume that $\bar{I}_{\psi}\left(  u\left(  f\right)  \right)
\geq \bar{I}_{\psi}\left(  u\left(  g\right)  \right)  $ for all $\psi \in
\tilde{V}$. Consider $\hat{\psi}\in B_{0}\left(  \Sigma \right)  $. It follows
that there exists $\left[  \psi \right]  $ in $V$ such that $\hat{\psi}%
\in \left[  \psi \right]  $. Similarly, consider $\psi^{\prime}\in \tilde{V}$
such that $\psi^{\prime}\in \left[  \psi \right]  $. It follows that $\hat{\psi
}\overset{Q}{=}\psi^{\prime}$. By Lemmas \ref{lem:def-sta} and
\ref{lem:upp-niv}\ and since $\succsim^{\ast}$ is objectively $Q$-coherent,
then $\bar{I}_{\hat{\psi}}=\bar{I}_{\psi^{\prime}}$, yielding that $\bar
{I}_{\hat{\psi}}\left(  u\left(  f\right)  \right)  \geq \bar{I}_{\hat{\psi}%
}\left(  u\left(  g\right)  \right)  $. Since $\hat{\psi}$ was arbitrarily
chosen $\bar{I}_{\psi}\left(  u\left(  f\right)  \right)  \geq \bar{I}_{\psi
}\left(  u\left(  g\right)  \right)  $ for all $\psi \in B_{0}\left(
\Sigma \right)  $. By construction, observe that there exists a bijection
$\tilde{f}:\tilde{V}\rightarrow V$. By Proposition \ref{pro:car}, we have that
there exists a bijection $f:V\rightarrow Q$. Define $\bar{f}=f\circ \tilde{f}$.
By Corollary \ref{cor:pro-can-rep-min}, if we define $\hat{I}_{q}=\bar
{I}_{\bar{f}^{-1}\left(  q\right)  }$ for all $q\in Q$, then we have that
$\hat{I}_{\bar{f}\left(  0\right)  }\leq \hat{I}_{q}$ for all $q\in Q$ and%
\begin{align*}
f  &  \succsim^{\ast}g\iff \bar{I}_{\psi}\left(  u\left(  f\right)  \right)
\geq \bar{I}_{\psi}\left(  u\left(  g\right)  \right)  \quad \forall \psi \in
B_{0}\left(  \Sigma \right)  \iff \bar{I}_{\psi}\left(  u\left(  f\right)
\right)  \geq \bar{I}_{\psi}\left(  u\left(  g\right)  \right)  \quad
\forall \psi \in \tilde{V}\\
&  \iff \hat{I}_{q}\left(  u\left(  f\right)  \right)  \geq \hat{I}_{q}\left(
u\left(  g\right)  \right)  \quad \forall q\in Q
\end{align*}
Since each $\hat{I}_{q}$ is a normalized concave niveloid, we have that\ for
each $q\in Q$\ there exists a function $c_{q}:\Delta \rightarrow \left[
0,\infty \right]  $ which is grounded, lower semicontinuous, convex and\ such
that%
\[
\hat{I}_{q}\left(  \varphi \right)  =\min_{p\in \Delta}\left \{  \int \varphi
dp+c_{q}\left(  p\right)  \right \}  \qquad \forall \varphi \in B_{0}\left(
\Sigma \right)
\]
Define $c:\Delta \times Q\rightarrow \left[  0,\infty \right]  $ by $c\left(
p,q\right)  =c_{q}\left(  p\right)  $ for all $\left(  p,q\right)  \in
\Delta \times Q$. Clearly, the $q$-sections of $c$ are grounded, lower
semicontinuous and convex and\ (\ref{eq:Bew-var}) holds. By Lemma
\ref{lem:del-Q} and (\ref{eq:Bew-var})\ and since $\succsim^{\ast}$ is
objectively $Q$-coherent, it follows that $\operatorname*{dom}c\left(
\cdot,q\right)  \subseteq \Delta^{\ll}\left(  Q\right)  $ for all $q\in Q$.
Finally, recall that%
\[
c\left(  p,q\right)  =\sup_{\varphi \in B_{0}\left(  \Sigma \right)  }\left \{
\hat{I}_{q}\left(  \varphi \right)  -\int \varphi dp\right \}  \qquad \forall
p\in \Delta,\forall q\in Q
\]
Since $\hat{I}_{\bar{f}\left(  0\right)  }\leq \hat{I}_{q}$ for all $q\in Q$,
we have that for each $q\in Q$%
\[
c\left(  p,\bar{f}\left(  0\right)  \right)  =\sup_{\varphi \in B_{0}\left(
\Sigma \right)  }\left \{  \hat{I}_{\bar{f}\left(  0\right)  }\left(
\varphi \right)  -\int \varphi dp\right \}  \leq \sup_{\varphi \in B_{0}\left(
\Sigma \right)  }\left \{  \hat{I}_{q}\left(  \varphi \right)  -\int \varphi
dp\right \}  =c\left(  p,q\right)  \quad \forall p\in \Delta
\]
Since $c\left(  \cdot,\bar{f}\left(  0\right)  \right)  $ is grounded, lower
semicontinuous and convex and $\bar{f}\left(  0\right)  \in Q$, this implies
that $c_{Q}\left(  \cdot \right)  =\min_{q\in Q}c\left(  \cdot,q\right)
=c\left(  \cdot,\bar{f}\left(  0\right)  \right)  $ is well defined and shares
the same properties, proving that $c$ is variational.\hfill$\blacksquare$

\subsubsection{Proof of Theorem \ref{thm:mai-rep}\label{app:fin-pro-mai}}

(i) implies (ii). We proceed by steps. Before starting, we make one
observation. By Lemma \ref{lem:inc-var}\ and since $\succsim^{\ast}$ is an
unbounded dominance relation which is objectively\ $Q$-coherent\ there
exist\ an onto affine function $u:X\rightarrow \mathbb{R}$ and a
variational\ $c:\Delta \times Q\rightarrow \left[  0,\infty \right]  $ such that
$\operatorname*{dom}c\left(  \cdot,q\right)  \subseteq \Delta^{\ll}\left(
Q\right)  $ for all $q\in Q$ (in particular, $\operatorname*{dom}c_{Q}\left(
\cdot \right)  \subseteq \cup_{q\in Q}\operatorname*{dom}c\left(  \cdot
,q\right)  \subseteq \Delta^{\ll}\left(  Q\right)  $)\ and\
\[
f\succsim^{\ast}g\iff \min_{p\in \Delta}\left \{  \int u\left(  f\right)
dp+c\left(  p,q\right)  \right \}  \geq \min_{p\in \Delta}\left \{  \int u\left(
g\right)  dp+c\left(  p,q\right)  \right \}  \quad \forall q\in Q
\]
We are left to show that $c_{Q}:\Delta \rightarrow \left[  0,\infty \right]  $ is
such that%
\begin{equation}
f\succsim g\Longleftrightarrow \min_{p\in \Delta}\left \{  \int u\left(
f\right)  dp+c_{Q}\left(  p\right)  \right \}  \geq \min_{p\in \Delta}\left \{
\int u\left(  g\right)  dp+c_{Q}\left(  p\right)  \right \}  \label{eq:rep-pro}%
\end{equation}
and $c_{Q}^{-1}\left(  0\right)  =Q$. To prove this we consider $c$\ as in the
proof of (i) implies (ii)\ of\ Lemma \ref{lem:inc-var}. This covers both cases
$\left \vert Q\right \vert =1$ and $\left \vert Q\right \vert >1$. In particular,
for each $q\in Q$ define $\hat{I}_{q}:B_{0}\left(  \Sigma \right)
\rightarrow \mathbb{R}$ by%
\[
\hat{I}_{q}\left(  \varphi \right)  =\min_{p\in \Delta}\left \{  \int \varphi
dp+c\left(  p,q\right)  \right \}  \qquad \forall \varphi \in B_{0}\left(
\Sigma \right)
\]
and recall that there exists $\hat{q}$($=\bar{f}\left(  0\right)  \in Q$\ when
$\left \vert Q\right \vert >1$) such that $c\left(  \cdot,\hat{q}\right)  \leq
c\left(  \cdot,q\right)  $, thus $\hat{I}_{\hat{q}}\leq \hat{I}_{q}$, for all
$q\in Q$.

\smallskip

\noindent \textit{Step 1. }$\succsim$\textit{ agrees with }$\succsim^{\ast}$
\textit{on }$X$. \textit{In particular, }$u:X\rightarrow \mathbb{R}$
\textit{represents} $\succsim^{\ast}$\textit{ and }$\succsim$.

\noindent \textit{Proof of the Step }Note that $\succsim^{\ast}$ and $\succsim$
restricted to $X$ are continuous weak orders that satisfy risk independence.
Moreover, by the observation above, $\succsim^{\ast}$ is represented by
$u$.\ By Herstein and Milnor (1953) and since $\succsim$ is non-trivial, it
follows that there exists a non-constant and affine function $v:X\rightarrow
\mathbb{R}$ that represents\ $\succsim$ on $X$. Since $\left(  \succsim^{\ast
},\succsim \right)  $ jointly satisfy consistency, it follows that for each
$x,y\in X$%
\[
u\left(  x\right)  \geq u\left(  y\right)  \Longrightarrow v\left(  x\right)
\geq v\left(  y\right)
\]
By Corollary B.3 of Ghirardato et al. (2004), $u\ $and $v$ are equal up to an
affine and positive transformation, hence the statement. We can set
$v=u$.\hfill$\square$

\smallskip

\noindent \textit{Step 2. There exists a normalized, monotone and continuous
functional\ }$I:B_{0}\left(  \Sigma \right)  \rightarrow \mathbb{R}%
\ $\textit{such that}%
\[
f\succsim g\Longleftrightarrow I\left(  u\left(  f\right)  \right)  \geq
I\left(  u\left(  g\right)  \right)
\]

\noindent \textit{Proof of the Step }By Cerreia-Vioglio et al. (2011a) and
since\ $\succsim$ is a rational preference relation, the statement
follows.\hfill$\square$

\smallskip

\noindent \textit{Step 3. }$I\left(  \varphi \right)  \leq \inf_{q\in Q}\hat
{I}_{q}\left(  \varphi \right)  $ \textit{for all }$\varphi \in B_{0}\left(
\Sigma \right)  $\textit{.}

\noindent \textit{Proof of the Step }Consider $\varphi \in B_{0}\left(
\Sigma \right)  $. Since each $\hat{I}_{q}$ is normalized and monotone and $u$
is onto, we have that $\hat{I}_{q}\left(  \varphi \right)  \in \left[
\inf_{s\in S}\varphi \left(  s\right)  ,\sup_{s\in S}\varphi \left(  s\right)
\right]  \subseteq \operatorname{Im}u\ $for all $q\in Q$. Since $\varphi \in
B_{0}\left(  \Sigma \right)  $, it follows that there exists $f\in \mathcal{F}$
such that $\varphi=u\left(  f\right)  $ and $x\in X$ such that $u\left(
x\right)  =\inf_{q\in Q}\hat{I}_{q}\left(  \varphi \right)  $. For each
$\varepsilon>0\ $there exists $x_{\varepsilon}\in X$ such that $u\left(
x_{\varepsilon}\right)  =u\left(  x\right)  +\varepsilon$. Since $\inf_{q\in
Q}\hat{I}_{q}\left(  \varphi \right)  =u\left(  x\right)  $, it follows that
for each $\varepsilon>0$ there exists $q\in Q$ such that $\hat{I}_{q}\left(
u\left(  f\right)  \right)  =\hat{I}_{q}\left(  \varphi \right)  <u\left(
x_{\varepsilon}\right)  =\hat{I}_{q}\left(  u\left(  x_{\varepsilon}\right)
\right)  $, yielding that $f\not \succsim ^{\ast}x_{\varepsilon}$. Since
$\left(  \succsim^{\ast},\succsim \right)  $ jointly satisfy caution, we have
that $x_{\varepsilon}\succsim f$ for all $\varepsilon>0$.\ By Step 2, this
implies that%
\[
u\left(  x\right)  +\varepsilon=u\left(  x_{\varepsilon}\right)  =I\left(
u\left(  x_{\varepsilon}\right)  \right)  \geq I\left(  u\left(  f\right)
\right)  =I\left(  \varphi \right)  \quad \forall \varepsilon>0
\]
that is, $\inf_{q\in Q}\hat{I}_{q}\left(  \varphi \right)  =u\left(  x\right)
\geq I\left(  \varphi \right)  $, proving the step.\hfill$\square$

\smallskip

\noindent \textit{Step 4. }$I\left(  \varphi \right)  \geq \inf_{q\in Q}\hat
{I}_{q}\left(  \varphi \right)  $ \textit{for all }$\varphi \in B_{0}\left(
\Sigma \right)  $\textit{.}

\noindent \textit{Proof of the Step }Consider $\varphi \in B_{0}\left(
\Sigma \right)  $. We use the same objects and notation of Step 3. Note that
for each $q^{\prime}\in Q$%
\[
\hat{I}_{q^{\prime}}\left(  u\left(  f\right)  \right)  =\hat{I}_{q^{\prime}%
}\left(  \varphi \right)  \geq \inf_{q\in Q}\hat{I}_{q}\left(  \varphi \right)
=u\left(  x\right)  =\hat{I}_{q^{\prime}}\left(  u\left(  x\right)  \right)
\]
that is, $f\succsim^{\ast}x$. Since $\left(  \succsim^{\ast},\succsim \right)
$ jointly satisfy consistency, we have that $f\succsim x$. By Step 2, this
implies that%
\[
I\left(  \varphi \right)  =I\left(  u\left(  f\right)  \right)  \geq I\left(
u\left(  x\right)  \right)  =u\left(  x\right)  =\inf_{q\in Q}\hat{I}%
_{q}\left(  \varphi \right)
\]
proving the step.\hfill$\square$

\smallskip

\noindent \textit{Step 5. }$I\left(  \varphi \right)  =\min_{p\in \Delta}\left \{
\int \varphi dp+c_{Q}\left(  p\right)  \right \}  $\  \textit{for all }%
$\varphi \in B_{0}\left(  \Sigma \right)  $\textit{.}

\noindent \textit{Proof of the Step }By Steps 3 and 4 and since $\hat{I}%
_{\hat{q}}\leq \hat{I}_{q}$ for all $q\in Q$, we have that%
\[
I\left(  \varphi \right)  =\min_{q\in Q}\hat{I}_{q}\left(  \varphi \right)
=\hat{I}_{\hat{q}}\left(  \varphi \right)  \qquad \forall \varphi \in B_{0}\left(
\Sigma \right)
\]
Since $c\left(  \cdot,\hat{q}\right)  =c_{Q}\left(  \cdot \right)  $, it
follows that for each $\varphi \in B_{0}\left(  \Sigma \right)  $%
\[
I\left(  \varphi \right)  =\hat{I}_{\hat{q}}\left(  \varphi \right)  =\min
_{p\in \Delta}\left \{  \int \varphi dp+c\left(  p,\hat{q}\right)  \right \}
=\min_{p\in \Delta}\left \{  \int \varphi dp+c_{Q}\left(  p\right)  \right \}
\]
proving the step.\hfill$\square$

\smallskip

\noindent \textit{Step 6. }$c_{Q}^{-1}\left(  0\right)  =Q$\textit{.}

\noindent \textit{Proof of the Step }By Steps 2 and 5, we have that
$V:\mathcal{F}\rightarrow \mathbb{R}$ defined by%
\[
V\left(  f\right)  =\min_{p\in \Delta}\left \{  \int u\left(  f\right)
dp+c_{Q}\left(  p\right)  \right \}
\]
represents $\succsim$. By Lemma \ref{lm:zero-set} and since $\succsim$ is
subjectively $Q$-coherent and $c_{Q}$ is well defined, grounded,\ lower
semicontinuous and convex, we can conclude that $c_{Q}^{-1}\left(  0\right)
=Q$.\hfill$\square$

Thus, (\ref{eq:rep-pro}) follows from Steps 2 and 5 while, by Step
6,\ $c_{Q}^{-1}\left(  0\right)  =Q$.\ This\ completes the proof.

\smallskip

(ii) implies (i). It is routine.

\smallskip

Next, assume that $c$ is uniquely null.\ Define the\ correspondence
$\Gamma:Q\rightrightarrows Q$ by
\[
\Gamma \left(  q\right)  =\left \{  p\in \Delta:c\left(  p,q\right)  =0\right \}
=\arg \min c_{q}%
\]
Since $c_{Q}\leq c_{q}$ for all $q\in Q$ and $c_{Q}^{-1}\left(  0\right)  =Q$,
we have that $\Gamma$ is well defined. Since $c_{q}$ is grounded, it follows
that $\Gamma \left(  q\right)  \not =\emptyset$\ for all $q\in Q$. Since $c$ is
uniquely null and $c_{q}$ is grounded, we have that $c_{q}^{-1}\left(
0\right)  $ is a singleton, that is,%
\[
c\left(  p,q\right)  =c\left(  p^{\prime},q\right)  =0\Longrightarrow
p=p^{\prime}%
\]
This implies that $\Gamma \left(  q\right)  $ is a singleton, therefore
$\Gamma$\ is a function. Since $c_{Q}^{-1}\left(  0\right)  =Q$,\ observe that%

\[
\cup_{q\in Q}\Gamma \left(  q\right)  =\cup_{q\in Q}\arg \min c_{q}=\arg \min
c_{Q}=Q
\]
that is, $\Gamma$ is surjective. Since $c$ is uniquely null, we have that
$c_{p}^{-1}\left(  0\right)  $ is at most a singleton, that is,%
\[
c\left(  p,q\right)  =c\left(  p,q^{\prime}\right)  =0\implies q=q^{\prime}%
\]
yielding that $\Gamma$ is injective. To sum up, $\Gamma \ $is a bijection.
Define $\tilde{c}:\Delta \times Q\rightarrow \left[  0,\infty \right]  $ by
$\tilde{c}\left(  p,q\right)  =c\left(  p,\Gamma^{-1}\left(  q\right)
\right)  $ for all $\left(  p,q\right)  \in \Delta \times Q$. Note that
$\tilde{c}\left(  \cdot,q\right)  $ is grounded, lower semicontinuous, convex
and $\operatorname*{dom}\tilde{c}\left(  \cdot,q\right)  \subseteq \Delta^{\ll
}\left(  Q\right)  $ for all $q\in Q$ and $\operatorname*{dom}\tilde{c}%
_{Q}\left(  \cdot \right)  \subseteq \Delta^{\ll}\left(  Q\right)  $. Next, we
show that $\tilde{c}_{Q}=c_{Q}$. Since $c_{Q}$\ is well defined, for each
$p\in \Delta$ there exists $q_{p}\in Q$\ such that%

\[
\tilde{c}\left(  p,\Gamma \left(  q_{p}\right)  \right)  =c\left(
p,q_{p}\right)  =\min_{q\in Q}c\left(  p,q\right)  \leq c\left(  p,q^{\prime
}\right)  =\tilde{c}\left(  p,\Gamma \left(  q^{\prime}\right)  \right)
\quad \forall q^{\prime}\in Q
\]
Since $\Gamma$ is a bijection, we have that $\tilde{c}\left(  p,\Gamma \left(
q_{p}\right)  \right)  \leq \tilde{c}\left(  p,q\right)  $ for all $q\in Q$.
Since $p$ was arbitrarily chosen, it follows that%
\[
c_{Q}\left(  p\right)  =\min_{q\in Q}c\left(  p,q\right)  =\tilde{c}\left(
p,\Gamma \left(  q_{p}\right)  \right)  =\min_{q\in Q}\tilde{c}\left(
p,q\right)  =\tilde{c}_{Q}\left(  p\right)  \quad \forall p\in \Delta
\]
To sum up, $\tilde{c}_{Q}=c_{Q}$ and $\tilde{c}_{Q}^{-1}\left(  0\right)
=c_{Q}^{-1}\left(  0\right)  =Q$. In turn, since $c_{Q}$ is grounded, lower
semicontinuous and convex, this implies that $\tilde{c}_{Q}$ is grounded,
lower semicontinuous and convex. Since $\Gamma$ is a bijection, we can
conclude that (\ref{eq:Bew-var-mai}) holds with $\tilde{c}$ in place of $c$
and (\ref{eq:var-var-mai}) holds with $\tilde{c}_{Q}$ in place of $c_{Q}$.

We are left to show that\ $\tilde{c}\left(  p,q\right)  =0$ if and only
if\ $p=q$. Since $c_{q}^{-1}\left(  0\right)  $ is a singleton for all $q\in
Q$ and $\Gamma$ is a bijection, if\ $\tilde{c}\left(  p,q\right)  =0$,
then\ $c\left(  p,\Gamma^{-1}\left(  q\right)  \right)  =0$, yielding
that\ $p=\Gamma \left(  \Gamma^{-1}\left(  q\right)  \right)  =q$. On the other
hand, $\tilde{c}\left(  q,q\right)  =c\left(  q,\Gamma^{-1}\left(  q\right)
\right)  =0$. We can conclude that $\tilde{c}\left(  p,q\right)  =0$ if and
only if $p=q$, proving that $\tilde{c}$ is a statistical distance.\hfill
$\blacksquare$

\subsubsection{Proof of Theorem \ref{thm:mai-rep-inc-fin},
Propositions\  \ref{pro:mai-rep-inc} and \ref{pro:com-no-cau}%
\label{app:thm-var-str}}

\noindent \textbf{Proof of Theorem \ref{thm:mai-rep-inc-fin}} We only prove (i)
implies (ii), the converse being routine.\footnote{The only exception is the
proof that the representation implies subjective $Q$-coherence. This is a
consequence of Theorem 2.4.18 in Zalinescu (2002) paired with Lemma 32 of
Maccheroni et al. (2006).} We proceed by steps.

\smallskip

\noindent \textit{Step 1. }$\succsim_{Q}^{\ast}$\textit{ agrees with }%
$\succsim_{Q^{\prime}}^{\ast}$ \textit{on }$X$\textit{ for all }$Q,Q^{\prime
}\in \mathcal{Q}$. \textit{In particular, there exists an affine and onto
function }$u:X\rightarrow \mathbb{R}$ \textit{representing}\ $\succsim
_{Q}^{\ast}$\textit{ for all }$Q\in \mathcal{Q}$.

\noindent \textit{Proof of the Step }Let $Q,Q^{\prime}\in \mathcal{Q}$ be such
that $Q\supseteq Q^{\prime}$. Note that $\succsim_{Q}^{\ast}$ and
$\succsim_{Q^{\prime}}^{\ast}$, restricted to $X$, satisfy weak order,
continuity and risk independence. By Herstein and Milnor (1953) and since
$\succsim_{Q}^{\ast}$ and $\succsim_{Q^{\prime}}^{\ast}$ are non-trivial,
there exist two non-constant affine functions $u_{Q},u_{Q^{\prime}%
}:X\rightarrow \mathbb{R}$\ which represent $\succsim_{Q}^{\ast}$ and
$\succsim_{Q^{\prime}}^{\ast}$, respectively. Since $\left \{  \succsim
_{Q}^{\ast}\right \}  _{Q\in \mathcal{Q}}$ is monotone in model ambiguity, we
have that%
\[
u_{Q}\left(  x\right)  \geq u_{Q}\left(  y\right)  \Longrightarrow
u_{Q^{\prime}}\left(  x\right)  \geq u_{Q^{\prime}}\left(  y\right)
\]
By Corollary B.3 of Ghirardato et al. (2004), $u_{Q}\ $and $u_{Q^{\prime}}$
are equal up to an affine and positive transformation. Next, fix $\bar{q}%
\in \Delta^{\sigma}$. Set $u=u_{\bar{q}}$. Given any other $q\in \Delta^{\sigma
}$,$\ $consider $\bar{Q}\in \mathcal{Q}$ such that $\bar{Q}\supseteq \left \{
\bar{q},q\right \}  $. By the previous part, it follows that $u_{\bar{Q}}$,
$u_{q}$\ and $u_{\bar{q}}$ are equal up to an affine and
positive\ transformation. Given that $q$ was arbitrarily chosen, we can set
$u=u_{q}$ for all $q\in Q$. Similarly, given a generic $Q\in \mathcal{Q}$,
select $q\in Q$. Since $Q\supseteq \left \{  q\right \}  $, it follows that we
can set $u=u_{Q}$. Since each $\succsim_{Q}^{\ast}$ is unbounded for all
$Q\in \mathcal{Q}$, we have that $u$ is onto.\hfill$\square$

\smallskip

\noindent \textit{Step 2. For each }$q\in \Delta^{\sigma}$\textit{ there exists
a normalized, monotone, translation invariant and concave functional\ }%
$I_{q}:B_{0}\left(  \Sigma \right)  \rightarrow \mathbb{R}\ $\textit{such that}%
\begin{equation}
f\succsim_{q}^{\ast}g\Longleftrightarrow I_{q}\left(  u\left(  f\right)
\right)  \geq I_{q}\left(  u\left(  g\right)  \right)  \label{eq:rep-swi-q}%
\end{equation}
\textit{Moreover, there exists a unique grounded, lower semicontinuous and
convex function }$c_{q}:\Delta \rightarrow \left[  0,\infty \right]
$\textit{\ such that}%
\begin{equation}
I_{q}\left(  \varphi \right)  =\min_{p\in \Delta}\left \{  \int \varphi
dp+c_{q}\left(  p\right)  \right \}  \qquad \forall \varphi \in B_{0}\left(
\Sigma \right)  \label{eq:niv-rep-sta}%
\end{equation}

\noindent \textit{Proof of the Step }Fix $q\in \Delta^{\sigma}$. Since
$\succsim_{q}^{\ast}$ is an unbounded dominance relation which is complete, we
have that $\succsim_{q}^{\ast}$ is a variational preference. By the proof of
Theorem 3 and Proposition 6 of Maccheroni et al. (2006) and Step\ 1, there
exists an onto and affine function $u_{q}:X\rightarrow \mathbb{R}$, which can
be set to be equal to $u$, and, given $u$, a unique grounded, lower
semicontinuous and convex function $c_{q}:\Delta \rightarrow \left[
0,\infty \right]  $ such that (\ref{eq:niv-rep-sta}) and (\ref{eq:rep-swi-q}%
)\ hold.\hfill$\square$

\smallskip

Define $c:\Delta \times \Delta^{\sigma}\rightarrow \left[  0,\infty \right]  $ by
$c\left(  p,q\right)  =c_{q}\left(  p\right)  $ for all $\left(  p,q\right)
\in \Delta \times \Delta^{\sigma}$.

\noindent \textit{Step 3. For each }$Q\in \mathcal{Q}$\textit{ we have that
}$f\succsim_{Q}^{\ast}g$\textit{ if and only if} $f\succsim_{q}^{\ast}%
g$\textit{ for all }$q\in Q$\textit{. In particular, we have that}%
\begin{equation}
f\succsim_{Q}^{\ast}g\iff \min_{p\in \Delta}\left \{  \int u\left(  f\right)
dp+c\left(  p,q\right)  \right \}  \geq \min_{p\in \Delta}\left \{  \int u\left(
g\right)  dp+c\left(  p,q\right)  \right \}  \quad \forall q\in Q
\label{eq:una-dom-rig}%
\end{equation}

\noindent \textit{Proof of the Step }Fix $Q\in \mathcal{Q}$.\ Since $\left \{
\succsim_{Q}^{\ast}\right \}  _{Q\in \mathcal{Q}}$ is monotone in model
ambiguity, we have that%
\[
f\succsim_{Q}^{\ast}g\implies f\succsim_{q}^{\ast}g\qquad \forall q\in Q
\]
Since $\left \{  \succsim_{Q}^{\ast}\right \}  _{Q\in \mathcal{Q}}$ is
$Q$-separable, we can conclude that $f\succsim_{Q}^{\ast}g$ if and only if
$f\succsim_{q}^{\ast}g$ for all $q\in Q$. By Step 2 and the definition of $c$,
(\ref{eq:una-dom-rig}) follows.\hfill$\square$

\smallskip

\noindent \textit{Step 4. }$\succsim_{Q}^{\ast}$\textit{ agrees with }%
$\succsim_{Q}$ \textit{on }$X$\textit{ for all }$Q\in \mathcal{Q}$\textit{.
Moreover, }$\succsim_{Q}$\textit{ is represented by the function }$u$\textit{
of Step 1.}

\noindent \textit{Proof of the Step }Fix $Q\in \mathcal{Q}$. Note that
$\succsim_{Q}^{\ast}$ and $\succsim_{Q}$, restricted to $X$, satisfy weak
order, continuity and risk independence. By Herstein and Milnor (1953) and
since\ $\succsim_{Q}$ is non-trivial, there exists\ a non-constant affine
function $v_{Q}$ which represents\ $\succsim_{Q}$. By Step 1, $\succsim
_{Q}^{\ast}$ is represented by $u$. Since $\left(  \succsim_{Q}^{\ast
},\succsim_{Q}\right)  $ jointly satisfy consistency, it follows that for each
$x,y\in X$%
\[
u\left(  x\right)  \geq u\left(  y\right)  \Longrightarrow v_{Q}\left(
x\right)  \geq v_{Q}\left(  y\right)
\]
By Corollary B.3 of Ghirardato et al. (2004), $v_{Q}\ $and $u$ are equal up to
an affine and positive transformation. So we can set $v_{Q}=u$, proving the
statement.\hfill$\square$

\smallskip

\noindent \textit{Step 5. For each }$Q\in \mathcal{Q}$\textit{ we have that}%
\begin{equation}
f\succsim_{Q}g\iff \inf_{p\in \Delta}\left \{  \int u\left(  f\right)
dp+\inf_{q\in Q}c\left(  p,q\right)  \right \}  \geq \inf_{p\in \Delta}\left \{
\int u\left(  g\right)  dp+\inf_{q\in Q}c\left(  p,q\right)  \right \}
\label{eq:rat-rep-var}%
\end{equation}
\noindent \textit{Proof of the Step }Fix $Q\in \mathcal{Q}$. By Cerreia-Vioglio
et al. (2011a) and since\ $\succsim_{Q}$ is a rational preference relation,
there exists a normalized, monotone and continuous functional $I_{Q}%
:B_{0}\left(  \Sigma \right)  \rightarrow \mathbb{R}$\ such that%
\begin{equation}
f\succsim_{Q}g\iff I_{Q}\left(  u\left(  f\right)  \right)  \geq I_{Q}\left(
u\left(  g\right)  \right)  \label{eq:rat-rep-Q}%
\end{equation}
By the same arguments in Steps 3 and 4 of Theorem \ref{thm:mai-rep}, we have
that $I_{Q}=\inf_{q\in Q}I_{q}$, yielding that%
\begin{align*}
I_{Q}\left(  \varphi \right)   &  =\inf_{q\in Q}\min_{p\in \Delta}\left \{
\int \varphi dp+c\left(  p,q\right)  \right \}  =\inf_{q\in Q}\inf_{p\in \Delta
}\left \{  \int \varphi dp+c\left(  p,q\right)  \right \} \\
&  =\inf_{p\in \Delta}\inf_{q\in Q}\left \{  \int \varphi dp+c\left(  p,q\right)
\right \}  =\inf_{p\in \Delta}\left \{  \int \varphi dp+\inf_{q\in Q}c\left(
p,q\right)  \right \}  \quad \forall \varphi \in B_{0}\left(  \Sigma \right)
\end{align*}
By (\ref{eq:rat-rep-Q}), this implies that (\ref{eq:rat-rep-var})
holds.\hfill$\square$

\smallskip

\noindent \textit{Step 6. }$c\left(  p,q\right)  =0$ \textit{if and only
if}\ $p=q$\textit{.}

\noindent \textit{Proof of the Step }By Steps 2 and\ 5, we have that
$\succsim_{q}^{\ast}$ coincides with $\succsim_{q}$ on $\mathcal{F}$ for all
$q\in \Delta^{\sigma}$. By Lemma \ref{lm:zero-set}\ and since $\succsim_{q}$ is
subjectively $\left \{  q\right \}  $-coherent, we have that
$\operatorname{argmin}c\left(  \cdot,q\right)  =\operatorname{argmin}%
c_{q}=\left \{  q\right \}  $.\hfill$\square$

\smallskip

\noindent \textit{Step 7. }$\operatorname*{dom}c\left(  \cdot,q\right)
\subseteq \Delta^{\ll}\left(  q\right)  \subseteq \Delta^{\ll}\left(  Q\right)
$\  \textit{for all }$q\in Q$\textit{\ and for all }$Q\in \mathcal{Q}$\textit{.}

\noindent \textit{Proof of the Step }By the previous part of the proof, we have
that $\succsim_{q}^{\ast}$ coincides with $\succsim_{q}$ on $\mathcal{F}$ for
all $q\in \Delta^{\sigma}$. By Lemma \ref{lem:del-Q} and since $\succsim
_{q}^{\ast}$ is objectively $\left \{  q\right \}  $-coherent, we can conclude
that $\operatorname*{dom}c\left(  \cdot,q\right)  \subseteq \Delta^{\ll}\left(
q\right)  \subseteq \Delta^{\ll}\left(  Q\right)  $\ for all $q\in Q$\ and for
all $Q\in \mathcal{Q}$.\hfill$\square$

\smallskip

\noindent \textit{Step 8. }$c$\  \textit{is jointly lower semicontinuous.}

\noindent \textit{Proof of the Step }Define the map $J:B_{0}\left(
\Sigma \right)  \times \Delta^{\sigma}\rightarrow \mathbb{R}$ by $J\left(
\varphi,q\right)  =I_{q}\left(  \varphi \right)  $ for all $q\in Q$.\ Observe
that, for each\ $\left(  p,q\right)  \in \Delta \times \Delta^{\sigma}$,%
\begin{equation}
c\left(  p,q\right)  =c_{q}\left(  p\right)  =\sup_{\varphi \in B_{0}\left(
\Sigma \right)  }\left \{  I_{q}\left(  \varphi \right)  -\int \varphi dp\right \}
=\sup_{\varphi \in B_{0}\left(  \Sigma \right)  }\left \{  J\left(
\varphi,q\right)  -\int \varphi dp\right \}  \label{eq:cos-fun-niv}%
\end{equation}
We begin by observing that $J\ $is lower semicontinuous in the second
argument. Note that for each $\varphi \in B_{0}\left(  \Sigma \right)  $ and for
each $q\in \Delta^{\sigma}$%
\[
J\left(  \varphi,q\right)  =I_{q}\left(  \varphi \right)  =u\left(
x_{f,q}\right)  \qquad \text{where }f\in \mathcal{F}\text{ is s.t. }%
\varphi=u\left(  f\right)
\]
Fix $\varphi \in B_{0}\left(  \Sigma \right)  $ and $t\in \mathbb{R}$. By the
axiom of lower semicontinuity, the set%
\[
\left \{  q\in \Delta^{\sigma}:J\left(  \varphi,q\right)  \leq t\right \}
=\left \{  q\in \Delta^{\sigma}:u\left(  x\right)  \geq u\left(  x_{f,q}\right)
\right \}  =\left \{  q\in \Delta^{\sigma}:x\succsim_{q}^{\ast}x_{f,q}\right \}
\]
is closed where $x\in X$ and $f\in \mathcal{F}$ are such that $u\left(
x\right)  =t$\ as well as $u\left(  f\right)  =\varphi$. Since $\varphi$ and
$t$ were arbitrarily chosen, this yields that $J$ is lower semicontinuous in
the second argument. Since $J$ is lower semicontinuous in the second
argument,\ the map $\left(  p,q\right)  \mapsto J\left(  \varphi,q\right)
-\int \varphi dp$, defined over $\Delta \times \Delta^{\sigma}$, is jointly lower
semicontinuous for all $\varphi \in B_{0}\left(  \Sigma \right)  $. By
(\ref{eq:cos-fun-niv}) and the definition of $c$, we conclude that $c$ is
jointly lower semicontinuous.\hfill$\square$

\smallskip

Step 1 proves that $u$ is affine and onto. Steps 2, 6, 7 and 8 prove that $c$
is a jointly lower semicontinuous divergence which is convex in the first
argument. Steps 1, 3, 5 and 8\ yield the representation of $\succsim_{Q}%
^{\ast}$ and $\succsim_{Q}$ for all $Q\in \mathcal{Q}$. As for uniqueness,
assume that the function $\tilde{c}:\Delta \times \Delta^{\sigma}\rightarrow
\left[  0,\infty \right]  $ is a divergence which is jointly lower
semicontinuous, convex in the first argument and that represents $\succsim
_{Q}^{\ast}$ and $\succsim_{Q}$ for all $Q\in \mathcal{Q}$. By Proposition 6 of
Maccheroni et al. (2006) and since $\operatorname{Im}u=\mathbb{R}$\ and
$\succsim_{q}^{\ast}$ is a variational preference for all $q\in \Delta^{\sigma
}$, it follows that $\tilde{c}\left(  \cdot,q\right)  =c\left(  \cdot
,q\right)  $ for all $q\in \Delta^{\sigma}$, yielding that $c=\tilde{c}$%
.\hfill$\blacksquare$

\smallskip

\noindent \textbf{Proof of Proposition\  \ref{pro:mai-rep-inc}} We only prove
(i) implies (ii), the converse being routine. We keep the notation of the
previous proof. Compared to Theorem \ref{thm:mai-rep-inc-fin}, we only need to
prove that $c$ is jointly convex. By Lemma \ref{lem:app-suff-div} in Appendix
\ref{app:sta-dis}, this will yield that $c$ is variational. Fix $\varphi \in
B_{0}\left(  \Sigma \right)  $, $q,q^{\prime}\in \Delta^{\sigma}$ and
$\lambda \in \left(  0,1\right)  $. By model hybridization aversion and since
$u$ is affine, we have that%
\begin{align*}
J\left(  \varphi,\lambda q+\left(  1-\lambda \right)  q^{\prime}\right)   &
=u\left(  x_{f,\lambda q+\left(  1-\lambda \right)  q^{\prime}}\right)  \leq
u\left(  \lambda x_{f,q}+\left(  1-\lambda \right)  x_{f,q^{\prime}}\right) \\
&  =\lambda u\left(  x_{f,q}\right)  +\left(  1-\lambda \right)  u\left(
x_{f,q^{\prime}}\right)  =\lambda J\left(  \varphi,q\right)  +\left(
1-\lambda \right)  J\left(  \varphi,q^{\prime}\right)
\end{align*}
where $f\in \mathcal{F}$ is such that $u\left(  f\right)  =\varphi$. Since
$\varphi$, $q$, $q^{\prime}$ and $\lambda$\ were arbitrarily chosen, this
yields that $J$ is convex in the second argument. Since $J$ is convex in the
second argument,\ the map $\left(  p,q\right)  \mapsto J\left(  \varphi
,q\right)  -\int \varphi dp$, defined over $\Delta \times \Delta^{\sigma}$, is
jointly convex for all $\varphi \in B_{0}\left(  \Sigma \right)  $. By
(\ref{eq:cos-fun-niv}) and the definition of $c$, we conclude that $c$ is
convex, proving the implication.\hfill$\blacksquare$

\smallskip

\noindent \textbf{Proof of Proposition\  \ref{pro:com-no-cau}\ }We only
prove\textbf{ }(i) implies (ii), being (ii) implies (i) routine. We keep the
same notation and terminology as in the proof and statement of Theorem
\ref{thm:mai-rep-inc-fin}. It is then immediate to notice that Steps 1--4 of
that proof continue to hold here. In particular, there exist an onto and
affine utility function and a function $c:\Delta \times \Delta^{\sigma
}\rightarrow \left[  0,\infty \right]  $, which is grounded, convex and lower
semicontinuous in the first argument, such that for each $Q\in \mathcal{Q}$ and
for each $f,g\in \mathcal{F}$%
\begin{equation}
f\succsim_{Q}^{\ast}g\iff \min_{p\in \Delta}\left \{  \int u\left(  f\right)
dp+c\left(  p,q\right)  \right \}  \geq \min_{p\in \Delta}\left \{  \int u\left(
g\right)  dp+c\left(  p,q\right)  \right \}  \quad \forall q\in
Q\label{eq:dom-rel-rep}%
\end{equation}
and for each $x,y\in X$%
\begin{equation}
x\succsim_{Q}^{\ast}y\iff x\succsim_{Q}y\iff u\left(  x\right)  \geq u\left(
y\right)  \label{eq:sam-u}%
\end{equation}
Fix $Q\in \mathcal{Q}$. Since $\succsim_{Q}$ is solvable, for each
$f\in \mathcal{F}$ there exists $x_{f,Q}\in X$ such that\ $x_{f,Q}\sim_{Q}f$.
Since $\operatorname{Im}u=\mathbb{R}$, define $I_{Q}:B_{0}\left(
\Sigma \right)  \rightarrow \mathbb{R}$ by $I_{Q}\left(  \varphi \right)
=u\left(  x_{f,Q}\right)  $ where $f\in \mathcal{F}$ is such that $u\left(
f\right)  =\varphi$. By (\ref{eq:sam-u}) and since $\succsim_{Q}$ is a
complete, transitive and monotone binary relation, we have that $I_{Q}$\ is
well defined and monotone. Moreover, by construction, we have that
$I_{Q}\left(  k1_{S}\right)  =k$ for all $k\in \mathbb{R}$. By (\ref{eq:sam-u})
and construction, note that%
\begin{equation}
I_{Q}\left(  u\left(  f\right)  \right)  \geq I_{Q}\left(  u\left(  g\right)
\right)  \iff u\left(  x_{f,Q}\right)  \geq u\left(  x_{g,Q}\right)  \iff
x_{f,Q}\succsim_{Q}x_{g,Q}\iff f\succsim_{Q}g\label{eq:I-Q}%
\end{equation}
Since $\succsim_{Q}^{\ast}$and $\succsim_{Q}$ jointly satisfy consistency for
all $Q\in \mathcal{Q}$,\footnote{By (\ref{eq:dom-rel-rep}) and (\ref{eq:I-Q}),
we have that $f\mapsto \min_{p\in \Delta}\left \{  \int u\left(  f\right)
dp+c\left(  p,q\right)  \right \}  $ and $f\mapsto I_{Q}\left(  u\left(
f\right)  \right)  $\ represent, respectively, $\succsim_{Q}^{\ast}$ and
$\succsim_{Q}$. Since $\succsim_{Q}^{\ast}$ and $\succsim_{Q}$ satisfy
consistency, we can conclude that there exists a (not necessarily strictly)
monotone function $h:\mathbb{R}\rightarrow \mathbb{R}$ such that $I_{Q}\left(
u\left(  f\right)  \right)  =h\left(  \min_{p\in \Delta}\left \{  \int u\left(
f\right)  dp+c\left(  p,q\right)  \right \}  \right)  $ for all $f\in
\mathcal{F}$. Since $I_{Q}$ is normalized and $\operatorname{Im}u=\mathbb{R}$,
we have that $h\left(  u\left(  x\right)  \right)  =u\left(  x\right)  $ for
all $x\in X$, proving that $h$ is the identity.} this implies that
$\succsim_{q}^{\ast}$and $\succsim_{q}$ coincide on $\mathcal{F}$\ for all
$q\in \Delta^{\sigma}$\ and%
\[
I_{q}\left(  \varphi \right)  =\min_{p\in \Delta}\left \{  \int \varphi
dp+c\left(  p,q\right)  \right \}  \qquad \forall \varphi \in B_{0}\left(
\Sigma \right)
\]
In particular, we have that Steps 6--8 of Theorem \ref{thm:mai-rep-inc-fin}
hold also in this case, proving that $c$ is a lower semicontinuous divergence,
which is convex in the first argument. Given $\varphi \in B_{0}\left(
\Sigma \right)  $, note that the map $q\mapsto I_{q}\left(  \varphi \right)  $
is such that $\min_{s\in S}\varphi \left(  s\right)  \leq I_{q}\left(
\varphi \right)  \leq \max_{s\in S}\varphi \left(  s\right)  $ for all $q\in Q$,
yielding that the map $q\mapsto I_{q}\left(  \varphi \right)  $ is an element
of $B\left(  Q\right)  $. Consider the set%
\[
M=\left \{  \tilde{\varphi}\in B\left(  Q\right)  :\exists \varphi \in
B_{0}\left(  \Sigma \right)  \text{ s.t. }\forall q\in Q,\tilde{\varphi}\left(
q\right)  =I_{q}\left(  \varphi \right)  \right \}
\]
Since $I_{q}\left(  k1_{S}\right)  =k$ for all $k\in \mathbb{R}$, we have that
$M$ contains all the constants $k1_{Q}\ $where $k\in \mathbb{R}$. Define
$\tilde{J}_{Q}:M\rightarrow \mathbb{R}$ by $\tilde{J}_{Q}\left(  \tilde
{\varphi}\right)  =I_{Q}\left(  \varphi \right)  $ where $\varphi \in
B_{0}\left(  \Sigma \right)  $ is such that $\tilde{\varphi}\left(  q\right)
=I_{q}\left(  \varphi \right)  $ for all $q\in Q$. Note that for each
$\varphi \in B_{0}\left(  \Sigma \right)  $ there exists $f\in \mathcal{F}$ such
that $u\left(  f\right)  =\varphi$. Assume that given $\tilde{\varphi}\in
M\ $there exist $\varphi,\psi \in B_{0}\left(  \Sigma \right)  $ such that
$\tilde{\varphi}\left(  q\right)  =I_{q}\left(  \varphi \right)  =I_{q}\left(
\psi \right)  $ for all $q\in Q$. Consider $f,g\in \mathcal{F}$ such that
$u\left(  f\right)  =\varphi$ and $u\left(  g\right)  =\psi$. It follows that
$I_{q}\left(  u\left(  f\right)  \right)  =I_{q}\left(  u\left(  g\right)
\right)  $ for all $q\in Q$. By (\ref{eq:dom-rel-rep}) and consistency, this
implies that $f\sim_{Q}^{\ast}g$ and $f\sim_{Q}g$. By (\ref{eq:I-Q}), it
follows that $I_{Q}\left(  \varphi \right)  =I_{Q}\left(  u\left(  f\right)
\right)  =I_{Q}\left(  u\left(  g\right)  \right)  =I_{Q}\left(  \psi \right)
$, proving that $\tilde{J}_{Q}$ is well defined. Next, assume that
$\tilde{\varphi},\tilde{\psi}\in M$\ are such that $\tilde{\varphi}\geq
\tilde{\psi}$. Let $\varphi,\psi \in B_{0}\left(  \Sigma \right)  $ be such that
$\tilde{\varphi}\left(  q\right)  =I_{q}\left(  \varphi \right)  $ and
$\tilde{\psi}\left(  q\right)  =I_{q}\left(  \psi \right)  $ for all $q\in Q$.
Consider $f,g\in \mathcal{F}$ such that $u\left(  f\right)  =\varphi$ and
$u\left(  g\right)  =\psi$. It follows that $I_{q}\left(  u\left(  f\right)
\right)  \geq I_{q}\left(  u\left(  g\right)  \right)  $ for all $q\in Q$. By
(\ref{eq:dom-rel-rep}) and consistency, this implies that $f\succsim_{Q}%
^{\ast}g$ and $f\succsim_{Q}g$. By (\ref{eq:I-Q}), it follows that
\[
\tilde{J}_{Q}\left(  \tilde{\varphi}\right)  =I_{Q}\left(  \varphi \right)
=I_{Q}\left(  u\left(  f\right)  \right)  \geq I_{Q}\left(  u\left(  g\right)
\right)  =I_{Q}\left(  \psi \right)  =\tilde{J}_{Q}\left(  \tilde{\psi}\right)
\]
proving that $\tilde{J}_{Q}$ is monotone. Moreover, by construction, we have
$\tilde{J}_{Q}\left(  k1_{Q}\right)  =I_{Q}\left(  k1_{S}\right)  =k$ for all
$k\in \mathbb{R}$, proving that $\tilde{J}_{Q}$ is normalized. By
(\ref{eq:I-Q}) and definition of $\tilde{J}_{Q}$, we can conclude that%
\begin{equation}
f\succsim_{Q}g\iff \tilde{J}_{Q}\left(  \min_{p\in \Delta}\left \{  \int u\left(
f\right)  dp+c\left(  p,\cdot \right)  \right \}  \right)  \geq \tilde{J}%
_{Q}\left(  \min_{p\in \Delta}\left \{  \int u\left(  g\right)  dp+c\left(
p,\cdot \right)  \right \}  \right)  \label{eq:J-til-Q-bis}%
\end{equation}
We next extend $\tilde{J}_{Q}$ to the entire set $B\left(  Q\right)  $. Define
$J_{Q}:B\left(  Q\right)  \rightarrow \mathbb{R}$ by%
\[
J_{Q}\left(  \tilde{\varphi}\right)  =\sup \left \{  \tilde{J}_{Q}\left(
\tilde{\psi}\right)  :M\ni \tilde{\psi}\leq \tilde{\varphi}\right \}
\text{\qquad}\forall \tilde{\varphi}\in B\left(  Q\right)
\]
It is routine to check that $J_{Q}$ in both cases extends $\tilde{J}_{Q}$ and
it is normalized and monotone. Moreover, by (\ref{eq:J-til-Q-bis}) it
satisfies (\ref{eq:uti-rep}), proving the implication. Uniqueness follows from
the same arguments of Theorem \ref{thm:mai-rep-inc-fin}.\hfill$\blacksquare$

\smallskip

\noindent \textbf{Proof of Corollary\  \ref{cor:com-no-cau+cau}\ }We only
prove\textbf{ }(ii) implies (iii), being (i) implies (ii) obvious and (iii)
implies (i) an immediate consequence of Theorem \ref{thm:mai-rep-inc-fin}. We
keep the same notation and terminology as in the proof and statement of
Proposition \ref{pro:com-no-cau}. By Proposition \ref{pro:com-no-cau}, there
exist an onto and affine utility function $u:X\rightarrow \mathbb{R}$ and a
lower semicontinuous divergence $c:\Delta \times \Delta^{\sigma}\rightarrow
\left[  0,\infty \right]  $, convex in $p$, such that for each $Q\in
\mathcal{Q}$ and for each $f,g\in \mathcal{F}$%
\[
f\succsim_{Q}^{\ast}g\iff \min_{p\in \Delta}\left \{  \int u\left(  f\right)
dp+c\left(  p,q\right)  \right \}  \geq \min_{p\in \Delta}\left \{  \int u\left(
g\right)  dp+c\left(  p,q\right)  \right \}  \quad \forall q\in Q
\]
and for each $x,y\in X$%
\[
x\succsim_{Q}^{\ast}y\iff x\succsim_{Q}y\iff u\left(  x\right)  \geq u\left(
y\right)
\]
Moreover, for each $Q\in \mathcal{Q}$ there exists a normalized and monotone
functional $I_{Q}:B_{0}\left(  \Sigma \right)  \rightarrow \mathbb{R}$ such that
$f\succsim_{Q}g$ if and only if $I_{Q}\left(  u\left(  f\right)  \right)  \geq
I_{Q}\left(  u\left(  g\right)  \right)  $. Fix $Q\in \mathcal{Q}$. By the same
arguments in Steps 3 and 4 of Theorem \ref{thm:mai-rep}, we have that%
\begin{align*}
I_{Q}\left(  \varphi \right)   &  =\inf_{q\in Q}\min_{p\in \Delta}\left \{
\int \varphi dp+c\left(  p,q\right)  \right \}  =\inf_{q\in Q}\inf_{p\in \Delta
}\left \{  \int \varphi dp+c\left(  p,q\right)  \right \} \\
&  =\inf_{p\in \Delta}\inf_{q\in Q}\left \{  \int \varphi dp+c\left(  p,q\right)
\right \}  =\inf_{p\in \Delta}\left \{  \int \varphi dp+\inf_{q\in Q}c\left(
p,q\right)  \right \}  \quad \forall \varphi \in B_{0}\left(  \Sigma \right)
\end{align*}
where the infima become\ minima since $c$ is lower semicontinuous. We can
conclude that%
\[
f\succsim_{Q}g\iff I_{Q}\left(  u\left(  f\right)  \right)  \geq I_{Q}\left(
u\left(  g\right)  \right)  \Longleftrightarrow \min_{p\in \Delta}\left \{  \int
u\left(  f\right)  dp+\min_{q\in Q}c\left(  p,q\right)  \right \}  \geq
\min_{p\in \Delta}\left \{  \int u\left(  g\right)  dp+\min_{q\in Q}c\left(
p,q\right)  \right \}
\]
proving the implication. Uniqueness follows from Proposition
\ref{pro:com-no-cau}.\hfill$\blacksquare$

\subsection{Other proofs\label{app:pro-oth}}

\noindent \textbf{Proof of Proposition \ref{prop:divergence-models} }Consider
first $\lambda \in \left(  0,\infty \right)  $. Note that\ $c\left(
\cdot,q\right)  =\lambda D_{\phi}\left(  \cdot||q\right)  $ is Shur convex
(with respect to $q$) for all $q\in Q$. Consider $A,B\in \Sigma$. Assume that
$q\left(  A\right)  \geq q\left(  B\right)  $ for all $q\in Q$. Let $q\in Q$.
Consider $x,y\in X$ such that $x\succ y$. It follows that%
\[
\int v\left(  u\left(  xAy\right)  \right)  dq\geq \int v\left(  u\left(
xBy\right)  \right)  dq
\]
for each $v:\mathbb{R}\rightarrow \mathbb{R}$ increasing and concave. By
Theorem 2 of Cerreia-Vioglio et al. (2012) and since $q$ was arbitrarily
chosen, it follows that%
\[
\min_{p\in \Delta}\left \{  \int u\left(  xAy\right)  dp+\lambda D_{\phi}\left(
p||q\right)  \right \}  \geq \min_{p\in \Delta}\left \{  \int u\left(  xBy\right)
dp+\lambda D_{\phi}\left(  p||q\right)  \right \}  \quad \forall q\in Q
\]
yielding that $xAy\succsim^{\ast}xBy$ and, in particular, $xAy\succsim xBy$.
If $\lambda=\infty$ instead, as pointed out in Section \ref{sect:basics}, we
have that $c\left(  \cdot,q\right)  =\lambda D_{\phi}\left(  \cdot||q\right)
=\delta_{\left \{  q\right \}  }\left(  \cdot \right)  $ for all $q\in Q$. This
implies that (\ref{eq:rep-bis}) takes the max-min form over the set $Q$, which
trivially implies bet-consistency.\footnote{The next result will indeed prove
a much more general fact.}\hfill$\blacksquare$

\smallskip

\noindent \textbf{Proof of Proposition \ref{prop:model-amb}} We prove the
\textquotedblleft only if\textquotedblright, the converse being obvious.
Define $\gtrsim^{\ast}$ by $f\gtrsim^{\ast}g$ if and only if $\int u\left(
f\right)  dq\geq \int u\left(  g\right)  dq$ for all $q\in Q$. By hypothesis,
the pair $\left(  \gtrsim^{\ast},\succsim \right)  $ satisfies consistency. Let
$f\not \gtrsim ^{\ast}x$. Then, there exists $q\in Q$ such that $u(x_{f}%
^{q})=\int u\left(  f\right)  dq<u\left(  x\right)  $. Hence, $x\succ
x_{f}^{q}$. Since $c_{Q}^{-1}\left(  0\right)  =Q$, by Lemma \ref{lm:zero-set}
we have that $x\succ f$. So, the pair $\left(  \gtrsim^{\ast},\succsim \right)
$ satisfies default to certainty. By Theorem 4 of Gilboa et al. (2010), this
pair admits the representation%
\[
f\gtrsim^{\ast}g\Longleftrightarrow \int u\left(  f\right)  dq\geq \int u\left(
g\right)  dq\quad \forall q\in Q
\]
and%
\[
f\succsim g\Longleftrightarrow \min_{q\in Q}\int u\left(  f\right)  dq\geq
\min_{q\in Q}\int u\left(  g\right)  dq
\]
Note that, in the notation of Gilboa et al. (2010), we have $C=Q$ because $C$
is unique up to closure and convexity and $Q$ is closed and convex.\hfill
$\blacksquare$

\smallskip

\noindent \textbf{Proof of Proposition \ref{prop:strong-dom-ch}} For each $q\in
Q$ define $I_{q}:B_{0}\left(  \Sigma \right)  \rightarrow \mathbb{R}$ by%
\[
I_{q}\left(  \varphi \right)  =\min_{p\in \Delta}\left \{  \int \varphi
dp+c\left(  p,q\right)  \right \}  \qquad \forall \varphi \in B_{0}\left(
\Sigma \right)
\]
Recall that $f\succ \hspace{-5pt}\succ^{\ast}g$ if and only if for
each\ $h,l\in \mathcal{F}$ there exists $\varepsilon>0$ such that%
\begin{equation}
\left(  1-\delta \right)  f+\delta h\succ^{\ast}\left(  1-\delta \right)
g+\delta l\qquad \forall \delta \in \left[  0,\varepsilon \right]
\label{eq:str-dom}%
\end{equation}
Moreover, given $h\in \mathcal{F}$, define $k_{h}=\inf_{s\in S}u\left(
h\left(  s\right)  \right)  $ and $k^{h}=\sup_{s\in S}u\left(  h\left(
s\right)  \right)  $.

\smallskip

\textquotedblleft Only if.\textquotedblright \ Assume that $f\succ \hspace
{-5pt}\succ^{\ast}g$. Let $\hat{\varepsilon}>0$. Consider $u\left(  x\right)
=k_{f}-\hat{\varepsilon}$ and\ $u\left(  y\right)  =k^{g}+\hat{\varepsilon}$.
By definition, there exists $\varepsilon>0$ such that $\left(  1-\delta
\right)  f+\delta x\succ^{\ast}\left(  1-\delta \right)  g+\delta y$ for all
$\delta \in \left[  0,\varepsilon \right]  $. Note that for each $q\in Q$ and for
each $\delta \in \left[  0,1\right]  $%
\begin{align*}
I_{q}\left(  u\left(  \left(  1-\delta \right)  f+\delta x\right)  \right)   &
=I_{q}\left(  \left(  1-\delta \right)  u\left(  f\right)  +\delta u\left(
x\right)  \right)  =I_{q}\left(  u\left(  f\right)  -\delta u\left(  f\right)
+\delta u\left(  x\right)  \right) \\
&  \leq I_{q}\left(  u\left(  f\right)  -\delta k_{f}+\delta \left(  k_{f}%
-\hat{\varepsilon}\right)  \right)  =I_{q}\left(  u\left(  f\right)  \right)
-\delta \hat{\varepsilon}%
\end{align*}
and%
\begin{align*}
I_{q}\left(  u\left(  \left(  1-\delta \right)  g+\delta y\right)  \right)   &
=I_{q}\left(  \left(  1-\delta \right)  u\left(  g\right)  +\delta u\left(
y\right)  \right)  =I_{q}\left(  u\left(  g\right)  -\delta u\left(  g\right)
+\delta u\left(  y\right)  \right) \\
&  \geq I_{q}\left(  u\left(  g\right)  -\delta k^{g}+\delta \left(  k^{g}%
+\hat{\varepsilon}\right)  \right)  =I_{q}\left(  u\left(  g\right)  \right)
+\delta \hat{\varepsilon}%
\end{align*}
It follows that for each $q\in Q$ and for each $\delta \in \left[
0,\varepsilon \right]  $%
\[
I_{q}\left(  u\left(  f\right)  \right)  -I_{q}\left(  u\left(  g\right)
\right)  -2\delta \hat{\varepsilon}\geq I_{q}\left(  u\left(  \left(
1-\delta \right)  f+\delta x\right)  \right)  -I_{q}\left(  u\left(  \left(
1-\delta \right)  g+\delta y\right)  \right)  \geq0
\]
If we set $\delta=\varepsilon>0$, then $I_{q}\left(  u\left(  f\right)
\right)  \geq I_{q}\left(  u\left(  g\right)  \right)  +2\varepsilon
\hat{\varepsilon}$ for all $q\in Q$,\ proving the statement.

\smallskip

\textquotedblleft If.\textquotedblright \ Let $f,g\in \mathcal{F}$. Assume there
exists $\varepsilon>0$ such that $I_{q}\left(  u\left(  f\right)  \right)
\geq I_{q}\left(  u\left(  g\right)  \right)  +\varepsilon$ for all $q\in Q$.
Consider $h,l\in \mathcal{F}$. Note that for each $q\in Q$ and for each
$\delta \in \left[  0,1\right]  $%
\begin{align*}
I_{q}\left(  u\left(  \left(  1-\delta \right)  f+\delta h\right)  \right)   &
=I_{q}\left(  \left(  1-\delta \right)  u\left(  f\right)  +\delta u\left(
h\right)  \right)  =I_{q}\left(  u\left(  f\right)  -\delta u\left(  f\right)
+\delta u\left(  h\right)  \right) \\
&  =I_{q}\left(  u\left(  f\right)  +\delta \left(  u\left(  h\right)
-u\left(  f\right)  \right)  \right) \\
&  \geq I_{q}\left(  u\left(  f\right)  +\delta \left(  k_{h}-k^{f}\right)
\right)  =I_{q}\left(  u\left(  f\right)  \right)  +\delta \left(  k_{h}%
-k^{f}\right)
\end{align*}
and%
\begin{align*}
I_{q}\left(  u\left(  \left(  1-\delta \right)  g+\delta l\right)  \right)   &
=I_{q}\left(  \left(  1-\delta \right)  u\left(  g\right)  +\delta u\left(
l\right)  \right)  =I_{q}\left(  u\left(  g\right)  -\delta u\left(  g\right)
+\delta u\left(  l\right)  \right) \\
&  =I_{q}\left(  u\left(  g\right)  +\delta \left(  u\left(  l\right)
-u\left(  g\right)  \right)  \right) \\
&  \leq I_{q}\left(  u\left(  g\right)  +\delta \left(  k^{l}-k_{g}\right)
\right)  =I_{q}\left(  u\left(  g\right)  \right)  +\delta \left(  k^{l}%
-k_{g}\right)
\end{align*}
It follows that for each $q\in Q$ and for each $\delta \in \left[  0,1\right]  $%
\begin{align*}
I_{q}\left(  u\left(  \left(  1-\delta \right)  f+\delta h\right)  \right)
-I_{q}\left(  u\left(  \left(  1-\delta \right)  g+\delta l\right)  \right)
&  \geq I_{q}\left(  u\left(  f\right)  \right)  +\delta \left(  k_{h}%
-k^{f}\right)  -I_{q}\left(  u\left(  g\right)  \right)  -\delta \left(
k^{l}-k_{g}\right) \\
&  \geq \varepsilon+\delta \hat{\varepsilon}%
\end{align*}
where $\hat{\varepsilon}=k_{h}-k^{f}-k^{l}+k_{g}$. We have two cases:

\begin{enumerate}
\item $\hat{\varepsilon}\geq0$. In this case, $I_{q}\left(  u\left(  \left(
1-\delta \right)  f+\delta h\right)  \right)  -I_{q}\left(  u\left(  \left(
1-\delta \right)  g+\delta l\right)  \right)  >0$ for all $\delta \in \left[
0,1\right]  $ and all $q\in Q$, proving (\ref{eq:str-dom}).

\item $\hat{\varepsilon}<0$. In this case, $I_{q}\left(  u\left(  \left(
1-\delta \right)  f+\delta h\right)  \right)  -I_{q}\left(  u\left(  \left(
1-\delta \right)  g+\delta l\right)  \right)  >0$ for all $\delta \in \left[
0,-\varepsilon/2\hat{\varepsilon}\right]  $ and all $q\in Q$, proving
(\ref{eq:str-dom}).
\end{enumerate}

This completes the proof of the result.\hfill$\blacksquare$

\smallskip

\noindent \textbf{Proof of Lemma \ref{lm:stat-dist-max}} \textquotedblleft
If.\textquotedblright \ Given $q\in Q$, if $c\left(  p,q\right)  =\infty$ for
all $p\notin Q$, then $c_{Q}\left(  p\right)  =\infty$ for all $p\notin Q$.
Since $c_{Q}\left(  p\right)  =0$ for all $p\in Q$, we conclude that
$c_{Q}\left(  p\right)  =\delta_{Q}\left(  p\right)  $ for all $p\in \Delta$.
\textquotedblleft Only if.\textquotedblright \ Conversely, for each $q\in Q$ we
have that\ $c\left(  p,q\right)  \geq c_{Q}\left(  p\right)  =\delta
_{Q}\left(  p\right)  =\infty$ for all $p\notin Q$.\hfill$\blacksquare$

\smallskip

\noindent \textbf{Proof of Proposition \ref{pro:c-ind}}\ Before starting, we
make two observations. First, observe that (i) and (ii) of Proposition
\ref{pro:c-ind} are particular cases of (i) and (ii) of Theorem
\ref{thm:mai-rep}. We thus adopt the same notation, terminology, and arguments
contained in the proof of this latter theorem. Second, consider an
unbounded\ dominance relation $\succsim^{\ast}$ and a rational preference
$\succsim$ that jointly satisfy consistency and caution. By consistency, it
follows that for each $f\in \mathcal{F}$ and for each $x\in X$%
\[
f\succsim^{\ast}x\implies f\succsim x
\]
By caution, we have that for each $f\in \mathcal{F}$ and for each $x\in X$%
\[
f\succ x\implies f\succsim^{\ast}x
\]
Moreover, by the same arguments of Step 1 of the proof of Theorem
\ref{thm:mai-rep}, $\succsim$ agrees with\ $\succsim^{\ast}$\ on\ $X$ and they
are represented by an onto affine utility function $u:X\rightarrow \mathbb{R}$.
This implies that if $f\sim x$, then there exists $y\in X$ such that
$x\succ \alpha x+\left(  1-\alpha \right)  y\succ y$ for all $\alpha \in \left(
0,1\right)  $, yielding that $f\succ \alpha x+\left(  1-\alpha \right)  y$ and
$f\succsim^{\ast}\alpha x+\left(  1-\alpha \right)  y$ for all $\alpha
\in \left(  0,1\right)  $. Since $\succsim^{\ast}$ satisfies continuity, we
have that $f\succsim^{\ast}x$. Thus, we can conclude that for each
$f\in \mathcal{F}$ and for each $x\in X$%
\begin{equation}
f\succsim^{\ast}x\iff f\succsim x \label{eq:as-amb-ave}%
\end{equation}
(i) implies (ii). By Theorem \ref{thm:mai-rep}, we have that there exist an
onto affine function $u:X\rightarrow \mathbb{R}$ and a variational
pseudo-statistical distance $c:\Delta \times Q\rightarrow \left[  0,\infty
\right]  $ such that, for all acts $f,g\in \mathcal{F}$,%
\begin{equation}
f\succsim^{\ast}g\Longleftrightarrow \min_{p\in \Delta}\left \{  \int u\left(
f\right)  dp+c\left(  p,q\right)  \right \}  \geq \min_{p\in \Delta}\left \{  \int
u\left(  g\right)  dp+c\left(  p,q\right)  \right \}  \qquad \forall q\in Q
\label{eq:Bew-var-cer-ind-pro}%
\end{equation}
and%
\[
f\succsim g\Longleftrightarrow \min_{p\in \Delta}\left \{  \int u\left(
f\right)  dp+\min_{q\in Q}c\left(  p,q\right)  \right \}  \geq \min_{p\in \Delta
}\left \{  \int u\left(  g\right)  dp+\min_{q\in Q}c\left(  p,q\right)
\right \}
\]
By (\ref{eq:as-amb-ave}) and since $\succsim^{\ast}$ satisfies c-independence,
we have that if $f\in \mathcal{F}$, $x,y\in X$, and $\alpha \in \left(
0,1\right]  $, then%
\[
f\succsim x\iff f\succsim^{\ast}x\iff \alpha f+\left(  1-\alpha \right)
y\succsim^{\ast}\alpha x+\left(  1-\alpha \right)  y\iff \alpha f+\left(
1-\alpha \right)  y\succsim \alpha x+\left(  1-\alpha \right)  y
\]
proving that $\succsim$ satisfies c-independence. By Propositions 6 and\ 19 of
Maccheroni et al. 2006 and since $u$ is onto, $c_{Q}:\Delta \rightarrow \left[
0,\infty \right]  $ is grounded, lower semicontinuous and convex and
$c_{Q}^{-1}\left(  0\right)  =Q$, we have that $c_{Q}=\delta_{Q}$,
proving\ (\ref{eq:var-var-cer-ind}). Since $c\left(  \cdot,q\right)  \geq
c_{Q}=\delta_{Q}$ for all $q\in Q$ , we have that $c\left(  p,q\right)  \geq
c_{Q}\left(  p\right)  =\delta_{Q}\left(  p\right)  =\infty$ for all
$p\not \in Q$,\ yielding that the $\min$ in (\ref{eq:Bew-var-cer-ind-pro}) can
be restricted to $Q$ and proving (\ref{eq:Bew-var-cer-ind-bis}).

\smallskip

(ii) implies (i). By the same arguments contained above and since
$c_{Q}=\delta_{Q}$, we have that the $\min$ in (\ref{eq:Bew-var-cer-ind-bis})
can be taken over $\Delta$. By (ii) implies (i) of Theorem \ref{thm:mai-rep},
the implication follows with the exception of proving that $\succsim^{\ast}$
satisfies c-independence. Since $\succsim$ is represented as in
(\ref{eq:var-var-cer-ind}), $\succsim$ satisfies c-independence. Since
$\succsim^{\ast}$ is an unbounded\ dominance relation and $\succsim$ is a
rational preference and jointly they satisfy consistency and caution,
(\ref{eq:as-amb-ave}) holds, yielding that if $f\in \mathcal{F}$, $x,y\in X$,
and $\alpha \in \left(  0,1\right]  $, then%
\[
f\succsim^{\ast}x\iff f\succsim x\iff \alpha f+\left(  1-\alpha \right)
y\succsim \alpha x+\left(  1-\alpha \right)  y\iff \alpha f+\left(
1-\alpha \right)  y\succsim^{\ast}\alpha x+\left(  1-\alpha \right)  y
\]
proving that $\succsim^{\ast}$ satisfies c-independence and the implication.

\smallskip

We prove the second part of the statement independently and with a different
technique in order to dispense with the assumption of $\left(  S,\Sigma
\right)  $ being a standard Borel space. We only need to prove the
\textquotedblleft only if\textquotedblright \ part, the \textquotedblleft
if\textquotedblright \ being trivial.

\smallskip

By Proposition 2 of Cerreia-Vioglio (2016) and since $\succsim^{\ast}$ is
unbounded, there exists a compact and convex set $C\subseteq \Delta$ and an
affine and onto map $u:X\rightarrow \mathbb{R}$\ such that%
\begin{equation}
f\succsim^{\ast}g\Longleftrightarrow \int u\left(  f\right)  dq\geq \int
u\left(  g\right)  dq\qquad \forall q\in C\label{eq:rep-Bew}%
\end{equation}
and%
\begin{equation}
f\succsim g\iff \min_{q\in C}\int u\left(  f\right)  dq\geq \min_{q\in C}\int
u\left(  g\right)  dq\label{eq:rep-GiSc}%
\end{equation}
By Lemma \ref{lm:zero-set}\ and since $\succsim$ is subjectively $Q$-coherent
and $\succsim^{\ast}$ and $\succsim$ coincide on $X$, we can conclude that
$C=Q$. If we set $c:\Delta \times Q\rightarrow \left[  0,\infty \right]  $ to be
$c\left(  p,q\right)  =\delta_{\left \{  q\right \}  }\left(  p\right)  $ for
all $\left(  p,q\right)  \in \Delta \times Q$, then it is immediate to see that
$c$ is a variational\ statistical distance. By (\ref{eq:rep-Bew}) and
(\ref{eq:rep-GiSc}) and since $C=Q$, the implication follows.\hfill
$\blacksquare$

\smallskip

\noindent \textbf{Proof of Proposition \ref{pro:lim-con}} We begin by making
two observations. It is well known that, given a bounded and measurable
$F:Q\rightarrow \mathbb{R}$,%
\begin{equation}
\lim_{\xi \rightarrow0^{+}}\phi_{\xi}^{-1}\left(  \int_{Q}\phi_{\xi}\left(
F\left(  q\right)  \right)  d\mu_{Q}\right)  =\min_{q\in \operatorname*{supp}%
\mu_{Q}}F\left(  q\right)  =\min_{q\in Q}F\left(  q\right)  \label{eq:wel-kno}%
\end{equation}
and%
\begin{equation}
\phi_{\xi}^{-1}\left(  \int_{Q}\phi_{\xi}\left(  F\left(  q\right)  \right)
d\mu_{Q}\right)  =\min_{\mu \ll \nu}\left \{  \int Fd\nu+\xi R(\nu||\mu)\right \}
\label{eq:Don-Var}%
\end{equation}
Fix $f\in \mathcal{F}$ and $\lambda \in \left(  0,\infty \right]  $. Since $c$ is
lower semicontinuous and each $f\in \mathcal{F}$ is finitely valued, if we set
$F_{\lambda}\left(  q\right)  =\min_{p\in \Delta}\left \{  u\left(  f\right)
dp+\lambda R\left(  p||q\right)  \right \}  $ for all $q\in Q$, it is immediate
to see that $F$ is bounded and measurable.

By (\ref{eq:wel-kno}), (\ref{eq:bayes-lim1}) follows. By Proposition 12 of
Maccheroni et al. (2006) and (\ref{eq:Don-Var})\ and since $\lim
_{\xi \rightarrow \infty}\xi R\left(  \nu||\mu \right)  =\infty$ if $\nu \neq \mu$
and $\lim_{\xi \rightarrow \infty}\xi R\left(  \nu||\mu \right)  =\infty$ if
$\nu=\mu$, (\ref{eq:bayes-lim2}) follows. By (\ref{eq:bayes-lim2}), we have
that%
\[
\lim_{\xi \rightarrow \infty}V_{Q}^{\lambda,\xi}\left(  f\right)  =\int
_{Q}\left(  \min_{p\in \Delta}\left \{  \int_{S}u\left(  f\left(  s\right)
\right)  dp\left(  s\right)  +\lambda R\left(  p||q\right)  \right \}  \right)
d\mu_{Q}\left(  q\right)
\]
By Proposition 12 of Maccheroni et al. (2006) and since $\lim_{\lambda
\rightarrow \infty}\lambda R\left(  p||q\right)  =\infty$ if $p\neq q$ and
$\lim_{\lambda \rightarrow \infty}\lambda R\left(  p||q\right)  =\infty$ if
$p=q$, we have that $\lim_{\lambda \rightarrow \infty}F_{\lambda}\left(
q\right)  =\int u\left(  f\right)  dq=F_{\infty}\left(  q\right)  $ for all
$q\in Q$. By the Lebesgue Dominanted Convergence Theorem and since $\left \{
F_{\lambda}\right \}  _{\lambda \in \left(  0,\infty \right)  }$ are uniformly
bounded, the second equality of (\ref{eq:bayes-lim3}) follows. The first has a
similar proof and we omit it.\hfill$\blacksquare$

\smallskip

\noindent \textbf{Proof of Lemma \ref{lem:bay-pri}} (i) Let $p\in \Delta$. Since
$c$ is lower semicontinuous, there exists\ $q_{p}\in Q$\ such that $c\left(
p,q_{p}\right)  =\min_{q\in Q}c\left(  p,q\right)  $, that is, $p\in
B_{c}\left(  q_{p},Q\right)  $. This proves that $\Delta \subseteq%
{\displaystyle \bigcup \limits_{q\in Q}}
B_{c}\left(  q,Q\right)  $, as desired (the other inclusion is trivial).

(ii) For each $q\in Q\ $we have that $0=c\left(  q,q\right)  \geq \min
_{\tilde{q}\in Q}c(q,\tilde{q})\geq0$. Thus, $c\left(  q,q\right)
=\min_{\tilde{q}\in Q}c(q,\tilde{q})$ and so $q\in B_{c}\left(  q,Q\right)  $.
It remains to show that $B_{c}\left(  q,Q\right)  \cap Q\subseteq \left \{
q\right \}  $. So, let $\bar{q}\in B_{c}\left(  q,Q\right)  \cap Q$. Then,
$c\left(  \bar{q},q\right)  =\min_{\tilde{q}\in Q}c\left(  \bar{q},\tilde
{q}\right)  $. Since $\bar{q}\in Q$, we have $\min_{\tilde{q}\in Q}c\left(
\bar{q},\tilde{q}\right)  =0$ and so $c\left(  \bar{q},q\right)  =0$, which
implies $\bar{q}=q$, as desired.

(iii) Let $q,q^{\prime}\in Q$ with $q\neq q^{\prime}$. In view of (ii), it is
enough to consider $p\in B_{c}\left(  q,Q\right)  \cap B_{c}\left(  q^{\prime
},Q\right)  \cap \Delta_{c,Q}$. Since $p\in \Delta_{c,Q}$, the map $c\left(
p,\cdot \right)  :Q\rightarrow \left[  0,\infty \right]  $ is proper and strictly
convex. Thus, $c\left(  p,q\right)  =c\left(  p,q^{\prime}\right)
=\min_{\tilde{q}\in Q}c(p,\tilde{q})<\infty$, which leads to the contradiction
$q=q^{\prime}$.\hfill$\blacksquare$

\section{Additional\ material\label{app:EM}}

In this appendix, we begin by proving few relevant properties of statistical
distances which we discussed in Section \ref{sect:basics}. We then discuss the
irrelevance of the convexity of the set $Q$ for the entropic model (cf.
Section \ref{app:non-cvx}). We conclude by providing the proofs of few
ancillary facts useful in obtaining and discussing our decision criterion (cf.
Sections \ref{app:anc-rep}\ and \ref{app:anc-ana}).

\subsection{Statistical distances and divergences\label{app:sta-dis}}

We here collect few properties of statistical distances. In order to
characterize variational statistical distances, we substantially need to prove
that the function $c_{Q}:\Delta \rightarrow \left[  0,\infty \right]  $,
defined\ by $c_{Q}\left(  p\right)  =\min_{q\in Q}c(p,q)$, is well defined,
grounded, lower semicontinuous and convex. This fact follows from the
following version of a well-known result (see, e.g., Fiacco and Kyparisis, 1986).

\begin{lemma}
\label{lem:min-c}Let $Q$ be a compact and convex subset of $\Delta^{\sigma}$.
If $c:\Delta \times Q\rightarrow \left[  0,\infty \right]  $ is a lower
semicontinuous and convex function such that there exist $\bar{p}\in \Delta$
and $\bar{q}\in Q$ such that $c\left(  \bar{p},\bar{q}\right)  =0$,
then\ $c_{Q}:\Delta \rightarrow \left[  0,\infty \right]  $ defined by%
\[
c_{Q}\left(  p\right)  =\min_{q\in Q}c\left(  p,q\right)  \qquad \forall
p\in \Delta
\]
is well defined, grounded, lower semicontinuous and convex.
\end{lemma}

\noindent \textbf{Proof} Since $c$ is lower semicontinuous and $Q$ is non-empty
and compact, $c_{Q}$ is well defined. Moreover, we have that $0\geq c\left(
\bar{p},\bar{q}\right)  \geq c_{Q}\left(  \bar{p}\right)  \geq0$, proving that
$c_{Q}$ is grounded. Even though $c\left(  p,q\right)  $ might be $\infty$ for
some $\left(  p,q\right)  \in \Delta \times Q$, by the same proof of the Maximum
Theorem (see, e.g., Lemma 17.30 in Aliprantis and Border, 2006), it follows
that $c_{Q}$ is lower semicontinuous. If $p_{1},p_{2}\in \Delta$, then define
$q_{1},q_{2}\in Q$\ to be such that%
\[
c\left(  p_{1},q_{1}\right)  =\min_{q\in Q}c\left(  p_{1},q\right)
=c_{Q}\left(  p_{1}\right)  \text{ and }c\left(  p_{2},q_{2}\right)
=\min_{q\in Q}c\left(  p_{2},q\right)  =c_{Q}\left(  p_{2}\right)
\]
Consider $\lambda \in \left(  0,1\right)  $. Define $p_{\lambda}=\lambda
p_{1}+\left(  1-\lambda \right)  p_{2}$\ and $q_{\lambda}=\lambda q_{1}+\left(
1-\lambda \right)  q_{2}\in Q$.\ Since $c$\ is jointly convex, it follows that%
\begin{align*}
c_{Q}\left(  p_{\lambda}\right)   &  =\min_{q\in Q}c\left(  p_{\lambda
},q\right)  \leq c\left(  p_{\lambda},q_{\lambda}\right)  \leq \lambda c\left(
p_{1},q_{1}\right)  +\left(  1-\lambda \right)  c\left(  p_{2},q_{2}\right) \\
&  =\lambda c_{Q}\left(  p_{1}\right)  +\left(  1-\lambda \right)  c_{Q}\left(
p_{2}\right)
\end{align*}
proving convexity.\hfill$\blacksquare$

\begin{lemma}
\label{lem:app-suff-div}Let $\mathcal{Q}$ consist of compact and convex
subsets of $\Delta^{\sigma}$. A lower semicontinuous and convex function
$c:\Delta \times \mathcal{S}\rightarrow \left[  0,\infty \right]  $ is a
variational statistical distance if and only if it satisfies the distance
property:%
\begin{equation}
c\left(  p,q\right)  =0\iff p=q \label{eq:app-fir-dis-pro-bis}%
\end{equation}

\end{lemma}

\noindent \textbf{Proof} We first prove the \textquotedblleft
If\textquotedblright \ part. By (\ref{eq:app-fir-dis-pro-bis}) and since $c$ is
lower semicontinuous, (c.i) and (c.ii) are satisfied. Fix $Q\in \mathcal{Q}$.
By (c.i) and since $c$ restricted to $\Delta \times Q$ is jointly lower
semicontinuous and convex, then we have that $p\mapsto \min_{q\in Q}c(p,q)$ is
well defined, grounded, lower semicontinuous and convex. By (c.i), it follows
that (C.i) is satisfied. By construction and since $p\mapsto \min_{q\in
Q}c(p,q)$ is lower semicontinuous and convex and $Q$ was arbitrarily chosen,
(C.ii), (C.iii), (C.iv) as well as (C.v) are satisfied, proving that $c$ is a
variational statistical distance. As for the \textquotedblleft Only
if\textquotedblright \ part, it is trivial since a statistical distance, by
definition, satisfies (\ref{eq:app-fir-dis-pro-bis}).\hfill$\blacksquare$

\smallskip

The next result shows, inter alia, that restricted $\phi$-divergences are
variational divergences.\footnote{Though a routine result, for the sake of
completeness, we provide a proof since we did not find one allowing for $S$
being\ infinite (see Topsoe, 2001, p. 178 for the finite case).} A piece of
notation and one of terminology: 1) we write $p\sim Q$ if there exists a
control measure $q\in Q$ such that $p\sim q$;\footnote{A probability $q\in Q$
is a \emph{control measure} of $Q$\ if $q^{\prime}\ll q$ for all $q^{\prime
}\in Q$. When $Q$\ is a compact and convex subset of $\Delta^{\sigma}$,\ $Q$
has a control measure (see, e.g., Maccheroni and Marinacci, 2001). Such a
measure might not be unique, yet any two control measures of $Q$ are
equivalent. So, the notion $p\sim Q$ is well defined and independent of the
chosen control measure.} 2) given a function $f:\Delta \rightarrow \left[
0,\infty \right]  $ we say it is \emph{strictly convex} if, given any
distinct\ $p,q\in \Delta$, we have $f\left(  \alpha p+\left(  1-\alpha \right)
q\right)  <\alpha f\left(  p\right)  +\left(  1-\alpha \right)  f\left(
q\right)  $ for all $\alpha \in \left(  0,1\right)  $ such that $\alpha
p+\left(  1-\alpha \right)  q\in \operatorname*{dom}f$.

\begin{lemma}
\label{lem:str-cvx}Let $\mathcal{Q}$ consist of compact and convex subsets of
$\Delta^{\sigma}$. A restricted $\phi$-divergence $D_{\phi}:\Delta
\times \mathcal{S}\rightarrow \left[  0,\infty \right]  $ is a variational
divergence. Moreover, for each $Q\in \mathcal{Q}$

\begin{enumerate}
\item[(i)] if $q\in Q$, then $D_{\phi}\left(  \cdot||q\right)  :\Delta
\rightarrow \left[  0,\infty \right]  $ is strictly convex;

\item[(ii)] if $p\in \Delta^{\sigma}$ and $p\sim Q$, then $D_{\phi}\left(
p||\cdot \right)  :Q\rightarrow \left[  0,\infty \right]  $ is strictly convex.
\end{enumerate}
\end{lemma}

\noindent \textbf{Proof} It is well known that on $\Delta \times \Delta^{\sigma}$
the function $D_{\phi}$ is jointly lower semicontinuous and convex and
satisfies the property%
\[
D_{\phi}\left(  p||q\right)  =0\iff p=q
\]
The same properties are preserved by\ $D_{\phi}$ restricted to $\Delta
\times \mathcal{S}$. By Lemma \ref{lem:app-suff-div}, it follows that $D_{\phi
}:\Delta \times \mathcal{S}\rightarrow \left[  0,\infty \right]  $ is a
variational statistical distance. Finally, by definition, we have that
$D_{\phi}\left(  p||q\right)  =\infty$ whenever $p\not \in \Delta^{\sigma
}\left(  q\right)  $, yielding that it is a variational divergence. We next
prove points (i) and (ii). Fix $Q\in \mathcal{Q}$.

\smallskip

\noindent(i). Consider $q\in Q$. Let $p^{\prime},p^{\prime \prime}\in \Delta$
and $\alpha \in \left(  0,1\right)  $ be such that $p^{\prime}\not =%
p^{\prime \prime}$\ and $D_{\phi}(\alpha p^{\prime}+\left(  1-\alpha \right)
p^{\prime \prime}||q)<\infty$. If either $D_{\phi}\left(  p^{\prime}||q\right)
$ or $D_{\phi}\left(  p^{\prime \prime}||q\right)  $ are not finite, we
trivially conclude that $D_{\phi}(\alpha p^{\prime}+\left(  1-\alpha \right)
p^{\prime \prime}||q)<\infty=\alpha D_{\phi}\left(  p^{\prime}||q\right)
+\left(  1-\alpha \right)  D_{\phi}\left(  p^{\prime \prime}||q\right)  $. Let
us then assume that both $D_{\phi}\left(  p^{\prime}||q\right)  $ and
$D_{\phi}\left(  p^{\prime \prime}||q\right)  $ are finite. By definition of
$D_{\phi}$ and since $\Delta^{\sigma}\left(  q\right)  $ is convex, this
implies that $p^{\prime},p^{\prime \prime}\in \Delta^{\sigma}\left(  q\right)  $
as well as $\alpha p^{\prime}+\left(  1-\alpha \right)  p^{\prime \prime}%
\in \Delta^{\sigma}\left(  q\right)  $. Since $p^{\prime}$ and $p^{\prime
\prime}$ are distinct, we have that $dp^{\prime}/dq$ and\ $dp^{\prime \prime
}/dq$ take different values on a set of strictly positive $q$-measure: call it
$\tilde{S}$. Since $\phi$ is strictly convex, it follows that%
\[
\phi \left(  \alpha \dfrac{dp^{\prime}}{dq}\left(  s\right)  +\left(
1-\alpha \right)  \dfrac{dp^{\prime \prime}}{dq}\left(  s\right)  \right)
<\alpha \phi \left(  \dfrac{dp^{\prime}}{dq}\left(  s\right)  \right)  +\left(
1-\alpha \right)  \phi \left(  \dfrac{dp^{\prime \prime}}{dq}\left(  s\right)
\right)  \quad \forall s\in \tilde{S}%
\]
By definition of $D_{\phi}$, this implies that%
\begin{align*}
D_{\phi}\left(  \alpha p^{\prime}+\left(  1-\alpha \right)  p^{\prime \prime
}||q\right)   &  =\int_{S}\phi \left(  \dfrac{d\left[  \alpha p^{\prime
}+\left(  1-\alpha \right)  p^{\prime \prime}\right]  }{dq}\left(  s\right)
\right)  dq\\
&  =\int_{S}\phi \left(  \alpha \dfrac{dp^{\prime}}{dq}\left(  s\right)
+\left(  1-\alpha \right)  \dfrac{dp^{\prime \prime}}{dq}\left(  s\right)
\right)  dq\\
&  =\int_{\tilde{S}}\phi \left(  \alpha \dfrac{dp^{\prime}}{dq}\left(  s\right)
+\left(  1-\alpha \right)  \dfrac{dp^{\prime \prime}}{dq}\left(  s\right)
\right)  dq\\
&  +\int_{S\backslash \tilde{S}}\phi \left(  \alpha \dfrac{dp^{\prime}}%
{dq}\left(  s\right)  +\left(  1-\alpha \right)  \dfrac{dp^{\prime \prime}}%
{dq}\left(  s\right)  \right)  dq\\
&  <\alpha \int_{S}\phi \left(  \dfrac{dp^{\prime}}{dq}\left(  s\right)
\right)  dq+\left(  1-\alpha \right)  \int_{S}\phi \left(  \dfrac{dp^{\prime
\prime}}{dq}\left(  s\right)  \right)  dq\\
&  =\alpha D_{\phi}\left(  p^{\prime}||q\right)  +\left(  1-\alpha \right)
D_{\phi}\left(  p^{\prime \prime}||q\right)
\end{align*}
We conclude that $D_{\phi}\left(  \cdot||q\right)  :\Delta \rightarrow \left[
0,\infty \right]  $ is strictly convex.

\smallskip

\noindent(ii). Before starting, we make three observations.

\smallskip

a. Since $Q$ is a non-empty, compact and convex subset of $\Delta^{\sigma}%
$,\ note that there exists $\bar{q}\in Q$ such that $q\ll \bar{q}$ for all
$q\in Q$. Since $p\sim Q$, we have that $p\sim \bar{q}$. This implies also that
$q\ll p$\ for all $q\in Q$.

\smallskip

b. If $q\sim p$, then $\left(  dp/dq\right)  ^{-1}$\ is well defined almost
everywhere (with respect to either $p$ or $q$) and can be chosen (after
defining arbitrarily the function over a set of zero measure) to be the
Radon-Nikodym derivative $dq/dp$.

\smallskip

c. Since $\phi$ is strictly convex, if we define $\phi^{\star}:\left(
0,\infty \right)  \rightarrow \left[  0,\infty \right)  $ by $\phi^{\star}\left(
x\right)  =x\phi \left(  1/x\right)  $ for all $x>0$, then also $\phi^{\star}$
is strictly convex. By point b, if $p\in \Delta^{\sigma}$ and $q\in Q$ are such
that $p\sim q$ and we define $\dot{p}=dp/dq$, then $p\left(  \left \{  \dot
{p}=0\right \}  \right)  =0=q\left(  \left \{  \dot{p}=0\right \}  \right)
$\ and%
\begin{align*}
D_{\phi}\left(  p||q\right)   &  =\int_{S}\phi \left(  \frac{dp}{dq}\right)
dq=\int_{\left \{  \dot{p}=0\right \}  }\phi \left(  \frac{dp}{dq}\right)
dq+\int_{\left \{  \dot{p}>0\right \}  }\phi \left(  \frac{dp}{dq}\right)  dq\\
&  =\int_{\left \{  \dot{p}>0\right \}  }\phi \left(  \frac{1}{\left(  \frac
{dp}{dq}\right)  ^{-1}}\right)  dq=\int_{\left \{  \dot{p}>0\right \}  }%
\phi^{\star}\left(  \frac{dq}{dp}\right)  \frac{dp}{dq}dq\\
&  =\int_{\left \{  \dot{p}>0\right \}  }\phi^{\star}\left(  \frac{dq}%
{dp}\right)  dp
\end{align*}
We can now prove the statement. Let $q^{\prime},q^{\prime \prime}\in Q$ and
$\alpha \in \left(  0,1\right)  $ be such that $q^{\prime}\not =q^{\prime \prime
}$\ and $D_{\phi}\left(  p||\alpha q^{\prime}+\left(  1-\alpha \right)
q^{\prime \prime}\right)  <\infty$. If either $D_{\phi}\left(  p||q^{\prime
}\right)  $ or $D_{\phi}\left(  p||q^{\prime \prime}\right)  $ are not finite,
we trivially conclude that $D_{\phi}\left(  p||\alpha q^{\prime}+\left(
1-\alpha \right)  q^{\prime \prime}\right)  <\infty=\alpha D_{\phi}\left(
p||q^{\prime}\right)  +\left(  1-\alpha \right)  D_{\phi}\left(  p||q^{\prime
\prime}\right)  $. Let us then assume that both $D_{\phi}\left(  p||q^{\prime
}\right)  $ and $D_{\phi}\left(  p||q^{\prime \prime}\right)  $ are finite. By
definition of $D_{\phi}$, we can conclude that $p\ll q^{\prime}$ and $p\ll
q^{\prime \prime}$. By point a, this yields\ that $q^{\prime}\sim p\sim
q^{\prime \prime}$ and $p\sim \alpha q^{\prime}+\left(  1-\alpha \right)
q^{\prime \prime}$. Since $q^{\prime}$ and $q^{\prime \prime}$ are distinct, we
have that $dq^{\prime}/dp$\ and $dq^{\prime \prime}/dp$ take different values
on a set of strictly positive $p$-measure. By point c, we have that%
\[
p\left(  \left \{  \frac{dp}{d\left[  \alpha q^{\prime}+\left(  1-\alpha
\right)  q^{\prime \prime}\right]  }=0\right \}  \right)  =p\left(  \left \{
\frac{dp}{dq^{\prime}}=0\right \}  \right)  =p\left(  \left \{  \frac
{dp}{dq^{\prime \prime}}=0\right \}  \right)  =0
\]
Thus, by point c and since $dq^{\prime}/dp$\ and $dq^{\prime \prime}/dp$ take
different values on a set of strictly positive $p$-measure and $\phi^{\star}$
is strictly convex, there exists a $p$-measure $1$ set $\tilde{S}$ such that%
\begin{align*}
D_{\phi}\left(  p||\alpha q^{\prime}+\left(  1-\alpha \right)  q^{\prime \prime
}\right)   &  =\int_{\tilde{S}}\phi^{\star}\left(  \dfrac{d\left[  \alpha
q^{\prime}+\left(  1-\alpha \right)  q^{\prime \prime}\right]  }{dp}\right)
dp\\
&  <\alpha \int_{\tilde{S}}\phi^{\star}\left(  \dfrac{dq^{\prime}}{dp}\right)
dp+\left(  1-\alpha \right)  \int_{\tilde{S}}\phi^{\star}\left(  \dfrac
{dq^{\prime \prime}}{dp}\right)  dp\\
&  =\alpha D_{\phi}\left(  p||q^{\prime}\right)  +\left(  1-\alpha \right)
D_{\phi}\left(  p||q^{\prime \prime}\right)
\end{align*}
proving point (ii).\hfill$\blacksquare$

\smallskip

Consider a finite set $Q=\left \{  q_{i}\right \}  _{i=1}^{n}$. Assume that for
each $q$ in the convex hull of $Q\ $there exists a unique collection $\left \{
\mu_{i}^{q}\right \}  _{i=1}^{n}\subseteq \mathbb{R}_{+}^{n}$ such that
$\sum_{i=1}^{n}\mu_{i}^{q}=1$ and $q=\sum_{i=1}^{n}\mu_{i}^{q}q_{i}$. Consider
a function $d:\Delta \times Q\rightarrow \left[  0,\infty \right]  $ which is
lower semicontinuous and convex in the first argument and such that $d\left(
p,q\right)  =0$ if and only if $p=q$. Define $c:\Delta \times \operatorname*{co}%
Q\rightarrow \left[  0,\infty \right]  $ by%
\begin{equation}
c\left(  p,q\right)  =\min_{\left(  p_{i}\right)  _{i=1}^{n}\in \Delta
^{n}:p=\sum_{i=1}^{n}\mu_{i}^{q}p_{i}}\sum_{i=1}^{n}d\left(  p_{i}%
,q_{i}\right)  \mu_{i}^{q}\qquad \forall \left(  p,q\right)  \in \Delta
\times \operatorname*{co}Q \label{eq:Lar-dis}%
\end{equation}

\begin{lemma}
\label{lem:Lar-dis} If $c$ is defined as in (\ref{eq:Lar-dis}), then $c$ is a
jointly lower semicontinuous and convex variational statistical distance.
Moreover, if $d\left(  p,q\right)  <\infty$ implies $p\ll q$, then $c$ is also
a divergence.
\end{lemma}

\noindent \textbf{Proof} We endow $\Delta^{n}$ with the product topology.
Clearly, $\Delta^{n}$ is compact. For each $p\in \Delta$ and $q\in
\operatorname*{co}Q$ set $\Psi \left(  p,q\right)  =\left \{  \left(
p_{i}\right)  _{i=1}^{n}\in \Delta^{n}:p=\sum_{i=1}^{n}\mu_{i}^{q}%
p_{i}\right \}  $. Note that $\Psi \left(  p,q\right)  $\ is non-empty, closed
(hence, compact) and convex. Since $d:\Delta \times Q\rightarrow \left[
0,\infty \right]  $ is lower semicontinuous in $p$, given $q$, we have that
$\left(  p_{i}\right)  _{i=1}^{n}\mapsto \sum_{i=1}^{n}d\left(  p_{i}%
,q_{i}\right)  \mu_{i}^{q}$ is lower semicontinuous. Since $\Psi \left(
p,q\right)  $ is compact, this implies that $c$ is well defined. Next, observe
that if $p=q$, then $\left(  \bar{p}_{i}\right)  _{i=1}^{n}\in \Delta^{n}$ such
that $\bar{p}_{i}=q_{i}$ for all $i\in \left \{  1,...,n\right \}  $ satisfies
$p=q=\sum_{i=1}^{n}\mu_{i}^{q}q_{i}=\sum_{i=1}^{n}\mu_{i}^{q}\bar{p}_{i}$,
that is, $\left(  \bar{p}_{i}\right)  _{i=1}^{n}\in \Psi \left(  p,q\right)
=\Psi \left(  q,q\right)  \ $and $0\leq c\left(  p,q\right)  \leq \sum_{i=1}%
^{n}d\left(  \bar{p}_{i},q_{i}\right)  \mu_{i}^{q}=0$. Vice versa, since
$d\geq0$, we have that if $c\left(  p,q\right)  =0$, then there exists
$\left(  \bar{p}_{i}\right)  _{i=1}^{n}\in \Psi \left(  p,q\right)  $ such that
$c\left(  p,q\right)  =\sum_{i=1}^{n}d\left(  \bar{p}_{i},q_{i}\right)
\mu_{i}^{q}=0$, yielding that $\bar{p}_{i}=q_{i}$ for all $i\in \left \{
1,...,n\right \}  $ such that $\mu_{i}^{q}>0$. Since $p=\sum_{i=1}^{n}\mu
_{i}^{q}\bar{p}_{i}=\sum_{i:\mu_{i}^{q}>0}^{{}}\mu_{i}^{q}\bar{p}_{i}$ and
$q=\sum_{i=1}^{n}\mu_{i}^{q}q_{i}=\sum_{i:\mu_{i}^{q}>0}^{{}}\mu_{i}^{q}q_{i}%
$, we can conclude that $p=q$. Consider $p,r\in \Delta$ as well as
$q,q^{\prime}\in \operatorname*{co}Q$ and $\lambda \in \left(  0,1\right)  $. Let
$\left(  \bar{p}_{i}\right)  _{i=1}^{n}\in \Psi \left(  p,q\right)  $ and
$\left(  \bar{r}_{i}\right)  _{i=1}^{n}\in \Psi \left(  r,q^{\prime}\right)  $
be such that $c\left(  p,q\right)  =\sum_{i=1}^{n}d\left(  \bar{p}_{i}%
,q_{i}\right)  \mu_{i}^{q}\ $and $c\left(  r,q^{\prime}\right)  =\sum
_{i=1}^{n}d\left(  \bar{r}_{i},q_{i}\right)  \mu_{i}^{q^{\prime}}$. For each
$i\in \left \{  1,...,n\right \}  $ set%
\[
\alpha_{i}=\left \{
\begin{array}
[c]{cc}%
\frac{\lambda \mu_{i}^{q}}{\lambda \mu_{i}^{q}+\left(  1-\lambda \right)  \mu
_{i}^{q^{\prime}}} & \text{if }\lambda \mu_{i}^{q}+\left(  1-\lambda \right)
\mu_{i}^{q^{\prime}}>0\\
\frac{1}{2} & \text{if }\lambda \mu_{i}^{q}+\left(  1-\lambda \right)  \mu
_{i}^{q^{\prime}}=0
\end{array}
\right.
\]
Clearly, we have that $\alpha_{i}\in \left[  0,1\right]  $ and%
\[
1-\alpha_{i}=\left \{
\begin{array}
[c]{cc}%
\frac{\left(  1-\lambda \right)  \mu_{i}^{q^{\prime}}}{\lambda \mu_{i}%
^{q}+\left(  1-\lambda \right)  \mu_{i}^{q^{\prime}}} & \text{if }\lambda
\mu_{i}^{q}+\left(  1-\lambda \right)  \mu_{i}^{q^{\prime}}>0\\
\frac{1}{2} & \text{if }\lambda \mu_{i}^{q}+\left(  1-\lambda \right)  \mu
_{i}^{q^{\prime}}=0
\end{array}
\right.  \quad \forall i\in \left \{  1,...,n\right \}
\]
Define $\left(  \hat{p}_{i}\right)  _{i=1}^{n}\in \Delta^{n}$ to be such that
$\hat{p}_{i}=\alpha_{i}\bar{p}_{i}+\left(  1-\alpha_{i}\right)  \bar{r}_{i}$
for all $i\in \left \{  1,...,n\right \}  $.\ Note that%
\[
\lambda q+\left(  1-\lambda \right)  q^{\prime}=\lambda \sum_{i=1}^{n}\mu
_{i}^{q}q_{i}+\left(  1-\lambda \right)  \sum_{i=1}^{n}\mu_{i}^{q^{\prime}%
}q_{i}=\sum_{i=1}^{n}\left[  \lambda \mu_{i}^{q}+\left(  1-\lambda \right)
\mu_{i}^{q^{\prime}}\right]  q_{i}%
\]
yielding that $\mu^{\lambda q+\left(  1-\lambda \right)  q^{\prime}}=\lambda
\mu^{q}+\left(  1-\lambda \right)  \mu^{q^{\prime}}$. Moreover, since
$\lambda \mu_{i}^{q}+\left(  1-\lambda \right)  \mu_{i}^{q^{\prime}}=0$ if and
only if $\mu_{i}^{q}=\mu_{i}^{q^{\prime}}=0$, we have that%
\begin{align*}
\sum_{i=1}^{n}\mu_{i}^{\lambda q+\left(  1-\lambda \right)  q^{\prime}}\hat
{p}_{i}  &  =\sum_{i=1}^{n}\mu_{i}^{\lambda q+\left(  1-\lambda \right)
q^{\prime}}\left(  \alpha_{i}\bar{p}_{i}+\left(  1-\alpha_{i}\right)  \bar
{r}_{i}\right) \\
&  =\sum_{i=1}^{n}\mu_{i}^{\lambda q+\left(  1-\lambda \right)  q^{\prime}%
}\alpha_{i}\bar{p}_{i}+\sum_{i=1}^{n}\mu_{i}^{\lambda q+\left(  1-\lambda
\right)  q^{\prime}}\left(  1-\alpha_{i}\right)  \bar{r}_{i}\\
&  =\lambda \sum_{i=1}^{n}\mu_{i}^{q}\bar{p}_{i}+\left(  1-\lambda \right)
\sum_{i=1}^{n}\mu_{i}^{q^{\prime}}\bar{r}_{i}=\lambda p+\left(  1-\lambda
\right)  r
\end{align*}
proving that $\left(  \hat{p}_{i}\right)  _{i=1}^{n}\in \Psi \left(  \lambda
p+\left(  1-\lambda \right)  r,\lambda q+\left(  1-\lambda \right)  q^{\prime
}\right)  $. Since $d$\ is convex in $p$, this implies that%
\begin{align*}
\lambda c\left(  p,q\right)  +\left(  1-\lambda \right)  c\left(  r,q^{\prime
}\right)   &  =\lambda \sum_{i=1}^{n}d\left(  \bar{p}_{i},q_{i}\right)  \mu
_{i}^{q}+\left(  1-\lambda \right)  \sum_{i=1}^{n}d\left(  \bar{r}_{i}%
,q_{i}\right)  \mu_{i}^{q^{\prime}}\\
&  =\sum_{i=1}^{n}\alpha_{i}d\left(  \bar{p}_{i},q_{i}\right)  \mu
_{i}^{\lambda q+\left(  1-\lambda \right)  q^{\prime}}+\sum_{i=1}^{n}\left(
1-\alpha_{i}\right)  d\left(  \bar{r}_{i},q_{i}\right)  \mu_{i}^{\lambda
q+\left(  1-\lambda \right)  q^{\prime}}\\
&  =\sum_{i=1}^{n}\left[  \alpha_{i}d\left(  \bar{p}_{i},q_{i}\right)
+\left(  1-\alpha_{i}\right)  d\left(  \bar{r}_{i},q_{i}\right)  \right]
\mu_{i}^{\lambda q+\left(  1-\lambda \right)  q^{\prime}}\\
&  \geq \sum_{i=1}^{n}d\left(  \alpha_{i}\bar{p}_{i}+\left(  1-\alpha
_{i}\right)  \bar{r}_{i},q_{i}\right)  \mu_{i}^{\lambda q+\left(
1-\lambda \right)  q^{\prime}}=\sum_{i=1}^{n}d\left(  \hat{p}_{i},q_{i}\right)
\mu_{i}^{\lambda q+\left(  1-\lambda \right)  q^{\prime}}\\
&  \geq c\left(  \lambda p+\left(  1-\lambda \right)  r,\lambda q+\left(
1-\lambda \right)  q^{\prime}\right)
\end{align*}
yielding that $c$ is jointly convex. Next, consider the map $\Gamma:\Delta
^{n}\times \operatorname*{co}Q\rightarrow \left[  0,\infty \right]  $ defined by%
\[
\Gamma \left(  \left(  p_{i}\right)  _{i=1}^{n},q\right)  =\sum_{i=1}%
^{n}d\left(  p_{i},q_{i}\right)  \mu_{i}^{q}\qquad \forall \left(  \left(
p_{i}\right)  _{i=1}^{n},q\right)  \in \Delta^{n}\times \operatorname*{co}Q
\]
We endow $\Delta^{n}\times \operatorname*{co}Q$\ with the product topology.
Consider a net $\left \{  \left(  \left(  p_{i,\alpha}\right)  _{i=1}%
^{n},q_{\alpha}\right)  \right \}  _{\alpha \in A}$ which converges to $\left(
\left(  p_{i}\right)  _{i=1}^{n},q\right)  $.\ Observe that $\left \{  \left(
\mu_{i}^{q_{\alpha}}\right)  _{i=1}^{n}\right \}  _{\alpha \in A}$ converges
pointwise\ to $\left(  \mu_{i}^{q}\right)  _{i=1}^{n}$, otherwise there would
exist a subnet $\left \{  \left(  \mu_{i}^{q_{\alpha_{\beta}}}\right)
_{i=1}^{n}\right \}  _{\beta \in B}$ which converges to $\left(  \bar{\mu}%
_{i}\right)  _{i=1}^{n}\not =\left(  \mu_{i}^{q}\right)  _{i=1}^{n}$. Since
$\left \{  q_{\alpha}\right \}  _{\alpha \in A}$ converges to $q$, this would
yield that%
\[
\sum_{i=1}^{n}\mu_{i}^{q}q_{i}=q=\lim_{\beta}q_{\alpha_{\beta}}=\lim_{\beta
}\sum_{i=1}^{n}\mu_{i}^{q_{\alpha_{\beta}}}q_{i}=\sum_{i=1}^{n}\bar{\mu}%
_{i}q_{i}%
\]
Since $\sum_{i=1}^{n}\bar{\mu}_{i}=1$, it follows that $\mu_{i}^{q}=\bar{\mu
}_{i}$ for all $i\in \left \{  1,...,n\right \}  $, a contradiction. Since $d$ is
lower semicontinuous, we can conclude that%
\begin{align*}
\Gamma \left(  \left(  p_{i}\right)  _{i=1}^{n},q\right)   &  =\sum_{i=1}%
^{n}d\left(  p_{i},q_{i}\right)  \mu_{i}^{q}\leq \sum_{i=1}^{n}\liminf_{\alpha
}d\left(  p_{i,\alpha},q_{i}\right)  \lim_{\alpha}\mu_{i}^{q_{\alpha}}\\
&  =\sum_{i=1}^{n}\liminf_{\alpha}d\left(  p_{i,\alpha},q_{i}\right)  \mu
_{i}^{q_{\alpha}}\leq \liminf_{\alpha}\sum_{i=1}^{n}d\left(  p_{i,\alpha}%
,q_{i}\right)  \mu_{i}^{q_{\alpha}}\\
&  =\liminf_{\alpha}\Gamma \left(  \left(  p_{i,\alpha}\right)  _{i=1}%
^{n},q_{\alpha}\right)
\end{align*}
proving that $\Gamma$ is lower semicontinuous. Let $t\in \mathbb{R}$. Consider
a net $\left \{  \left(  p_{\alpha},q_{\alpha}\right)  \right \}  _{\alpha \in
A}\in \Delta \times \operatorname*{co}Q$ that converges to $\left(  p,q\right)  $
and such that $c\left(  p_{\alpha},q_{\alpha}\right)  \leq t$ for all
$\alpha \in A$. For each $\alpha \in A$, consider\ $\left(  \bar{p}_{i,\alpha
}\right)  _{i=1}^{n}\in \Psi \left(  p_{\alpha},q_{\alpha}\right)  $ such that
$c\left(  p_{\alpha},q_{\alpha}\right)  =\sum_{i=1}^{n}d\left(  \bar
{p}_{i,\alpha},q_{i}\right)  \mu_{i}^{q_{\alpha}}=\Gamma \left(  \left(
\bar{p}_{i,\alpha}\right)  _{i=1}^{n},q_{\alpha}\right)  $. Since $\Delta^{n}$
is compact and the previous part of the proof, there exists a subnet $\left \{
\left(  \bar{p}_{i,\alpha_{\beta}}\right)  _{i=1}^{n}\right \}  _{\beta \in B}$
which converges to $\left(  \bar{p}_{i}\right)  _{i=1}^{n}\in \Delta^{n}$
while\ $\left \{  \left(  \mu_{i}^{q_{\alpha_{\beta}}}\right)  _{i=1}%
^{n}\right \}  _{\beta \in B}$ converges to $\left(  \mu_{i}^{q}\right)
_{i=1}^{n}$. Since $\left \{  p_{\alpha_{\beta}}\right \}  _{\beta \in B}$
converges to $p$, it follows that $p=\lim_{\beta}p_{\alpha_{\beta}}%
=\lim_{\beta}\sum_{i=1}^{n}\mu_{i}^{q_{\alpha_{\beta}}}\bar{p}_{i,\alpha
_{\beta}}=\sum_{i=1}^{n}\mu_{i}^{q}\bar{p}_{i}$, yielding that $\left(
\bar{p}_{i}\right)  _{i=1}^{n}\in \Psi \left(  p,q\right)  $. By definition of
$c$ and since $\Gamma$ is lower semicontinuous, this implies that%
\begin{align*}
c\left(  p,q\right)   &  \leq \sum_{i=1}^{n}d\left(  \bar{p}_{i},q_{i}\right)
\mu_{i}^{q}=\Gamma \left(  \left(  \bar{p}_{i}\right)  _{i=1}^{n},q\right)
\leq \liminf_{\beta}\Gamma \left(  \left(  \bar{p}_{i,\alpha_{\beta}}\right)
_{i=1}^{n},q_{\alpha_{\beta}}\right) \\
&  =\liminf_{\beta}c\left(  p_{\alpha_{\beta}},q_{\alpha_{\beta}}\right)  \leq
t
\end{align*}
proving that $c$ is jointly lower semicontinuous. By Lemma
\ref{lem:app-suff-div}, if $\mathcal{Q}$ is a collection of compact and convex
subsets of $\Delta^{\sigma}$ such that $\mathcal{S}=\operatorname{co}Q$, then
we can conclude that $c$ is a variational statistical distance.\ Finally,
assume that $d\left(  p,q\right)  <\infty$ implies $p\ll q$. Consider $\left(
p,q\right)  \in \Delta \times \operatorname*{co}Q$ and assume that $c\left(
p,q\right)  <\infty$. Let $\left(  \bar{p}_{i}\right)  _{i=1}^{n}\in
\Psi \left(  p,q\right)  $ be such that $c\left(  p,q\right)  =\sum_{i=1}%
^{n}d\left(  \bar{p}_{i},q_{i}\right)  \mu_{i}^{q}$. Since $c\left(
p,q\right)  <\infty$, we have that $d\left(  \bar{p}_{i},q_{i}\right)
<\infty$ for all $i\in \left \{  1,...,n\right \}  $ such that $\mu_{i}^{q}>0$,
proving that $\bar{p}_{i}\ll q_{i}$ for all $i\in \left \{  1,...,n\right \}  $
such that $\mu_{i}^{q}>0$. Next, consider $A\in \Sigma$ such that $q\left(
A\right)  =0$. Since $q=\sum_{i=1}^{n}\mu_{i}^{q}q_{i}$, we have that
$q_{i}\left(  A\right)  =0$ for all $i\in \left \{  1,...,n\right \}  $ such that
$\mu_{i}^{q}>0$, yielding that $\bar{p}_{i}\left(  A\right)  =0$ for all
$i\in \left \{  1,...,n\right \}  $ such that $\mu_{i}^{q}>0$. Since
$p=\sum_{i=1}^{n}\mu_{i}^{q}\bar{p}_{i}=\sum_{i:\mu_{i}^{q}>0}\mu_{i}^{q}%
\bar{p}_{i}$, this implies that $p\left(  A\right)  =0$, that is, $p\ll q$,
yielding that $c$ is a divergence.\hfill$\blacksquare$

\subsection{Non-convex set of structured models\label{app:non-cvx}}

Let us consider two decision makers who adopt criterion
(\ref{eq:entropic-criterion}), the first one posits a, possibly non-convex,
set of structured models $Q$ and the second one posits its closed convex hull
$\overline{\operatorname*{co}}\,Q$. So, the second decision maker considers
also all the mixtures of structured models posited by the first decision
maker. Next we show that their preferences over acts actually agree. We deal
with the case $\lambda \in(0,\infty)$, being $\lambda=\infty$ trivial. It is
thus without loss of generality to assume that the set of posited structured
models is convex, as it was assumed in mostly of the main text. Before doing
so we prove formula (\ref{eq:dual-entropy}). Observe that given a compact
subset $Q\subseteq \Delta^{\sigma}$, be that convex or not, we have%

\begin{align*}
\min_{p\in \Delta}\left \{  \int u\left(  f\right)  dp+\lambda \min_{q\in
Q}R\left(  p||q\right)  \right \}   &  =\min_{p\in \Delta}\min_{q\in Q}\left \{
\int u\left(  f\right)  dp+\lambda R\left(  p||q\right)  \right \} \\
&  =\min_{q\in Q}\min_{p\in \Delta}\left \{  \int u\left(  f\right)  dp+\lambda
R\left(  p||q\right)  \right \} \\
&  =\min_{q\in Q}\phi_{\lambda}^{-1}\left(  \int \phi_{\lambda}\left(  u\left(
f\right)  \right)  dq\right)
\end{align*}
where $\phi_{\lambda}\left(  t\right)  =-e^{-\frac{1}{\lambda}t}$ for all
$t\in \mathbb{R}$ where $\lambda>0$.

\begin{proposition}
\label{prop:non-cvx}If $Q\subseteq \Delta^{\sigma}$ is compact, then for each
$f\in \mathcal{F}$%
\[
\min_{p\in \Delta}\left \{  \int u\left(  f\right)  dp+\lambda \min_{q\in
Q}R\left(  p||q\right)  \right \}  =\min_{p\in \Delta}\left \{  \int u\left(
f\right)  dp+\lambda \min_{q\in \overline{\operatorname*{co}}\,Q}R\left(
p||q\right)  \right \}
\]

\end{proposition}

\noindent \textbf{Proof} First observe that $\overline{\operatorname*{co}%
}\,Q\subseteq \Delta^{\sigma}$. Indeed, since $Q$ is a compact subset of
$\Delta^{\sigma}$, the set function $\nu:\Sigma \rightarrow \left[  0,1\right]
$, defined by $\nu \left(  E\right)  =\min_{q\in Q}q\left(  E\right)  $ for all
$E\in \Sigma$\ is an exact capacity which is continuous at $S$. This implies
that $Q\subseteq \operatorname*{core}\nu \subseteq \Delta^{\sigma}$, yielding
that $\overline{\operatorname*{co}}\,Q\subseteq \operatorname*{core}%
\nu \subseteq \Delta^{\sigma}$. Given what we have shown before we can conclude
that%
\begin{align*}
\min_{p\in \Delta}\left \{  \int u\left(  f\right)  dp+\lambda \min_{q\in
Q}R\left(  p||q\right)  \right \}   &  =\min_{q\in Q}\phi_{\lambda}^{-1}\left(
\int \phi_{\lambda}\left(  u\left(  f\right)  \right)  dq\right) \\
&  =\phi_{\lambda}^{-1}\left(  \min_{q\in Q}\left(  \int \phi_{\lambda}\left(
u\left(  f\right)  \right)  dq\right)  \right) \\
&  =\phi_{\lambda}^{-1}\left(  \min_{q\in \overline{\operatorname*{co}}%
\,Q}\left(  \int \phi_{\lambda}\left(  u\left(  f\right)  \right)  dq\right)
\right) \\
&  =\min_{q\in \overline{\operatorname*{co}}\,Q}\phi_{\lambda}^{-1}\left(
\int \phi_{\lambda}\left(  u\left(  f\right)  \right)  dq\right) \\
&  =\min_{p\in \Delta}\left \{  \int u\left(  f\right)  dp+\lambda \min
_{q\in \overline{\operatorname*{co}}\,Q}R\left(  p||q\right)  \right \}
\end{align*}
proving the statement.\hfill$\blacksquare$

\bigskip

After (\ref{eq:rob-mea-var-bis}), we claimed that the Gini criterion is a
monotone version of the max-min mean-variance criterion. To be more precise,
given a probability $q\in \Delta^{\sigma}$ and a weight $1/2\lambda>0$ for the
variance, the mean-variance criterion is not monotone over its entire domain,
but it is normalized, translation invariant, and monotone in an area
containing the constant functions (see Theorem 24 of Maccheroni et al., 2006).
At the same time, the variational preference with cost function the Gini index
$\lambda \chi^{2}(\cdot||q)$ is monotone and coincides with the mean-variance
criterion over such an area. A similar argument, \emph{mutatis mutandis},
holds for the max-min mean-variance criterion and our formula
(\ref{eq:rob-mea-var}). This allows us to see the corresponding variational
criteria as a monotonization of the corresponding mean-variance ones.

\subsection{Main theorems: ancillary results\label{app:anc-rep}}

We begin by proving the two ancillary variational lemmas.

\smallskip

\noindent \textbf{Proof of\ Lemma \ref{lm:zero-set}}\ We actually prove that
(i)$\Longrightarrow$(ii)$\Longleftrightarrow$(iii), with equivalence when
$\succsim$ is unbounded.

\smallskip

(i) implies (ii). Let $f\in \mathcal{F}$. It is enough to observe that
$c\left(  \bar{p}\right)  =0$ implies%
\[
V\left(  x_{f}^{\bar{p}}\right)  =u\left(  x_{f}^{\bar{p}}\right)  =\int
u\left(  f\right)  d\bar{p}+c\left(  \bar{p}\right)  \geq \min_{p\in \Delta
}\left \{  \int u\left(  f\right)  dp+c\left(  p\right)  \right \}  =V\left(
f\right)
\]
yielding that $x_{f}^{\bar{p}}\succsim f$.

\smallskip

(ii) implies (iii). Assume that $x_{f}^{\bar{p}}\succsim f$\ for all
$f\in \mathcal{F}$. Since $\succsim$ is complete and transitive, it follows
that if $x\succ x_{f}^{\bar{p}}$, then $x\succ f$.

\smallskip

(iii) implies (ii). By contradiction, suppose that there exists $f\in
\mathcal{F}$ such that $f\succ x_{f}^{\bar{p}}$. Let\ $x_{f}\in X$ be such
that $x_{f}\sim f$. This implies that $x_{f}\succ x_{f}^{\bar{p}}$ and so
$x_{f}\succ f$, a contradiction.

\smallskip

(ii) implies (i). Let $\succsim$ be unbounded. Assume that $x_{f}^{\bar{p}%
}\succsim f$\ for all $f\in \mathcal{F}$, i.e., $V\left(  f\right)  \leq \int
u\left(  f\right)  d\bar{p}$ for all $f\in \mathcal{F}$. So, $\bar{p}$
corresponds to a SEU preference that is less ambiguity averse than $\succsim$.
By Lemma 32 of Maccheroni et al. (2006), we can conclude that $c\left(
\bar{p}\right)  =0$.\hfill$\blacksquare$

\smallskip

\noindent \textbf{Proof of\ Lemma \ref{lem:del-Q}} We begin by observing that
in proving the two implications, $Q$ being either compact or convex plays no role.

\smallskip

(i) implies (ii). Let $p\in \Delta \backslash \Delta^{\ll}\left(  Q\right)  $. It
follows that there exists $A\in \Sigma$ such that $q\left(  A\right)  =0$ for
all $q\in Q$ as well as $p\left(  A\right)  >0$. Define $I:B_{0}\left(
\Sigma \right)  \rightarrow \mathbb{R}$ by $I\left(  \varphi \right)  =\min
_{p\in \Delta}\left \{  \int \varphi dp+c\left(  p\right)  \right \}  $ for all
$\varphi \in B_{0}\left(  \Sigma \right)  $. Since $u$ is unbounded, for each
$\lambda \in \mathbb{R}$ there exists $x_{\lambda}\in X$ such that $u\left(
x_{\lambda}\right)  =\lambda$. Similarly, there exists $y\in X$ such that
$u\left(  y\right)  =0$. For each $\lambda \in \mathbb{R}$ define $f_{\lambda
}=x_{\lambda}Ay$. By construction, we have that $f_{\lambda}\overset{Q}{=}y$
for all $\lambda \in \mathbb{R}$. This implies that $I\left(  \lambda
1_{A}\right)  =V\left(  f_{\lambda}\right)  =V\left(  y\right)  =I\left(
0\right)  =0$ for all $\lambda \in \mathbb{R}$.\ By Maccheroni et al. (2006) and
since $u$ is unbounded and $p\left(  A\right)  >0$, we have that%
\[
c\left(  p\right)  =\sup_{\varphi \in B_{0}\left(  \Sigma \right)  }\left \{
I\left(  \varphi \right)  -\int \varphi dp\right \}  \geq \sup_{\lambda
\in \mathbb{R}}\left \{  I\left(  \lambda1_{A}\right)  -\lambda p\left(
A\right)  \right \}  =\infty
\]
Since $p$ was arbitrarily chosen, it follows that $\operatorname*{dom}%
c\subseteq \Delta^{\ll}\left(  Q\right)  $.

\smallskip

(ii) implies (i). Assume that $\operatorname*{dom}c\subseteq \Delta^{\ll
}\left(  Q\right)  $. If $f\overset{Q}{=}g$, then $u\left(  f\right)
\overset{Q}{=}u\left(  g\right)  $. This implies that $u\left(  f\right)
\overset{p}{=}u\left(  g\right)  $ for all $p\in \Delta^{\ll}\left(  Q\right)
$ and, in particular,%
\begin{align*}
V\left(  f\right)   &  =\min_{p\in \Delta}\left \{  \int u\left(  f\right)
dp+c\left(  p\right)  \right \}  =\min_{p\in \Delta^{\ll}\left(  Q\right)
}\left \{  \int u\left(  f\right)  dp+c\left(  p\right)  \right \} \\
&  =\min_{p\in \Delta^{\ll}\left(  Q\right)  }\left \{  \int u\left(  g\right)
dp+c\left(  p\right)  \right \}  =\min_{p\in \Delta}\left \{  \int u\left(
g\right)  dp+c\left(  p\right)  \right \}  =V\left(  g\right)
\end{align*}
proving that $f\sim g$.\hfill$\blacksquare$

\smallskip

\noindent \textbf{Proof of Lemma \ref{lem:def-sta}} We begin by showing that
$\succeq^{\ast}$ is well defined and does not depend on the representing
elements of\ $\psi$\ and\ $\varphi$. Assume that $f_{1},f_{2},g_{1},g_{2}%
\in \mathcal{F}$ are such that $u\left(  f_{i}\right)  =\varphi$ and $u\left(
g_{i}\right)  =\psi$ for all $i\in \left \{  1,2\right \}  $. It follows that
$u\left(  f_{1}\left(  s\right)  \right)  =u\left(  f_{2}\left(  s\right)
\right)  $ and $u\left(  g_{1}\left(  s\right)  \right)  =u\left(
g_{2}\left(  s\right)  \right)  $ for all $s\in S$. By Lemma \ref{lem:HM},
this implies that $f_{1}\left(  s\right)  \sim^{\ast}f_{2}\left(  s\right)  $
and $g_{1}\left(  s\right)  \sim^{\ast}g_{2}\left(  s\right)  $ for all $s\in
S$. Since $\succsim^{\ast}$ is a preorder that satisfies monotonicity, this
implies that $f_{1}\sim^{\ast}f_{2}$ and $g_{1}\sim^{\ast}g_{2}$. Since
$\succsim^{\ast}$ is a preorder, if $f_{1}\succsim^{\ast}g_{1}$, then%
\[
f_{2}\succsim^{\ast}f_{1}\succsim^{\ast}g_{1}\succsim^{\ast}g_{2}\implies
f_{2}\succsim^{\ast}g_{2}%
\]
that is, $f_{1}\succsim^{\ast}g_{1}$ implies $f_{2}\succsim^{\ast}g_{2}$.
Similarly, we can prove that $f_{2}\succsim^{\ast}g_{2}$ implies
$f_{1}\succsim^{\ast}g_{1}$. In other words, $f_{1}\succsim^{\ast}g_{1}$ if
and only if $f_{2}\succsim^{\ast}g_{2}$, proving that $\succeq^{\ast}$ is well
defined and does not depend on the representing elements of\ $\psi
$\ and\ $\varphi$. It is immediate to prove that $\succeq^{\ast}$ is a
preorder. We next prove properties 1--5.

\begin{enumerate}
\item Consider\ $\varphi,\psi \in B_{0}\left(  \Sigma \right)  $ and
$k\in \mathbb{R}$. Assume that $\varphi \succeq^{\ast}\psi$. Let $f,g\in
\mathcal{F}$ and $x,y\in X$ be such that $u\left(  f\right)  =2\varphi$,
$u\left(  g\right)  =2\psi$,\ $u\left(  x\right)  =0$ and $u\left(  y\right)
=2k$. Since $u$ is affine, it follows that%
\begin{align*}
u\left(  \frac{1}{2}f+\frac{1}{2}x\right)   &  =\frac{1}{2}u\left(  f\right)
+\frac{1}{2}u\left(  x\right)  =\varphi \succeq^{\ast}\psi \\
&  =\frac{1}{2}u\left(  g\right)  +\frac{1}{2}u\left(  x\right)  =u\left(
\frac{1}{2}g+\frac{1}{2}x\right)
\end{align*}
proving that $\frac{1}{2}f+\frac{1}{2}x\succsim^{\ast}\frac{1}{2}g+\frac{1}%
{2}x$. Since $\succsim^{\ast}$ satisfies weak c-independence and $u$ is
affine, we have that $\frac{1}{2}f+\frac{1}{2}y\succsim^{\ast}\frac{1}%
{2}g+\frac{1}{2}y$, yielding that%
\begin{align*}
\varphi+k  &  =\frac{1}{2}u\left(  f\right)  +\frac{1}{2}u\left(  y\right)
=u\left(  \frac{1}{2}f+\frac{1}{2}y\right)  \succeq^{\ast}u\left(  \frac{1}%
{2}g+\frac{1}{2}y\right) \\
&  =\frac{1}{2}u\left(  g\right)  +\frac{1}{2}u\left(  y\right)  =\psi+k
\end{align*}

\item Consider $\varphi,\psi \in B_{0}\left(  \Sigma \right)  $ and $\left \{
k_{n}\right \}  _{n\in%
\mathbb{N}
}\subseteq \mathbb{R}$ such that\ $k_{n}\uparrow k$ and $\varphi-k_{n}%
\succeq^{\ast}\psi$ for all $n\in%
\mathbb{N}
$. We have two cases:

\begin{enumerate}
\item $k>0$. Consider $f,g,h\in \mathcal{F}$ such that%
\[
u\left(  f\right)  =\varphi \text{, }u\left(  g\right)  =\varphi-k\text{ and
}u\left(  h\right)  =\psi
\]
Since $k>0$ and $k_{n}\uparrow k$, there exists $\bar{n}\in%
\mathbb{N}
$ such that $k_{n}>0$ for all $n\geq \bar{n}$. Define $\lambda_{n}=1-k_{n}/k$
for all $n\in%
\mathbb{N}
$. It follows that $\lambda_{n}\in \left[  0,1\right]  $ for all $n\geq \bar{n}%
$. Since $u$ is affine, for each $n\geq \bar{n}$%
\[
u\left(  \lambda_{n}f+\left(  1-\lambda_{n}\right)  g\right)  =\lambda
_{n}u\left(  f\right)  +\left(  1-\lambda_{n}\right)  u\left(  g\right)
=\varphi-k_{n}\succeq^{\ast}\psi=u\left(  h\right)
\]
yielding that $\lambda_{n}f+\left(  1-\lambda_{n}\right)  g\succsim^{\ast}h$
for all $n\geq \bar{n}$. Since $\succsim^{\ast}$ satisfies continuity and
$\lambda_{n}\rightarrow0$, we have that $g\succsim^{\ast}h$, that is,%
\[
\varphi-k=u\left(  g\right)  \succeq^{\ast}u\left(  h\right)  =\psi
\]

\item $k\leq0$. Since $\left \{  k_{n}\right \}  _{n\in%
\mathbb{N}
}$ is convergent, $\left \{  k_{n}\right \}  _{n\in%
\mathbb{N}
}$ is bounded. Thus, there exists\ $h>0$ such that $k_{n}+h>0$ for all $n\in%
\mathbb{N}
$. Moreover, $k_{n}+h\uparrow k+h>0$. By point 1, we also have that
$\varphi-\left(  k_{n}+h\right)  =\left(  \varphi-k_{n}\right)  -h\succeq
^{\ast}\psi-h$\ for all $n\in%
\mathbb{N}
$. By subpoint a, we can conclude that $\left(  \varphi-k\right)
-h=\varphi-\left(  k+h\right)  \succeq^{\ast}\psi-h$. By point 1, we obtain
that $\varphi-k\succeq^{\ast}\psi$.
\end{enumerate}

\item Consider\ $\varphi,\psi \in B_{0}\left(  \Sigma \right)  $ such that
$\varphi \geq \psi$. Let $f,g\in \mathcal{F}$ be such that $u\left(  f\right)
=\varphi$ and $u\left(  g\right)  =\psi$. It follows that $u\left(  f\left(
s\right)  \right)  \geq u\left(  g\left(  s\right)  \right)  $ for all $s\in
S$. By Lemma \ref{lem:HM}, this implies that $f\left(  s\right)
\succsim^{\ast}g\left(  s\right)  $\ for all $s\in S$. Since $\succsim^{\ast}$
satisfies monotonicity, this implies that $f\succsim^{\ast}g$, yielding that
$\varphi=u\left(  f\right)  \succeq^{\ast}u\left(  g\right)  =\psi$.

\item Consider\ $k,h\in \mathbb{R}$ and $\varphi \in B_{0}\left(  \Sigma \right)
$. We first assume that $k>h$ and $k=0$. By point 3, we have that
$\varphi=\varphi+k\succeq^{\ast}\varphi+h$. By contradiction, assume that
$\varphi \not \succ ^{\ast}\varphi+h$. It follows that $\varphi \sim^{\ast
}\varphi+h$, yielding that $I=\left \{  w\in \mathbb{R}:\varphi \sim^{\ast
}\varphi+w\right \}  $ is a non-empty set which contains $0$ and $h$. We next
prove that $I\ $is an unbounded interval, that is, $I=\mathbb{R}$. First,
consider $w_{1},w_{2}\in I$. Without loss of generality, assume that
$w_{1}\geq w_{2}$. By point 3 and since $w_{1},w_{2}\in I$, we have that for
each $\lambda \in \left(  0,1\right)  $%
\[
\varphi \succeq^{\ast}\varphi+w_{1}\succeq^{\ast}\varphi+\left(  \lambda
w_{1}+\left(  1-\lambda \right)  w_{2}\right)  \succeq^{\ast}\varphi
+w_{2}\succeq^{\ast}\varphi
\]
proving that $\varphi \sim^{\ast}\varphi+\left(  \lambda w_{1}+\left(
1-\lambda \right)  w_{2}\right)  $, that is, $\lambda w_{1}+\left(
1-\lambda \right)  w_{2}\in I$. Next, we observe that $I\cap \left(
-\infty,0\right)  \not =\emptyset \not =I\cap \left(  0,\infty \right)  $. Since
$h\in I$ and $h<0$, we have that $I\cap \left(  -\infty,0\right)
\not =\emptyset$. Since $I$ is an interval and $0,h\in I$, we have that
$h/2\in I$. By point 1 and since $\varphi \sim^{\ast}\varphi+h/2$, we have that
$\varphi-h/2\sim^{\ast}\left(  \varphi+h/2\right)  -h/2=\varphi$, proving that
$0<-h/2\in I\cap \left(  0,\infty \right)  $. By definition of $I$, note that if
$w\in I\backslash \left \{  0\right \}  $, then $\varphi+w\sim^{\ast}\varphi$. By
point 1 and since $w/2\in I$ and $\succeq^{\ast}$ is a preorder, we have that
$\left(  \varphi+w\right)  +w/2\sim^{\ast}\varphi+w/2\sim^{\ast}\varphi$, that
is, $\frac{3}{2}w,\frac{1}{2}w\in I$. Since $I$ is an interval, we have that
either $\left[  \frac{3}{2}w,\frac{1}{2}w\right]  \subseteq I$ if $w<0$ or
$\left[  \frac{1}{2}w,\frac{3}{2}w\right]  \subseteq I$ if $w>0$. This will
help us in proving that $I$ is unbounded from below and above. By
contradiction, assume that $I$ is bounded from below and define $m=\inf I$.
Since $I\cap \left(  -\infty,0\right)  \not =\emptyset$, we have that $m<0$.
Consider $\left \{  w_{n}\right \}  _{n\in%
\mathbb{N}
}\subseteq I\cap \left(  -\infty,0\right)  $ such that $w_{n}\downarrow m$.
Since $\left[  \frac{3}{2}w_{n},\frac{1}{2}w_{n}\right]  \subseteq I$ for all
$n\in%
\mathbb{N}
$, it follows that $m\leq \frac{3}{2}w_{n}$\ for all $n\in%
\mathbb{N}
$. By passing to the limit, we obtain that $m\leq \frac{3}{2}m<0$, a
contradiction. By contradiction, assume that $I$ is bounded from above and
define $M=\sup I$. Since $I\cap \left(  0,\infty \right)  \not =\emptyset$, we
have that $M>0$. Consider $\left \{  w_{n}\right \}  _{n\in%
\mathbb{N}
}\subseteq I\cap \left(  0,\infty \right)  $ such that $w_{n}\uparrow M$. Since
$\left[  \frac{1}{2}w_{n},\frac{3}{2}w_{n}\right]  \subseteq I$ for all $n\in%
\mathbb{N}
$, it follows that $M\geq \frac{3}{2}w_{n}$\ for all $n\in%
\mathbb{N}
$. By passing to the limit, we obtain that $M\geq \frac{3}{2}M>0$, a
contradiction. To sum up, $I$ is a non-empty\ unbounded interval, that is,
$I=\mathbb{R}$. This implies that $\varphi \sim^{\ast}\varphi+w$ for all
$w\in \mathbb{R}$. In particular, select $w_{1}=\left \Vert \varphi \right \Vert
_{\infty}+1$ and $w_{2}=-\left \Vert \varphi \right \Vert _{\infty}-1$. Since
$\succeq^{\ast}$ is a preorder, we have that $\varphi+w_{1}\sim^{\ast}%
\varphi+w_{2}$. Moreover, $\varphi+w_{1}\geq1>-1\geq \varphi+w_{2}$. By point
3, this implies that $\varphi+w_{1}\succeq^{\ast}1\succeq^{\ast}%
-1\succeq^{\ast}\varphi+w_{2}$. Since $\succeq^{\ast}$ is a preorder and
$\varphi+w_{1}\sim^{\ast}\varphi+w_{2}$, we can conclude that $1\sim^{\ast}%
-1$. Note also that there exist $x,y\in X$ such that $u\left(  x\right)  =1$
and $u\left(  y\right)  =-1$.\ By Lemma \ref{lem:HM}, this implies that
$x\succ^{\ast}y$. By definition of $\succeq^{\ast}$ and since $u\left(
x\right)  =1\sim^{\ast}-1=u\left(  y\right)  $, we also have that
$y\succsim^{\ast}x$, a contradiction. Thus, we proved that if $k>h$ and $k=0$,
then $\varphi+k\succ^{\ast}\varphi+h$. Assume simply that $k>h$. This implies
that $0>h-k$ and $\varphi \succ^{\ast}\varphi+\left(  h-k\right)  $. By point
1, we can conclude that $\varphi+k\succ^{\ast}\varphi+\left(  h-k\right)
+k=\varphi+h$.

\item Consider\ $\varphi,\psi,\xi \in B_{0}\left(  \Sigma \right)  $ and
$\lambda \in \left(  0,1\right)  $. Assume that $\varphi \succeq^{\ast}\xi$ and
$\psi \succeq^{\ast}\xi$. Let $f,g,h\in \mathcal{F}$ be such that $u\left(
f\right)  =\varphi$, $u\left(  g\right)  =\psi$ and $u\left(  h\right)  =\xi$.
By assumption and definition of $\succeq^{\ast}$, we have that $f\succsim
^{\ast}h$ and $g\succsim^{\ast}h$. Since $\succsim^{\ast}$ satisfies convexity
and $u$ is affine,\ this implies that $\lambda f+\left(  1-\lambda \right)
g\succsim^{\ast}h$, yielding that $\lambda \varphi+\left(  1-\lambda \right)
\psi=\lambda u\left(  f\right)  +\left(  1-\lambda \right)  u\left(  g\right)
=u\left(  \lambda f+\left(  1-\lambda \right)  g\right)  \succeq^{\ast}u\left(
h\right)  =\xi$.
\end{enumerate}

Points 1--5 prove the first part of the statement. Finally, consider
$\varphi,\psi \in B_{0}\left(  \Sigma \right)  $. Note that there exist a
partition $\left \{  A_{i}\right \}  _{i=1}^{n}\subseteq \Sigma$ of $S$ and
$\left \{  \alpha_{i}\right \}  _{i=1}^{n}$ and $\left \{  \beta_{i}\right \}
_{i=1}^{n}$ in $\mathbb{R}$ such that%
\[
\varphi=\sum_{i=1}^{n}\alpha_{i}1_{A_{i}}\text{ and }\psi=\sum_{i=1}^{n}%
\beta_{i}1_{A_{i}}%
\]
Note that $\left \{  s\in S:\varphi \left(  s\right)  \not =\psi \left(
s\right)  \right \}  =\cup_{i\in \left \{  1,...,n\right \}  :\alpha_{i}%
\not =\beta_{i}}A_{i}$. Since $\varphi \overset{Q}{=}\psi$, we have that
$q\left(  A_{i}\right)  =0$ for all $q\in Q$ and for all $i\in \left \{
1,...,n\right \}  $ such that $\alpha_{i}\not =\beta_{i}$. Since $u$ is
unbounded, define $\left \{  x_{i}\right \}  _{i=1}^{n}\subseteq X$\ to be such
that $u\left(  x_{i}\right)  =\alpha_{i}$ for all $i\in \left \{
1,...,n\right \}  $. Since $u$ is unbounded, define $\left \{  y_{i}\right \}
_{i=1}^{n}\subseteq X$ to be such that $y_{i}=x_{i}$ for all $i\in \left \{
1,...,n\right \}  $ such that $\alpha_{i}=\beta_{i}$ and $u\left(
y_{i}\right)  =\beta_{i}$ otherwise. Define $f,g:S\rightarrow X$ by $f\left(
s\right)  =x_{i}$ and $g\left(  s\right)  =y_{i}$ for all $s\in A_{i}$ and for
all $i\in \left \{  1,...,n\right \}  $. It is immediate to see that
$f\overset{Q}{=}g$ as well as $u\left(  f\right)  =\varphi$ and $u\left(
g\right)  =\psi$. Since $\succsim^{\ast}$ is objectively $Q$-coherent, we have
that $f\sim^{\ast}g$, yielding that $\varphi \sim^{\ast}\psi$ and proving the
second part of the statement.\hfill$\blacksquare$

\smallskip

\noindent \textbf{Proof of Lemma \ref{lem:upp-niv} }Consider $\varphi \in
B_{0}\left(  \Sigma \right)  $. Define $C_{\varphi}=\left \{  k\in
\mathbb{R}:\varphi-k\in U\left(  \psi \right)  \right \}  $. Note that
$C_{\varphi}$ is non-empty. Indeed, if we set $k=-\left \Vert \varphi
\right \Vert _{\infty}-\left \Vert \psi \right \Vert _{\infty}$, then we obtain
that $\varphi-k=\varphi+\left \Vert \varphi \right \Vert _{\infty}+\left \Vert
\psi \right \Vert _{\infty}\geq0+\left \Vert \psi \right \Vert _{\infty}\geq \psi \in
U\left(  \psi \right)  $. By property 4 of Lemma \ref{lem:upp-con}, we can
conclude that $\varphi-k\in U\left(  \psi \right)  $, that is, $k\in
C_{\varphi}$. Since $U\left(  \psi \right)  $ is convex, it follows that
$C_{\varphi}$ is an interval. Since $\varphi \in B_{0}\left(  \Sigma \right)  $,
note that there exists $\hat{k}\in \mathbb{R}$ such that $\psi \geq \varphi
-\hat{k}$. It follows that $\psi \succeq^{\ast}\varphi-\hat{k}$. In particular,
we can conclude that $\psi \succ^{\ast}\varphi-\left(  \hat{k}+\varepsilon
\right)  $ for all $\varepsilon>0$. This yields that $C_{\varphi}$ is bounded
from above. Finally, assume that $\left \{  k_{n}\right \}  _{n\in%
\mathbb{N}
}\subseteq C_{\varphi}$ and $k_{n}\uparrow k$. By property 2 of Lemma
\ref{lem:upp-con}, we can conclude that $k\in C_{\varphi}$. To sum up,
$C_{\varphi}$ is a non-empty bounded from above interval of $\mathbb{R}$ that
satisfies the property%
\begin{equation}
\left \{  k_{n}\right \}  _{n\in%
\mathbb{N}
}\subseteq C_{\varphi}\text{ and }k_{n}\uparrow k\implies k\in C_{\varphi}
\label{eq:C-phi-clo}%
\end{equation}
The first part yields that $\sup \left \{  k\in \mathbb{R}:\varphi-k\in U\left(
\psi \right)  \right \}  =\sup C_{\varphi}\in \mathbb{R}$ is well defined. By
(\ref{eq:C-phi-clo}), we also have that $\sup C_{\varphi}\in C_{\varphi}$,
that is, $\sup C_{\varphi}=\max C_{\varphi}$, proving that $I_{\psi}$ is well
defined.\ Next, we prove that $I_{\psi}$ is a concave niveloid. We first show
that\ $I_{\psi}$ is monotone and translation invariant. By Proposition 2\ of
Cerreia-Vioglio et al. (2014), this implies that $I_{\psi}$ is a
niveloid.\ Rather than proving\ monotonicity, we prove that $I_{\psi}$ is
$\succeq^{\ast}$-consistent.\footnote{Since\ if $\varphi_{1}\geq \varphi_{2}$,
then $\varphi_{1}\succeq^{\ast}\varphi_{2}$, it follows that $\succeq^{\ast}%
$-consistency implies monotonicity.}\ Consider $\varphi_{1},\varphi_{2}\in
B_{0}\left(  \Sigma \right)  $ such that $\varphi_{1}\succeq^{\ast}\varphi_{2}%
$. By the properties of $\succeq^{\ast}$ and definition of $I_{\psi}$, we have
that%
\[
\varphi_{1}-I_{\psi}\left(  \varphi_{2}\right)  \succeq^{\ast}\varphi
_{2}-I_{\psi}\left(  \varphi_{2}\right)  \text{ and }\varphi_{2}-I_{\psi
}\left(  \varphi_{2}\right)  \in U\left(  \psi \right)
\]
and, in particular, $\varphi_{2}-I_{\psi}\left(  \varphi_{2}\right)
\succeq^{\ast}\psi$. Since $\succeq^{\ast}$ is a preorder, this implies that
$\varphi_{1}-I_{\psi}\left(  \varphi_{2}\right)  \succeq^{\ast}\psi$, that is,
$\varphi_{1}-I_{\psi}\left(  \varphi_{2}\right)  \in U\left(  \psi \right)  $
and $I_{\psi}\left(  \varphi_{2}\right)  \in C_{\varphi_{1}}$, proving\ that
$I_{\psi}\left(  \varphi_{1}\right)  \geq I_{\psi}\left(  \varphi_{2}\right)
$. We next prove translation invariance.\ Consider $\varphi \in B_{0}\left(
\Sigma \right)  $ and $k\in \mathbb{R}$. By definition of $I_{\psi}$, we can
conclude that%
\[
\left(  \varphi+k\right)  -\left(  I_{\psi}\left(  \varphi \right)  +k\right)
=\varphi-I_{\psi}\left(  \varphi \right)  \in U\left(  \psi \right)
\]
This implies that $I_{\psi}\left(  \varphi \right)  +k\in C_{\varphi+k}$ and,
in particular, $I_{\psi}\left(  \varphi+k\right)  \geq I_{\psi}\left(
\varphi \right)  +k$. Since $k$ and $\varphi$ were arbitrarily chosen, we have
that%
\[
I_{\psi}\left(  \varphi+k\right)  \geq I_{\psi}\left(  \varphi \right)
+k\qquad \forall \varphi \in B_{0}\left(  \Sigma \right)  ,\forall k\in \mathbb{R}%
\]
This yields that $I_{\psi}\left(  \varphi+k\right)  =I_{\psi}\left(
\varphi \right)  +k$ for all $\varphi \in B_{0}\left(  \Sigma \right)  $ and for
all $k\in \mathbb{R}$.\footnote{Observe that if $\varphi \in B_{0}\left(
\Sigma \right)  $ and $k\in \mathbb{R}$, then $-k\in%
\mathbb{R}
$\ and%
\[
I_{\psi}\left(  \varphi \right)  =I_{\psi}\left(  \left(  \varphi+k\right)
-k\right)  \geq I_{\psi}\left(  \varphi+k\right)  -k
\]
yielding that $I_{\psi}\left(  \varphi+k\right)  \leq I_{\psi}\left(
\varphi \right)  +k$.}

We move\ to prove that\ $I_{\psi}$ is concave. Consider $\varphi_{1}%
,\varphi_{2}\in B_{0}\left(  \Sigma \right)  $ and $\lambda \in \left(
0,1\right)  $. By definition of $I_{\psi}$, we have that%
\[
\varphi_{1}-I_{\psi}\left(  \varphi_{1}\right)  \in U\left(  \psi \right)
\text{ and }\varphi_{2}-I_{\psi}\left(  \varphi_{2}\right)  \in U\left(
\psi \right)
\]
Since $U\left(  \psi \right)  $ is convex, we have that%
\begin{align*}
&  \left(  \lambda \varphi_{1}+\left(  1-\lambda \right)  \varphi_{2}\right)
-\left(  \lambda I_{\psi}\left(  \varphi_{1}\right)  +\left(  1-\lambda
\right)  I_{\psi}\left(  \varphi_{2}\right)  \right) \\
&  =\lambda \left(  \varphi_{1}-I_{\psi}\left(  \varphi_{1}\right)  \right)
+\left(  1-\lambda \right)  \left(  \varphi_{2}-I_{\psi}\left(  \varphi
_{2}\right)  \right)  \in U\left(  \psi \right)
\end{align*}
yielding that $\lambda I_{\psi}\left(  \varphi_{1}\right)  +\left(
1-\lambda \right)  I_{\psi}\left(  \varphi_{2}\right)  \in C_{\lambda
\varphi_{1}+\left(  1-\lambda \right)  \varphi_{2}}$ and, in particular,
$I_{\psi}\left(  \lambda \varphi_{1}+\left(  1-\lambda \right)  \varphi
_{2}\right)  \geq \lambda I_{\psi}\left(  \varphi_{1}\right)  +\left(
1-\lambda \right)  I_{\psi}\left(  \varphi_{2}\right)  $.

Finally, since $\psi \in U\left(  \psi \right)  $, note that $0\in C_{\psi}$ and
$I_{\psi}\left(  \psi \right)  \geq0$. By definition of $I_{\psi}$, if
$I_{\psi}\left(  \psi \right)  >0$, then $\psi-I_{\psi}\left(  \psi \right)  \in
U\left(  \psi \right)  $, a contradiction with property 3 of Lemma
\ref{lem:upp-con}.

\smallskip

1. It is routine to check that $\bar{I}_{\psi}$ is a normalized concave
niveloid which is $\succeq^{\ast}$-consistent.

\smallskip

2. Clearly, we have that if $\psi \sim^{\ast}\psi^{\prime}$, then $U\left(
\psi \right)  =U\left(  \psi^{\prime}\right)  $, yielding that $I_{\psi
}=I_{\psi^{\prime}}$ and, in particular, $I_{\psi}\left(  0\right)
=I_{\psi^{\prime}}\left(  0\right)  $ as well as $\bar{I}_{\psi}=\bar{I}%
_{\psi^{\prime}}$. The point trivially follows.\hfill$\blacksquare$

\smallskip

\noindent \textbf{Proof of Proposition \ref{pro:car}\ }We begin by observing
that:%
\[
\left \vert ca\left(  \Sigma \right)  \right \vert \leq \left \vert ca_{+}\left(
\Sigma \right)  \times ca_{+}\left(  \Sigma \right)  \right \vert =\left \vert
ca_{+}\left(  \Sigma \right)  \right \vert =\left \vert \left(  0,\infty \right)
\times \Delta^{\sigma}\right \vert =\left \vert \Delta^{\sigma}\right \vert
\]
The first inequality holds because the map $g:ca\left(  \Sigma \right)
\rightarrow ca_{+}\left(  \Sigma \right)  \times ca_{+}\left(  \Sigma \right)
$, defined by $\mu \mapsto \left(  \mu^{+},\mu^{-}\right)  $, is injective. By
Theorem 1.4.5 of Srivastava (1998) and since $\Sigma$ is non-trivial, we have
that $ca_{+}\left(  \Sigma \right)  $ is infinite, yielding that a bijection
justifying the first equality exists. As to the second equality, the map
$g:ca_{+}\left(  \Sigma \right)  \backslash \left \{  0\right \}  \rightarrow
\left(  0,\infty \right)  \times \Delta^{\sigma}$, defined by $\mu \mapsto \left(
\mu \left(  S\right)  ,\mu/\mu \left(  S\right)  \right)  $, is a bijection and
so $\left \vert ca_{+}\left(  \Sigma \right)  \backslash \left \{  0\right \}
\right \vert =\left \vert \left(  0,\infty \right)  \times \Delta^{\sigma
}\right \vert $. By Theorem 1.3.1 of Srivastava (1998), we can conclude that
$\left \vert ca_{+}\left(  \Sigma \right)  \right \vert =\left \vert ca_{+}\left(
\Sigma \right)  \backslash \left \{  0\right \}  \right \vert =\left \vert \left(
0,\infty \right)  \times \Delta^{\sigma}\right \vert $. As to the last equality,
by Theorem 1.4.5 and Exercise 1.5.1 of Srivastava (1998), being $\left \vert
\left(  0,\infty \right)  \right \vert =\left \vert \left(  0,1\right)
\right \vert \leq \left \vert \Delta^{\sigma}\right \vert $, we have $\left \vert
\Delta^{\sigma}\right \vert \leq \left \vert \left(  0,\infty \right)
\times \Delta^{\sigma}\right \vert =\left \vert \left(  0,1\right)  \times
\Delta^{\sigma}\right \vert \leq \left \vert \Delta^{\sigma}\times \Delta^{\sigma
}\right \vert =\left \vert \Delta^{\sigma}\right \vert $, yielding that
$\left \vert \left(  0,\infty \right)  \times \Delta^{\sigma}\right \vert
=\left \vert \Delta^{\sigma}\right \vert $.

We conclude that $\left \vert ca\left(  \Sigma \right)  \right \vert
\leq \left \vert \Delta^{\sigma}\right \vert $, that is, there exists an
injective map $g:ca\left(  \Sigma \right)  \rightarrow \Delta^{\sigma}$. Since
$Q$ is a compact and convex subset of $\Delta^{\sigma}$, there exists $\bar
{q}\in Q$ such that $q\ll \bar{q}$ for all $q\in Q$.\ We define $h:V\rightarrow
ca\left(  \Sigma \right)  $ by%
\[
h\left(  \left[  \psi \right]  \right)  \left(  A\right)  =\int_{A}\psi
d\bar{q}\qquad \forall A\in \Sigma
\]
Note that $h$ is well defined. For, if $\psi^{\prime}\in \left[  \psi \right]
$, that is, $\psi \overset{Q}{=}\psi^{\prime}$, then $\psi \overset{\bar{q}}%
{=}\psi^{\prime}$, yielding that $\int_{A}\psi d\bar{q}=\int_{A}\psi^{\prime
}d\bar{q}$ for all $A\in \Sigma$. Similarly, $h\left(  \left[  \psi \right]
\right)  =h\left(  \left[  \psi^{\prime}\right]  \right)  $ implies that
$\psi \overset{\bar{q}}{=}\psi^{\prime}$. Since $q\ll \bar{q}$ for all $q\in
Q$,\ this implies that $\psi \overset{Q}{=}\psi^{\prime}$ and $\left[
\psi \right]  =\left[  \psi^{\prime}\right]  $, proving $h$ is injective. This
implies that $\tilde{f}=g\circ h$ is a well defined injective function from
$V$ to $\Delta^{\sigma}$. Clearly, we have that $\left \vert \Delta^{\sigma
}\right \vert \geq \left \vert \tilde{f}\left(  V\right)  \right \vert
\geq \left \vert \left[  0,1\right]  \right \vert $. Since $\left(
S,\Sigma \right)  $ is a standard Borel space and $Q$ is convex and $\left \vert
Q\right \vert \geq2$, we also have that $\left \vert \left[  0,1\right]
\right \vert \geq \left \vert \Delta^{\sigma}\right \vert \geq \left \vert
Q\right \vert \geq \left \vert \left[  0,1\right]  \right \vert $. This implies
that $\left \vert V\right \vert =\left \vert \tilde{f}\left(  V\right)
\right \vert =\left \vert Q\right \vert $, proving the statement.\hfill
$\blacksquare$

\subsection{Analysis of the decision criterion: missing
proofs\label{app:anc-ana}}

The proof of Proposition \ref{prop:comput} follows from the following lemma.
Here, as usual, $\phi$ is extended to $\mathbb{R}$ by setting $\phi \left(
t\right)  =+\infty$ if $t\notin \left[  0,\infty \right)  $. In particular,
$\phi^{\ast}$ is non-decreasing.

\begin{lemma}
For each $Q\subseteq \Delta^{\sigma}$ and each $\lambda \in(0,\infty)$,%
\[
\inf_{p\in \Delta}\left \{  \int u\left(  f\right)  dp+\lambda \inf_{q\in
Q}D_{\phi}(p||q)\right \}  =\lambda \inf_{q\in Q}\sup_{\eta \in \mathbb{R}%
}\left \{  \eta-\int \phi^{\ast}\left(  \eta-\frac{u\left(  f\right)  }{\lambda
}\right)  dq\right \}
\]
for all $u:X\rightarrow \mathbb{R}$ and all $f:S\rightarrow X$ such that
$u\circ f$ is bounded and $\Sigma$-measurable.
\end{lemma}

\noindent \textbf{Proof} By Theorem 4.2 of Ben-Tal and Teboulle (2007), for
each $q\in \Delta^{\sigma}$ it holds%
\[
\inf_{p\in \Delta}\left \{  \int \xi dp+D_{\phi}(p||q)\right \}  =\sup_{\eta
\in \mathbb{R}}\left \{  \eta-\int \phi^{\ast}\left(  \eta-\xi \right)
dq\right \}
\]
for all $\xi \in L^{\infty}\left(  q\right)  $. Then, if $u\circ f$ is bounded
and measurable, from $u\circ f\in L^{\infty}\left(  q\right)  $ for all
$q\in \Delta^{\sigma}$, it follows that%
\begin{align*}
\inf_{p\in \Delta}\left \{  \int u\left(  f\right)  dp+\lambda D_{\phi
}(p||q)\right \}   &  =\lambda \inf_{p\in \Delta}\left \{  \int \frac{u\left(
f\right)  }{\lambda}dp+D_{\phi}(p||q)\right \} \\
&  =\lambda \sup_{\eta \in \mathbb{R}}\left \{  \eta-\int \phi^{\ast}\left(
\eta-\frac{u\left(  f\right)  }{\lambda}\right)  dq\right \}
\end{align*}
for all $\lambda>0$, as desired. By taking the $\inf$ over $Q$ on both sides
of the equation, the statement follows.\hfill$\blacksquare$

\smallskip

\noindent \textbf{Proof of Proposition \ref{prop:comput}} In view of the last
lemma, it is enough to observe that, if $f:S\rightarrow X$ is simple and
measurable, then $u\circ f$ is simple and $\Sigma$-measurable for all
$u:X\rightarrow \mathbb{R}$ and the infima are achieved.\hfill$\blacksquare$

\bigskip

\noindent \textbf{Proof of Proposition \ref{prop:lim}} First, note that
$\min_{q\in Q}R\left(  p||q\right)  =0$ if and only if $p\in Q$. Indeed, we
have that%
\[
\min_{q\in Q}R\left(  p||q\right)  =0\iff \exists \bar{q}\in Q\text{ s.t.
}R\left(  p||\bar{q}\right)  =0\iff \exists \bar{q}\in Q\text{ s.t. }p=\bar{q}%
\]
Define $\lambda_{n}=n$ for all $n\in%
\mathbb{N}
$. For each\ $n\in%
\mathbb{N}
$, we have $\lambda_{n}\min_{q\in Q}R\left(  p||q\right)  =0$ if and only if
$p\in Q$. So, for each $p\in \Delta$,
\[
\lim_{n}\lambda_{n}\min_{q\in Q}R\left(  p||q\right)  =\left \{
\begin{array}
[c]{cc}%
0 & \text{if }p\in Q\\
+\infty & \text{if }p\not \in Q
\end{array}
\right.
\]
Since $\lambda_{n}\min_{q\in Q}R\left(  p||q\right)  =0$ for each $n\in%
\mathbb{N}
$ if and only if $p\in Q$, by Proposition 12 of Maccheroni et al. (2006) we
have%
\[
\lim_{n}\min_{p\in \Delta}\left \{  \int u\left(  f\right)  dp+\lambda_{n}%
\min_{q\in Q}R\left(  p||q\right)  \right \}  =\min_{q\in Q}\int u\left(
f\right)  dq\quad \forall f\in \mathcal{F}%
\]
Finally, by (\ref{eq:com-amb}), we have that for each $f\in \mathcal{F}$%
\begin{align*}
\min_{q\in Q}\int u\left(  f\right)  dq  &  \leq \lim_{n}\min_{p\in \Delta
}\left \{  \int u\left(  f\right)  dp+\lambda_{n}\min_{q\in Q}R\left(
p||q\right)  \right \} \\
&  \leq \lim_{\lambda \uparrow \infty}\min_{p\in \Delta}\left \{  \int u\left(
f\right)  dp+\lambda \min_{q\in Q}R\left(  p||q\right)  \right \}  \leq
\min_{q\in Q}\int u\left(  f\right)  dq
\end{align*}
yielding the statement.\hfill$\blacksquare$

\smallskip

\noindent \textbf{Proof of Proposition \ref{pro:adm}} (i) Let $\hat{f}\in F$ be
optimal. By (\ref{eq:dom-rat}), if there is $g\in F$ such that $g\succ
\hspace{-5pt}\succ_{Q}^{\ast}\hat{f}$, then $g\succ_{Q}\hat{f}$, a
contradiction with $\hat{f}$ being optimal. We conclude that $\hat{f}$ is
weakly admissible. A similar argument proves that there is no $g\in F$ such
that $g\succ_{Q}^{\ast}\hat{f}$ when (\ref{eq:strict-pref}) holds.

\smallskip

(ii) Suppose $\hat{f}\in F$ is the unique optimal act, that is, $\hat{f}%
\succ_{Q}f$ for all $f\in F\backslash \left \{  \hat{f}\right \}  $. If $g\in F$
is such that $g\succ_{Q}^{\ast}\hat{f}$, then $g\not =\hat{f}$\ and
$g\succsim_{Q}\hat{f}$. In turn, this implies $g\succsim_{Q}\hat{f}\succ_{Q}%
g$, a contradiction. We conclude that $\hat{f}$ is admissible.\hfill
\hfill \hfill \hfill$\blacksquare$

\smallskip

\noindent \textbf{Proof of Proposition \ref{pro:com-sta-val}} Since $Q\subseteq
Q^{\prime}$, it follows that $\min_{q\in Q}c\left(  p,q\right)  \geq \min_{q\in
Q^{\prime}}c\left(  p,q\right)  $ for all $p\in \Delta$.\ We thus have%
\[
\min_{p\in \Delta}\left \{  \int u\left(  f\right)  dp+\min_{q\in Q}c\left(
p,q\right)  \right \}  \geq \min_{p\in \Delta}\left \{  \int u\left(  f\right)
dp+\min_{q\in Q^{\prime}}c\left(  p,q\right)  \right \}  \qquad \forall f\in F
\]
yielding that $v\left(  Q\right)  \geq v\left(  Q^{\prime}\right)  $. Next,
fix $Q$ and assume that the $\sup$ in (\ref{eq:dec-pro}) is achieved. Let
$\bar{f}\in F$ be such that%
\[
\min_{p\in \Delta}\left \{  \int u\left(  \bar{f}\right)  dp+\min_{q\in
Q}c\left(  p,q\right)  \right \}  =v\left(  Q\right)
\]
By contradiction, assume that $\bar{f}\in F/F_{Q}^{\ast}$. By Proposition
\ref{prop:strong-dom-ch} and since $\bar{f}\not \in F_{Q}^{\ast}$ and $\bar
{f}\in F$, there exists $g\in F$ such that $g\succ \hspace{-5pt}\succ_{Q}%
^{\ast}\bar{f}$, that is, there exists $\varepsilon>0$ such that%
\[
\min_{p\in \Delta}\left \{  \int u\left(  g\right)  dp+c\left(  p,q\right)
\right \}  \geq \min_{p\in \Delta}\left \{  \int u\left(  \bar{f}\right)
dp+c\left(  p,q\right)  \right \}  +\varepsilon \qquad \forall q\in Q
\]
Since $g$ is finite-valued, this implies that $v\left(  Q\right)  <\infty
$\ and%
\begin{align*}
v\left(  Q\right)   &  \geq \min_{p\in \Delta}\left \{  \int u\left(  g\right)
dp+\min_{q\in Q}c\left(  p,q\right)  \right \}  =\min_{p\in \Delta}\min_{q\in
Q}\left \{  \int u\left(  g\right)  dp+c\left(  p,q\right)  \right \} \\
&  \geq \inf_{q\in Q}\min_{p\in \Delta}\left \{  \int u\left(  g\right)
dp+c\left(  p,q\right)  \right \}  \geq \inf_{q\in Q}\min_{p\in \Delta}\left \{
\int u\left(  \bar{f}\right)  dp+c\left(  p,q\right)  \right \}  +\varepsilon \\
&  \geq \min_{p\in \Delta}\min_{q\in Q}\left \{  \int u\left(  \bar{f}\right)
dp+c\left(  p,q\right)  \right \}  +\varepsilon=\min_{p\in \Delta}\left \{  \int
u\left(  \bar{f}\right)  dp+\min_{q\in Q}c\left(  p,q\right)  \right \}
+\varepsilon \\
&  =v\left(  Q\right)  +\varepsilon
\end{align*}
a contradiction.\hfill$\blacksquare$

\end{document}